\documentclass{article}

\usepackage{amssymb}

\usepackage{setspace}

\usepackage{cite}
\usepackage{amsmath,amssymb,amsfonts,bm, mleftright}
\usepackage{algorithm}
\usepackage{algpseudocode}         
\usepackage{graphicx}
\usepackage{textcomp}
\usepackage{xcolor}
\usepackage{amsthm}
\usepackage{epstopdf}
\usepackage{array}
\usepackage{tabularx} 
\usepackage{setspace}
\usepackage{autobreak}
\usepackage{multirow}
\usepackage{booktabs}
\usepackage{comment}
\usepackage{siunitx}

\usepackage{caption}
\usepackage{subcaption}
\usepackage{hyperref}
\usepackage{url}
\usepackage{array}
\newcolumntype{P}[1]{>{\small\centering\arraybackslash}p{#1}}
\newcolumntype{M}[1]{>{\small\centering\arraybackslash}m{#1}}
\hypersetup{
    colorlinks,
    linkcolor={black!50!black},
    citecolor={black!50!black},
    urlcolor={blue!80!blue}
}
\newcommand{\s}{\mathbf{s}}
\newcommand{\y}{\mathbf{y}}
\newcommand{\w}{\mathbf{w}}
\newcommand{\A}{\mathbf{A}}

\usepackage{nicefrac, xfrac}
\usepackage{adjustbox}
\usepackage{float}
\usepackage{multicol}
\usepackage{multirow}
\newcommand{\R}{\mathbb{R}}

\usepackage{lineno}
\usepackage{arxiv}

\usepackage[utf8]{inputenc} 
\usepackage[T1]{fontenc}   
\usepackage{hyperref}       
\usepackage{url}           
\usepackage{booktabs}      
\usepackage{amsfonts}       
\usepackage{nicefrac}       
\usepackage{microtype}      
\usepackage{lipsum}
\usepackage{graphicx}
\graphicspath{ {./images/} }

\newcommand\blfootnote[1]{%
  \begingroup
  \renewcommand\thefootnote{}\footnote{#1}%
  \addtocounter{footnote}{-1}%
  \endgroup
}

\title{Efficient Physics-Based Learned Reconstruction Methods 
for Real-Time 3D Near-Field MIMO Radar Imaging}
\author{
 Irfan Manisali\\
        Department of Electrical and Electronics Engineering\\
        Middle East Technical University (METU)\\
        Ankara, 06800, Turkey \\
        \texttt{irfan.manisali@metu.edu.tr} \\
        \And
        Okyanus Oral\\
        Department of Electrical and Electronics Engineering\\
        Middle East Technical University (METU)\\
        Ankara, 06800, Turkey \\
        \texttt{ookyanus@metu.edu.tr} \\
	\And
        Figen S. Oktem \\
        Department of Electrical and Electronics Engineering\\
        Middle East Technical University (METU)\\
        Ankara, 06800, Turkey \\
        \texttt{figeno@metu.edu.tr} \\
}

\begin{document}
\maketitle

\blfootnote{Funding: This work was supported by the Scientific and Technological Research Council of Turkey (TUBITAK) under Grant 120E505 (1001 Research Program).}

\blfootnote{Accepted for publication in Digital Signal Processing. © 2023. This manuscript version is made available under the CC-BY-NC-ND 4.0 license \href{https://creativecommons.org/licenses/by-nc-nd/4.0/}{https://creativecommons.org/licenses/by-nc-nd/4.0/}}

\title{Efficient Physics-Based Learned Reconstruction Methods
for Real-Time 3D Near-Field MIMO Radar Imaging}

\begin{abstract}
Near-field multiple-input multiple-output (MIMO) radar imaging systems have recently gained significant attention. These systems generally reconstruct the three-dimensional (3D) complex-valued reflectivity distribution of the scene using sparse measurements. Consequently, imaging quality highly relies on the image reconstruction approach. Existing analytical reconstruction approaches suffer from either high computational cost or low image quality. In this paper, we develop novel non-iterative deep learning-based reconstruction methods for real-time near-field MIMO imaging. The goal is to achieve high image quality with low computational cost at compressive settings. The developed approaches have two stages. In the first approach, physics-based initial stage performs adjoint operation to back-project the measurements to the image-space, and deep neural network (DNN)-based second stage converts the 3D backprojected measurements to a magnitude-only reflectivity image. Since scene reflectivities often have random phase, DNN processes directly the magnitude of the adjoint result. As DNN, 3D U-Net is used to jointly exploit range and cross-range correlations. To comparatively evaluate the significance of exploiting physics in a learning-based approach, two additional approaches that replace the physics-based first stage with fully connected layers are also developed as purely learning-based methods. The performance is also analyzed by changing the DNN architecture for the second stage to include complex-valued processing (instead of magnitude-only processing), 2D convolution kernels (instead of 3D), and ResNet architecture (instead of U-Net). Moreover, we  develop a synthesizer to generate large-scale dataset for training with 3D extended targets. We illustrate the performance through experimental data and extensive simulations. The results show the effectiveness of the developed physics-based learned reconstruction approach compared to commonly used approaches in terms of both run-time and image quality at highly compressive settings. Our source codes and dataset are made available at \href{https://github.com/METU-SPACE-Lab/Efficient-Learned-3D-Near-Field-MIMO-Imaging}{https://github.com/METU-SPACE-Lab/Efficient-Learned-3D-Near-Field-MIMO-Imaging} upon publication to advance research in this field.
\end{abstract}


\section{Introduction}

Near-field radar imaging systems are of interest in diverse fields such as airport security, surveillance, concealed weapon detection, through-wall imaging, and medicine~\cite{ullah2023multistatic,zhang2023MIMO,mamandipoor2019millimeter, zhuge2010sparse, klemm2010microwave, sheen2001three,anadol2018uwb,li2015near}. Earlier near-field radar imaging systems have operated in monostatic mode using collocated transmitter and receiver antennas~ \cite{sheen2001three, ahmed2012advanced,mamandipoor2019millimeter}. In these monostatic systems, a large number of transceiver antennas are required to achieve high image quality with fine range and cross-range resolutions. This gives rise to undesirably long acquisition time, as well as high hardware complexity and cost~\cite{yanik2019near,hu2017matrix}.

Recently, sparse multiple-input multiple-output (MIMO) arrays (i.e. multistatic arrays) that contain spatially distributed transmit and receive antennas have gained more attention since they enable high resolution with reduced cost, acquisition time, and hardware complexity. The number of antennas in MIMO systems can be significantly reduced compared to the monostatic case while maintaining high image quality~\cite{zhuge2010sparse,zhuge2012study,yanik2019near, anadol2018uwb, ahmed2012advanced,kocamis2017optimal, hu2017matrix, zhang2023MIMO}.

Near-field radar imaging systems, operating in monostatic or multistatic mode, are computational imaging systems that reconstruct the complex-valued 3D reflectivity distribution of the scene from the acquired radar data. As a result, the imaging performance highly depends on the underlying image reconstruction method. Various analytical reconstruction methods exist for this purpose. These methods can be grouped into two categories: traditional direct inversion methods and iterative regularized reconstruction methods. 

Traditional direct inversion methods aim to obtain a direct solution for the forward (observation) model equation without using any prior information. These methods generally involve back projecting the measurements to the object plane using a coherent summation
and a filter-like operation~\cite{jin2017deep, liu2017mimo}. These direct inversion approaches are related to applying the adjoint of the forward operator to the measurements~\cite{marks2017fourier}. Back-projection (delay-and-sum), filtered back-projection, Kirchhoff migration methods, and their variants
can be given as examples of this class of methods, in addition to the Fourier-based alternative techniques such as range migration and phase shift migration to accelerate the direct inversion process~\cite{zhuge2010modified,wang2023efficient, alvarez2016fourier, zhuge2012three, tan2018omega, marks2017fourier, liu2017mimo, yang2021mimo}. 
These traditional methods generally have low computational complexity but they suffer from reconstruction artifacts when observations are noisy and/or limited (as acquired with sparse MIMO arrays).

Regularized reconstruction methods on the other hand incorporate additional prior information (such as sparsity) into the reconstruction process. 
With the advent of compressed sensing (CS) theory \cite{candes2008introduction}, 
sparsity-based reconstruction methods are the most commonly used model-based iterative inversion 
methods 
in various imaging problems including radar imaging, 
both for far-field and monostatic imaging settings \cite{wei2013sparse, potter2010sparsity, guven2016augmented, ma2014mimo, guo2015microwave,  ma2018multiple,  huang2018tensor, li2015near}, 
as well as for multistatic and near-field settings \cite{zhang2015generalized, oktem2019sparsity,cheng2017near,miran2021sparse}.
Although sparsity-based 
methods provide better reconstruction quality than the traditional direct inversion methods at compressive settings, they suffer from high computational cost 
which is undesirable in real-time applications. 
Moreover, regularized methods in the near-field radar imaging literature generally enforce smoothness or sparsity on the complex-valued reflectivity distribution~\cite{li2015near,oktem2019sparsity}. These methods are therefore built on the assumption that the phase and magnitude of the scene reflectivity exhibit local correlations. However, for many applications, the phase of the reflectivity at a particular point can be more accurately modeled as random and uncorrelated with the phase at other points~\cite{Munson1984offset,cetin2001feature}. This is because phase shift can occur when imaging rough surfaces and also at the air/target interface due to the electrical properties of materials~\cite{Munson1984offset}. As a result, enforcing regularization on the magnitude of the 
 scene reflectivity can yield better 
 performance than enforcing it directly on the complex-valued reflectivity~\cite{cetin2001feature,alver2021plug, guven2016augmented}. 

Recently, reconstruction techniques that exploit deep learning have emerged as an alternative to the analytical ones~\cite{lopez2021deep}. These methods are shown to simultaneously 
achieve high reconstruction quality and low computational cost for various imaging problems~\cite{lopez2021deep,jin2017deep,ongie2020, lucas2018using}. 
Deep learning-based reconstruction approaches in the literature can be grouped into three main classes: 1) learning-based direct inversion, 2) plug-and-play regularization, 3) learned iterative reconstruction based on unrolling.

Learning-based direct inversion methods are aimed to perform the reconstruction directly from the measurements using a deep neural network. Hence the neural network is trained to learn the direct mapping from the observations to the desired image solely using training data. Although these methods yield the state-of-the-art performance for simpler inverse problems like denoising~\cite{lucas2018using}, 
they can not provide successful results whenever the observation model is complex, the unknown image does not look alike observations, or there is not a lot of training data available. For this reason, commonly an efficient approximate inverse of the forward model is first applied analytically and these initial images are provided to the network as a warm start. Subsequently a deep neural network is employed to refine this initial reconstruction~\cite{ongie2020,jin2017deep, lucas2018using,lopez2021deep}. This type of physics-based learning approaches are hybrid approaches that combine the neural networks with analytical methods, and they have been successfully applied to various 2D and real-valued reconstruction problems in imaging such as deconvolution, super-resolution, tomography, and phase-retrieval~\cite{lopez2021deep,ongie2020,jin2017deep, lucas2018using, isil2019deep, sinha2017lensless}. An important advantage of learning-based direct inversion methods is their low computational complexity due to their feed-forward (non-iterative) nature, which is ideal for real-time imaging applications. 

On the other hand, plug-and-play regularization and unrolling-based deep-learning methods are iterative methods. The key idea in these approaches is generally to replace the hand-crafted analytical priors in model-based regularized reconstruction methods with data-driven deep priors. In plug-and-play methods, a deep prior is first learned from training data and then utilized for the regularization of a model-based iterative inversion method. Although plug-and-play methods can perform better than learning-based direct inversion methods in terms of image quality, flexibility, and generalizability, these approaches generally require higher memory usage and computational complexity due to their iterative nature and requiring computation of the forward (system) operator and its adjoint at every iteration.
In unrolling-based reconstruction, iterative methods that utilize proximal operators or deep-priors, such as plug-and-play, are unrolled into an end-to-end trainable network to further improve the imaging quality~\cite{ongie2020,aggarwal2018modl,Yang2020ADMM-CSNet,adler2018learned}.
Therefore, similar to the plug-and-play methods, these learned iterative reconstruction methods generally require high computational complexity when they involve computation with large sensing matrices to evaluate the forward operator and its adjoint. Moreover, unlike plug-and-play and direct inversion methods, unrolling-based approaches require the computation of the forward and adjoint operators also during training. This causes substantial increase in
training time and complexity, possibly making it impractical for 3D large-scale reconstruction problems with no fast forward solver. These approaches have also been applied to various 2D and real-valued reconstruction problems in imaging including deconvolution, super-resolution, image inpainting, tomography, and phase retrieval~\cite{ongie2020, lucas2018using, adler2017solving, aggarwal2018modl,Yang2020ADMM-CSNet}. 

In the near-field MIMO radar imaging case, we encounter a 3D complex-valued reconstruction problem where the unknown 3D image has random phase in various applications. Due to these differences, existing approaches for 2D and real-valued imaging problems are not directly applicable to this inverse problem. Despite this, deep learning-based reconstruction methods have not been studied much in the literature for near-field radar imaging. Most of the proposed methods are for far-field settings in SAR/ISAR or MIMO radar imaging~\cite{alver2021plug, hu2020inverse, peng2019generating, wang2021single, weiss2021joint, Gao2019Enhanced, sun2021photonics, yang2020isar,mu2020deepimaging}. In the near-field and MIMO radar imaging context, there are few works for learning-based approaches~\cite{cheng2020compressive, wang2021rmist}, but to the best of our knowledge, there is no DNN-based reconstruction approach developed and shown to be successful for imaging 3D extended targets with random phase.

In this paper, we develop three novel deep learning-based methods to reconstruct the 3D scene reflectivity from the near-field observations of a MIMO imaging radar. The main goal is to achieve high image quality with low computational cost so that the developed method can be used in real-time applications. Learning-based direct inversion methods enable such capabilities. For this reason, the developed approaches are based on learned direct reconstruction. 

The first approach has a two-stage structure that consists of an adjoint operation followed by a 3D U-Net architecture. The adjoint stage exploits the physical observation model of the system and back project the measurements to the reconstruction
space. The second stage employs a 3D deep neural network which is trained to convert the backprojected measurements to a magnitude-only reflectivity image. Due to the random phase nature of the scene reflectivities in various applications, the second stage processes directly the magnitude of the adjoint result from the first stage. As DNN, a 3D U-Net architecture that makes use of multi-resolution learning is used to jointly exploit the correlations along range and cross-range dimensions. 

For comparison, a second approach is also developed which replaces the physics-based first stage with a fully connected neural network. In this two-stage structure, the reconstruction is performed directly from the radar measurements using only neural networks, and the observation model is not used. Moreover, to investigate the effect of physics-based initialization on the training of a purely learning-based method, we also present another approach that replaces the adjoint operation in the first stage with a fully connected layer to perform multiplication with a trainable matrix which is learned through transfer learning from the adjoint matrix. These architectures are trained end-to-end to learn the direct mapping between the radar measurements and the reflectivity magnitudes.

We illustrate the performance of the developed
methods using both simulated and experimental data. For training the DNNs, a large synthetic dataset 
consisting of 3D extended targets is randomly generated and used. 
We also add random phase to our synthetically generated dataset to analyze the performance on a more realistic
scenario that takes into account the complex-valued and random phase nature of scene reflectivities,
which have been mostly neglected in the earlier works. 
For performance analysis, we first consider various compressive observation scenarios using simulated data, and
compare the performance with backprojection and
sparsity-based CS reconstructions. 
We also investigate the effect of measurement SNR. Moreover, the performance is comparatively analyzed by changing the DNN architecture for the second stage to include complex-valued processing (instead of magnitude-only
processing), 2D convolution kernels (instead of 3D), and ResNet architecture (instead of U-Net).
Furthermore, we investigate the resolution achieved with our reconstruction approach at compressive MIMO imaging settings and compare this to the expected theoretical resolution for the conventional (non-compressive) settings.
Lastly, we illustrate the performance with experimental data to demonstrate applicability to real-world targets and measurements.
The results show the effectiveness of the developed physics-based approach compared to commonly used approaches in terms of both computation time and image quality when observations are limited. 

The main contributions of this paper can be summarized as follows:
\begin{itemize}
\item Development of three novel 3D image reconstruction methods for near-field MIMO radar imaging using deep learning-based direct inversion and 3D convolutional neural networks
\item Development of a synthesizer to generate 
3D scenes with extended targets 
to obtain large data for training the neural networks
\item Comprehensive experiments on synthetic 3D scenes with quantitative and qualitative analysis by considering different compression and noise levels in the observations 
\item Comparative performance evaluation by changing magnitude-only processing with complex-valued processing, 3D convolution kernels with their 2D counterparts, and U-Net architecture with ResNet
\item Resolution analysis at compressive imaging settings with sparse MIMO arrays
\item Performance analysis with experimental data, and comparison with the commonly used direct inversion and regularized reconstruction methods 
\end{itemize}

Compared with the previous works for near-field MIMO radar imaging, the developed physics-based approach offers promising imaging performance with reduced computation time at compressive settings. In fact, compared to back-projection and sparsity-based methods, the developed physics-based approach with adjoint operation achieves the best reconstruction quality both visually and quantitatively, while also enabling fast reconstruction. Since all adjustable parameters of the developed approaches are learned end-to-end, there is also no need for computationally intensive parameter tuning unlike regularized (i.e. sparsity-based) reconstruction methods.

Different than the related learning-based works in near-field MIMO radar imaging~\cite{cheng2020compressive, wang2021rmist}, our method is a DNN-based approach developed for imaging 3D extended targets. In particular, the work in \cite{cheng2020compressive} develops a deep learning-based direct inversion method to reconstruct isolated point scatterers from near-field MIMO radar data. In this approach, the magnitude and phase of the backprojection images are separately processed using 2D-convolutional layer blocks. Because the used network performs 2D processing only (using 2D kernels that work on the cross-range dimensions), this method can not exploit range correlations unlike our approach, which is particularly important for reconstructing extended 3D targets. Moreover, both training and testing are performed only with unrealistic scenes that consist of isolated point scatterers. Hence this method is not applicable for imaging complex extended (distributed) targets as encountered in practice. The other learning-based work in \cite{wang2021rmist} is an unrolling-based method developed for near-field MIMO-SAR imaging. But unlike our approach, this method is not a DNN-based approach and only learns the step-size and soft-threshold parameters of the unrolled iterative shrinkage threshold algorithm. 

Related learning-based works in far-field SAR/ISAR imaging~\cite{Gao2019Enhanced,yang2020isar, mu2020deepimaging, hu2020inverse} are developed for 2D complex-valued reconstruction problems based on the refinement of initial images with 2D network architectures. On the other hand, our approaches are developed for reconstruction of complex-valued 3D extended targets using 3D network architectures. While the work in \cite{jing2022enhanced} also develops a 3D reconstruction method for millimeter-wave imaging using a network architecture involving successive 3D convolutional blocks, this approach has been trained with only point scatterers similar to the earlier SAR/ISAR works. That is, in none of the earlier related approaches, the training is performed with 3D extended targets involving random phase. In contrast, we develop a synthesizer to generate large-scale dataset specifically for this purpose. Moreover, all of these earlier works are based on the refinement of initial images with DNNs, which shares the same idea with one of the approaches developed in this paper although our approach differs with its usage of 3D U-Net that enables multi-resolution learning. The other two approaches developed are purely learning-based methods which are not studied before. We, for the first time, comparatively evaluate the performance of these different type of approaches in a fair setting both using simulated and experimental data. Furthermore,  
compared to the learning-based works~\cite{Gao2019Enhanced, jing2022enhanced, yang2020isar,mu2020deepimaging} that refine the intermediate reconstructions in complex-valued form by either processing real and imaginary parts as two independent channels or using complex-valued CNNs, our approaches simply process only the magnitudes of intermediate reconstructions by taking into account the random phase nature of the reflectivities in various applications.
To the best of our knowledge, there is no work that compared the performance of complex-valued refinement with magnitude-only refinement 
to reconstruct reflectivities involving random phase, and showed the superiority of one of them to the other. For this reason, we also perform this comparative evaluation that has been missing in the earlier works.

The paper is organized as follows. We present the observation model in Section~\ref{section:Observation Model}, and discuss the inverse problem and the related analytical image reconstruction methods in Section~\ref{section:Related Work}. The developed deep learning-based methods are presented in Section~\ref{section:Image Reconstruction}. Section~\ref{section:Numerical Results} demonstrates the performance of the developed methods under various observation scenarios using both simulated data and experimental measurements, provides extensive comparisons with different network architectures, and presents resolution analysis for MIMO imaging with sparse arrays. Some preliminary results have also been presented in \cite{irfan2023_eusipco}. In Section~\ref{section:Conclusions}, conclusions and final remarks are provided.

\section{Observation Model}
\label{section:Observation Model}

\begin{figure}[tbh!]
	\centering
	\includegraphics[width=0.8\linewidth]{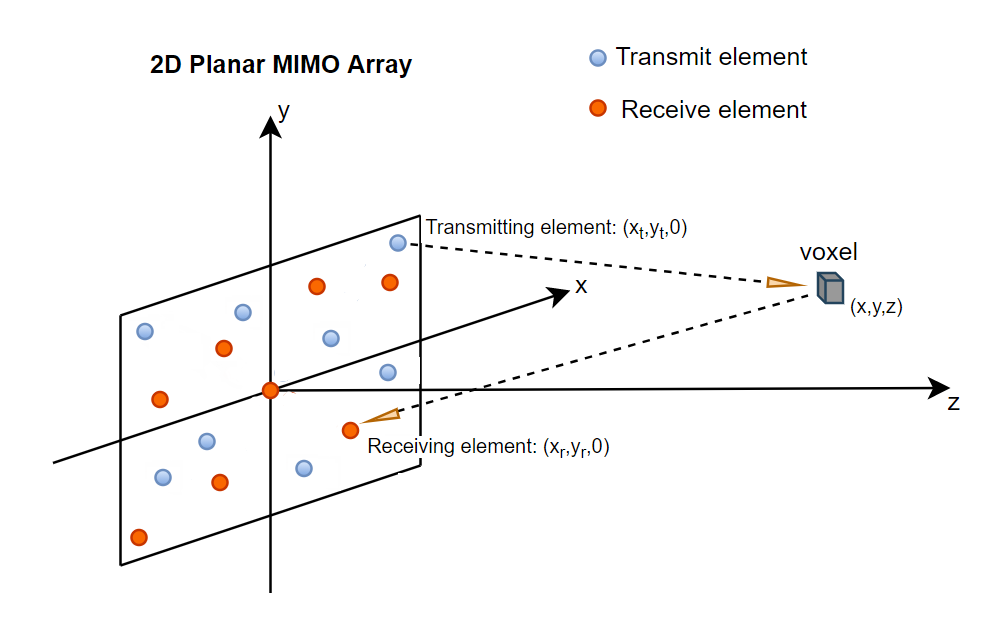}
	\caption{Near-field MIMO Radar Imaging System.}
	\label{fig:RadarImagingSystem}
\end{figure}

In this section, we present the observation model for near-field MIMO radar imaging. A sample observation geometry is given in Fig. \ref{fig:RadarImagingSystem}. As illustrated, a planar MIMO array is located at $z=0$ and contains spatially distributed transmit and receive antennas. Each transmit antenna with location $(x_{t}, y_{t}, 0)$ illuminates the scene that is in the near-field of the array. Using Born approximation for the scattered field, we can express the time-domain signal acquired by the receive antenna located at $(x_{r}, y_{r}, 0)$ due to a single scatterer at $(x,y,z)$ with reflectivity $s(x,y,z)$ as follows \cite{zhuge2012three,oktem2019sparsity}:
\begin{equation}
	r\left(x_{t}, y_{t}, x_{r}, y_{r}, t\right)= \frac{1}{4 \pi d_{t} d_{r}} \, s(x, y, z) \, p\left(t-\frac{d_{t}}{c}-\frac{d_{r}}{c}\right) 
	\label{eq:ForwardModel_time}
\end{equation}
where the distances
	$d_{t}=\left[(x_{t}-x)^{2}+(y_{t}-y)^{2}+z^{2}\right]^{1/2}$ and
	$d_{r}=\left[(x_{r}-x)^{2}+(y_{r}-y)^{2}+z^{2}\right]^{1/2}$, $p(t)$ is the transmitted radar signal, and $c$ denotes the speed of the light.
Hence $r\left(x_{t}, y_{t}, x_{r}, y_{r}, t\right)$ represents the time-domain measurement acquired with the transmitter at $(x_{t}, y_{t}, 0)$ and the receiver at $(x_{r}, y_{r}, 0)$, with $d_{t}$ and $d_{r}$ corresponding to the distances from the used transmitter and receiver to the point scatterer at $(x,y,z)$.

After taking Fourier transform, the received signal due to a single scatterer can be expressed in the temporal frequency domain as:
\begin{equation}
	r\left(x_{t}, y_{t}, x_{r}, y_{r}, k\right)= \frac{1}{4 \pi d_{t} d_{r}} \, s(x, y, z) p(k) \, e^{-j k ( d_{t}+ d_{r} )} 
\end{equation}
where $p(k)$ now represents the Fourier transform of the transmitted pulse with $k = 2\pi f/c$ denoting the frequency-wavenumber and $f$ denoting the temporal frequency. Then the total received signal $\tilde{r}\left(x_{t}, y_{t}, x_{r}, y_{r}, k\right)$ due to an extended target can be written as
\begin{equation}
	\tilde{r}\left(x_{t}, y_{t}, x_{r}, y_{r}, k\right)= \iiint \frac{1}{4 \pi d_{t} d_{r}} \, s(x, y, z) \, p(k) \,
	e^{-j k ( d_{t}+ d_{r} )} \,dx\,dy\,dz
	\label{eq:ForwardModel}
\end{equation}
where $s(x, y, z)$ represents the complex-valued 3D reflectivity function of the scene.

A discrete forward model is required to perform the image reconstruction on a computer using digitally acquired measurements. For this reason, we convert the continuous forward model in Eq.~\ref{eq:ForwardModel} to a discrete model by replacing the 3D continuous reflectivity distribution with its discretized version in terms of voxels. The size of the voxels is determined based on the desired cross-range and down-range resolutions of the imaging system. Through lexicographic ordering, the vector $\s \in C^{N}$ is formed using the voxel values of the discretized reflectivity function. Similarly the noisy measurements are put into the vector $\y \in C^{M}$ using the discrete measurements obtained with different transmitter-receiver pairs and frequency steps. We can then express the linear relation between the measurement vector $\y$ and the complex-valued image vector $\s$ as follows: 
\begin{align}
	\y = \A\s + \w
	\label{eq:ForwardModelMatrixForm}
\end{align}
Here $\w$ is the complex-valued noise vector and $\A \in C^{M \times N}$ is the observation (system) matrix with $N$ denoting the number of image voxels and $M$ denoting the number of radar measurements, given by the multiplication of the number of frequency steps, and transmit and receive antennas. In a general compressive setting, the system matrix $\A$ is rectangular with $M \ll N$. From the continuous observation model in Eq. \ref{eq:ForwardModel}, we can obtain the $(m,n)$th element of the system matrix that denotes the contribution of the $n$th voxel to the $m$th measurement as follows:
\begin{align}
	A_{m, n}=\frac{p\left(k_{m}\right) 
	e^{-j k_{m} \left( d_{t_{m}}^{(n)} + \, d_{r_{m}}^{(n)}\right)}}{4 \pi d_{t_{m}}^{(n)} d_{r_{m}}^{(n)}}
	\label{eq:MatrixElements}
\end{align}
Note that the measurement index $m$ represents the locations of the transmitter and receiver antennas, and the frequency, $k_{m}$, used in the $m$th measurement. Furthermore, $d_{t_{m}}^{(n)}$ and $d_{r_{m}}^{(n)}$ denote the distances from the center of the $n$th voxel to the transmitting and receiving antenna used in this measurement, respectively. 

\section{Inverse Problem} 
\label{section:Related Work}

In the inverse problem for MIMO radar imaging, the goal is to reconstruct the unknown reflectivity field of the scene, $\s$, from the limited radar measurements, $\y$. This linear inverse problem is inherently ill-posed in a compressive imaging setting. 

There are various approaches for solving such linear inverse problems including traditional direct inversion methods, regularized iterative methods, and deep learning-based reconstruction methods. In this work, we develop deep learning-based non-iterative reconstruction methods with the aim of achieving high image quality at low computational cost. For performance comparison, we also consider commonly used approaches from analytical direct inversion and regularized reconstruction methods. Here we discuss the merits and drawbacks of the existing methods in the literature and the motivation of the developed methods. 

Conventional reconstruction methods in radar imaging are based on direct solution of the continuous observation model. These direct inversion methods are related to applying the adjoint of the forward operator to the measurements~\cite{marks2017fourier,jin2017deep}, and implemented either in the time domain or in the frequency domain, frequently using FFT. For monostatic radar systems with collocated transmitter and receiver antennas, commonly used methods are backprojection (also known as diffraction stack migration or delay-and-sum), Kirchoff migration, range migration, and their variants~\cite{sheen2001three,ahmed2012advanced}. These direct inversion methods have also been extended for MIMO (multistatic) arrays with spatially distributed transmit and receive antennas~\cite{zhuge2010sparse,zhuge2010modified, zhuge2012three,  tan2018omega,liu2017mimo}. The multistatic imaging configuration makes the FFT-based imaging more challenging and requires a multidimensional interpolation process~\cite{zhuge2012three,alvarez2016fourier}.

In our experiments, for performance comparison, the backprojection (BP) algorithm is chosen from these traditional methods due to its common use. Considering the forward model in Eq. \eqref{eq:ForwardModel} with constant $p(k)$, the three-dimensional reflectivity distribution of the scene can be reconstructed with the frequency-domain BP algorithm~\cite{liu2017mimo,anadol2018uwb} using $\hat{s}_{n}=\frac{1}{M} \sum_{m=1}^M  y_{m} \, e^{j k_{m} \left( d_{t_{m}}^{(n)} + d_{r_{m}}^{(n)} \right)}$. Here $\hat{\s}$ is the reconstructed complex-valued vector for the sampled 3D reflectivity image. That is, $\hat{s}_{n}$ represents the $n$th voxel of the reconstructed scene reflectivity and $y_{m}$ is the $m$th measurement. As defined before, $d_{t_{m}}^{(n)}$ and $d_{r_{m}}^{(n)}$ respectively represent the distances from the $n$th voxel to the transmit and receive antennas used in the $m$th measurement. Note that BP operation is similar to the adjoint of the forward operator, with the only difference being the additional scaling in the adjoint operator. Such direct inversion methods have low computational complexity but as a drawback, they can not offer state-of-the-art reconstruction performance. As well-known, the reconstruction quality and resolution degrade in the presence of noise or limited data (as acquired with sparse MIMO arrays). 

Different than direct inversion methods, regularized iterative reconstruction methods incorporate additional prior information (such as sparsity) into the reconstruction process to eliminate uniqueness and noise amplification issues arising due to limited data and measurement noise. With the advent of compressed sensing (CS) theory \cite{candes2008introduction}, sparsity-based reconstruction is the most commonly used analytical regularization approach and has been widely studied in various imaging problems~\cite{tropp2010computational,afonso2010fast}, including radar imaging both for far-field and monostatic imaging settings \cite{wei2013sparse, potter2010sparsity, guven2016augmented, ma2014mimo, guo2015microwave,  ma2018multiple,  huang2018tensor}, as well as for multistatic and near-field settings \cite{zhang2015generalized, oktem2019sparsity,cheng2017near,miran2021sparse}. 

To enforce sparsity \cite{tropp2010computational,oktem2018computational}, the inverse problem can be formulated as the following regularized least-squares problem:
\begin{align}
	\min _{\s}\|\y-\A \s\|_{2}^2+\lambda \mathcal{R}(\s) 
	\label{eq:sparseRecovery}
\end{align}
Here the first term ensures that the solution is consistent with the forward model and the measured data (under the assumption of i.i.d. Gaussian noise), and the second term, $\mathcal{R}(\s)$, controls how well the solution matches the prior knowledge on sparsity, with the parameter $\lambda$ trading off between these two terms. In radar imaging, commonly used sparsity-based regularizers are $l_1$-norm penalty $\mathcal{R}(\s)=\| \s\|_{1}$ and total variation (TV) penalty $\mathcal{R}(\s)=\|\Phi \s\|_{1}$ where $\Phi$ is the discrete gradient operator (i.e. finite-difference operation), operating on the complex-valued reflectivity field of the scene. It is well known that $l_1$-norm penalty provides a good sparsifying penalty for scenes involving point-like targets whereas TV provides superior results for extended targets with significant structure or piecewise-constant characteristics. Moreover, it has been observed that for scene reflectivities with random phase nature, enforcing total variation penalty on the magnitude of the reflectivity, i.e. $\mathcal{R}(\s)=\|\, \Phi |\s| \,\|_{1}$, yields improved reconstruction compared to regular TV~\cite{cetin2001feature,guven2016augmented}. 

For near-field MIMO radar imaging, various sparsity-based reconstruction algorithms have been developed to solve the optimization problem in Eq.~\eqref{eq:sparseRecovery} with sparsity inducing regularizer $\mathcal{R}(\s)$. These algorithms are generally adapted from sparsity-based reconstruction algorithms developed for two-dimensional image restoration problems and mainly differ from each other in their efficiency in terms of computation time and memory usage. For example, Cheng et al.~\cite{cheng2017near} adapted the split augmented lagrangian shrinkage algorithm (SALSA) \cite{afonso2010fast} and Oktem~\cite{oktem2019sparsity} adapted the half-quadratic regularization approach~\cite{geman1995nonlinear}. There have been also some efforts to reduce the computational cost and memory usage of such sparsity-based reconstruction algorithms by exploiting the special structure of the forward problem~\cite{li2015near,miran2021sparse}. Although sparsity-based methods provide better reconstruction quality than the traditional direct inversion methods, they suffer from higher computational cost due to their iterative nature and requiring computation of the forward scattering operator and its adjoint at every iteration. They also require parameter tuning to achieve good reconstruction under different observation scenarios. Although sparsity-based methods offer better image quality than the traditional direct inversion methods, their high computational cost is undesirable in real-time applications. 

In our experiments, for performance comparison, we use a sparsity-based CS method~\cite{oktem2019sparsity} that is based on half-quadratic regularization approach. As sparsifying regularizer, this method uses 3D total variation operating on the scene reflectivity (i.e. $\mathcal{R}(\s)=\|\, \Phi \s \,\|_{1}$) to exploit correlations along both range and cross-range dimensions. 

Recently, deep learning-based reconstruction methods have emerged to overcome the limitations of the analytical reconstruction methods. These methods are shown to simultaneously achieve high reconstruction quality and low computational cost for various inverse problems in imaging~\cite{lopez2021deep,jin2017deep,ongie2020}. In this work, we developed three novel deep learning-based reconstruction methods for near-field MIMO radar imaging with the goal of achieving high image quality with lower computational complexity than the existing iterative CS reconstruction algorithms. To comparatively evaluate the reconstruction performance and computational cost of the developed methods, we compare with the BP and TV-based reconstruction.

\section{Developed Deep Learning-based Reconstruction Methods}
\label{section:Image Reconstruction}

\begin{figure}[tbh!]
	\centering
	\includegraphics[width=\linewidth]{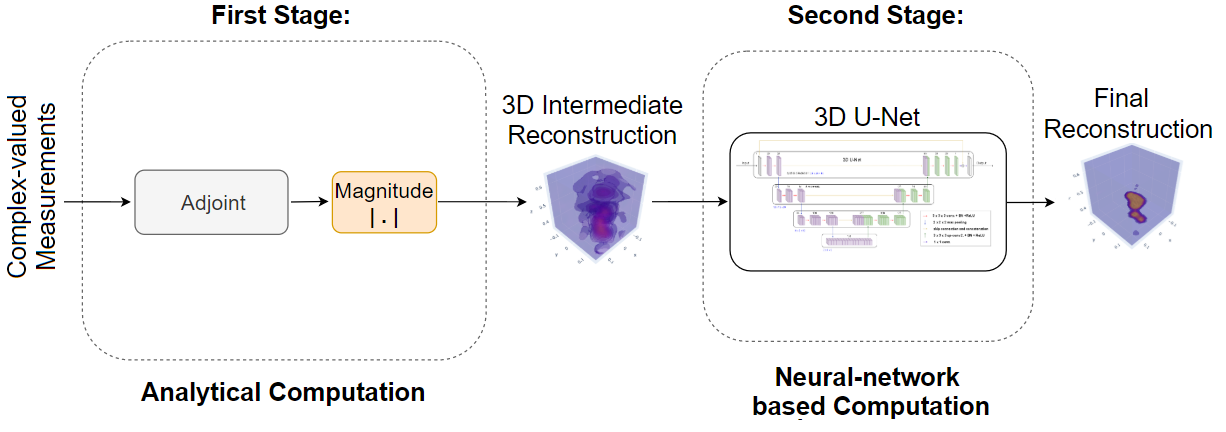}
	\caption{Developed DNN-based Two-Stage (Deep2S) Reconstruction Approach.}
	\label{fig:ProposedMethod}
\end{figure}

The developed deep learning-based methods for near-field MIMO radar imaging are presented in this section. The main goal is to achieve high image quality with low computational cost so that the method can be used in real-time applications. Learning-based direct inversion methods enable such capabilities. For this reason, the developed approaches are based on learned direct reconstruction~\cite{ongie2020,jin2017deep,lucas2018using,kulkarni2016reconnet,zhu2018image}. The first approach is a physics-based learned reconstruction method with a two-stage structure as shown in Fig.~\ref{fig:ProposedMethod}. Here the first stage is an adjoint operation that exploits the physics-based model to provide an initial reconstruction, and the second stage is a 3D U-Net denoiser for refinement. For comparison, a second approach is also developed as shown in Fig.~\ref{fig:DIS}, which replaces the physics-based first stage with a fully connected neural network. Hence this approach aims to perform the reconstruction directly from the radar measurements using only deep neural networks and does not use the physics-based model. In what follows, the details of these two DNN-based approaches are presented.

\subsection{First Approach: DNN-based Two-Stage (Deep2S) Reconstruction}

The first approach has two consecutive stages as shown in Fig. \ref{fig:ProposedMethod} where the first stage performs an analytical computation whereas the second stage is a neural network. In the first stage, adjoint of the forward scattering operator is used to back project the measurements to the object space. This intermediate result has worse quality compared to a sparsity-based reconstruction like total-variation but the adjoint operation has the benefit of fast computation due to its non-iterative nature. Subsequently, a deep neural network is employed to improve this 3D intermediate result. This second stage employs a 3D DNN which is trained to convert the backprojected measurements to a magnitude-only reflectivity image. Although scene reflectivities are complex-valued, in most applications they have random phase nature; hence it is generally sufficient to reconstruct the magnitude of the scene reflectivity. Because of this, the magnitude of the backprojected measurements from the first stage is input to the DNN in the second stage. To jointly exploit range and cross-range correlations for extended 3D targets, a 3D U-Net architecture is used as DNN.

For training the network, the simulated radar measurements are first passed through the adjoint operation stage. The magnitude of the intermediate result obtained with this adjoint operation is then input to the second stage. The DNN in the second stage is trained using these magnitudes together with the corresponding ground-truth reflectivity magnitudes, which form our training input and output respectively. After training, we process our radar measurements in the test dataset with the proposed method to reconstruct the reflectivity magnitude of the unknown scene. In what follows, we provide the details of our approach.

\subsubsection{First Stage (Adjoint Operation)}

Since it is generally a difficult task for a network to learn the direct mapping from the measurement space to the 3D object space~\cite{ongie2020,jin2017deep, lucas2018using}, we first apply the adjoint operator to the measurements to provide the network in the second stage a warm start.
The adjoint operation encapsulates the physical model of the near-field MIMO imaging system and has the benefit of fast computation due to its non-iterative nature. The adjoint operation is applied to the discrete radar measurements by using the hermitian of the system matrix as follows:
\begin{align}
	\hat{\s} =  \A^{H}\y
\end{align}
This gives us a 3D intermediate result, $\mathnormal{\hat{\s}}$, in the object (reconstruction) space. The magnitude of these intermediate results are input to the DNN in the second stage after normalization to the range $[0,1]$ (since the largest possible value for reflectivity magnitude is 1 in practice). This normalization enables Deep2S to have robust performance for scenes with different dynamic ranges.

The physics-based first stage simplifies the learning process of the 3D U-Net architecture in the second stage by back-projecting the radar measurements to the object space using the adjoint operator. In this way, the network can be trained to improve the normalized reflectivity magnitude obtained from the first stage by directly performing the refinement in the object space. Note that adjoint operation is similar to the BP algorithm, with the only difference being the additional scaling in the adjoint operator. Hence, just like BP, adjoint operation provides a fast approximate solution, which is then enhanced by the denoiser in the second stage. The motivation of using adjoint operation for direct inversion in the first stage, instead of the conventional BP, is twofold. Firstly, as illustrated in Section~\ref{section:Numerical Results}, for all tested cases, the quality of the images obtained with the adjoint operation is similar to (even slightly better than) the BP reconstruction (see for example, Figures \ref{fig:result_different_algorithms} and \ref{fig:YarovoyResults} and Table~\ref{tab:result1}). Secondly, the structure of our approach is motivated by \cite{jin2017deep} which demonstrates that various direct and iterative inversion approaches involve adjoint operation followed by filtering plus pointwise nonlinearities, which suggests that CNNs may offer an alternative solution to process the adjoint output.

\subsubsection{Second Stage (3D U-Net)}

Motivated by the success of 2D U-Net architectures~\cite{ronneberger2015u} in several inverse problems~\cite{ongie2020,jin2017deep}, we develop a four-level 3D U-Net architecture for the second stage of our approach. Unlike the 2D DNN used in earlier works~\cite{cheng2020compressive}, this 3D U-Net architecture exploits multichannel 3D filters to jointly capture the correlations along both range and cross-range directions of a 3D extended target. Three-dimensional convolution kernels, max pooling, and up-sampling operations are used in our network instead of their two-dimensional counterparts.

\begin{figure}[tbh!]
	\centering
	\includegraphics[width=\linewidth]{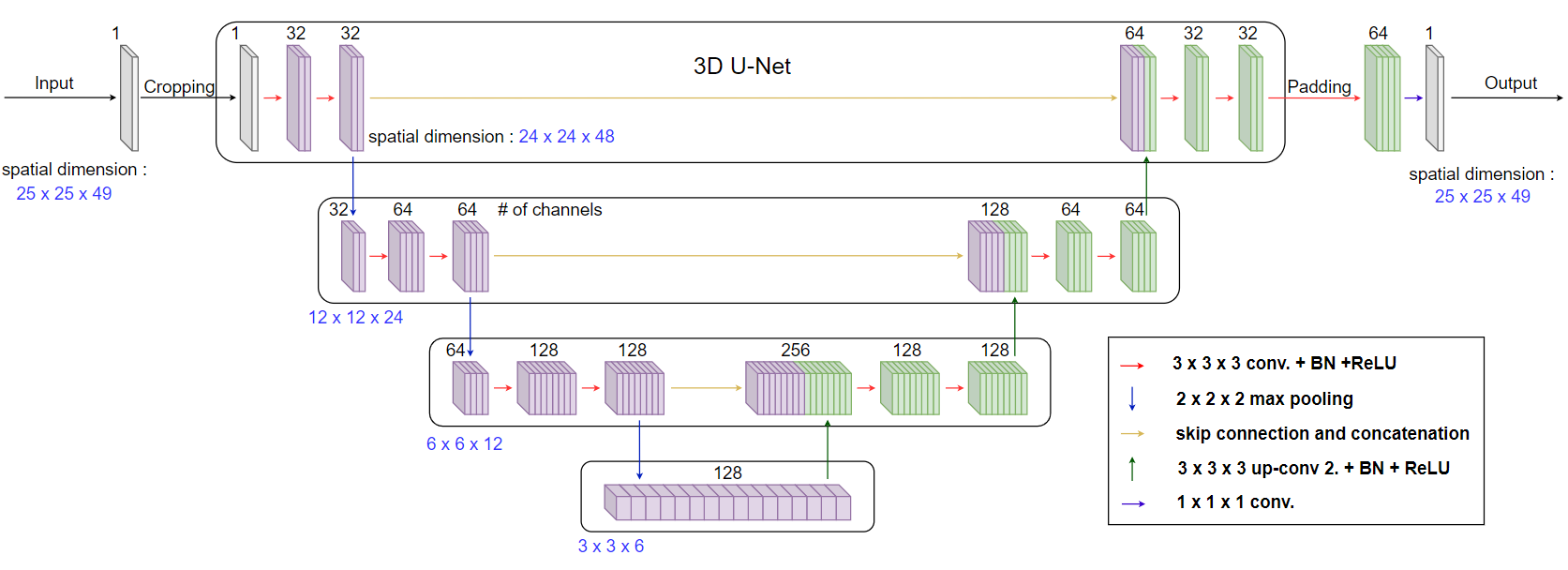}
	\caption{Architecture of the 3D U-Net.}
	\label{fig:U_Net}
\end{figure}

The 3D U-Net architecture is illustrated in Fig. \ref{fig:U_Net}. As shown, this architecture has an encoding and decoding path. The encoding and decoding paths respectively contain repeated application of 3D convolution, and 3D upconvolution blocks. 
The number of channels are indicated above each block in the figure and involve filters of size $3 \times 3 \times 3$. These convolutional layers are followed by batch normalization (BN) and rectified linear unit (ReLU). While the encoding path contains $2 \times 2 \times 2$ max-pooling layers with strides of two to decrease the spatial size, the decoding path involves upconvolutions with strides of 2 in each dimension to increase the spatial size. Because the input of the network is chosen to be of size $25 \times 25 \times 49$ voxels in x, y, and z directions, cropping is first performed to arrive at the size $24 \times 24 \times 48$ so that spatial size reduction can be performed through $2 \times 2 \times 2$ max-pooling operations. In the decoding path, there are also concatenations with the cropped feature maps from the encoding path as shown with the padding arrow in Fig. \ref{fig:U_Net}. Hence both the input and output of the network are of size $25 \times 25 \times 49$. The total number of parameters in this 3D U-Net architecture is 2,873,153. 

The 3D U-Net architecture has three properties suited for our imaging problem. Firstly, due to the encoding/decoding procedure, the effective receptive field of the network increases. In our problem, our main purpose is the refinement of the input image from the first stage. Having a large receptive field over the input image can improve the quality of the output image~\cite{johnson2016perceptual}. Secondly, the 3D U-Net employs multichannel filters. By this way it can better extract the feature maps of its input. This increases the dimension of the latent representation of our input images, which increases the expressive power of the network~\cite{krizhevsky2012imagenet}. Thirdly, the 3D U-Net architecture can capture the correlation along both range and cross-range directions of an extended target through 3D operations, unlike the existing 2D approaches in the literature.

In the end, a feed-forward approach is obtained while incorporating the physics-based knowledge of the MIMO imaging system through the hermitian of the system matrix. The Deep2S approach has low computational complexity as desired. The computational cost of the algorithm is dominated by the 3D U-Net.

\subsection{Second Approach: DNN-based Direct Inversion (DeepDI)}

\begin{figure}[tbh!]
	\centering
	\includegraphics[width=0.8\linewidth]{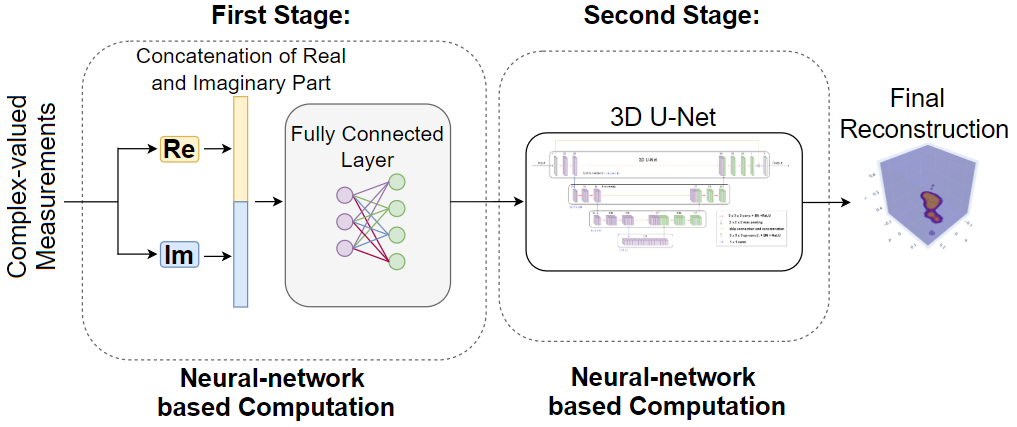}
	\caption{Developed DNN-based Direct Inversion (DeepDI) Approach.}
	\label{fig:DIS}
\end{figure}

To compare with the developed physics-based approach, a purely DNN-based approach is also proposed to perform direct inversion from the measurements as shown in Fig.~\ref{fig:DIS}. In this approach, former adjoint stage is replaced with a fully connected neural network. As a result, the observation model is not used. The second stage contains a 3D U-Net as before to improve the intermediate reconstruction. Hence the approach contains two consecutive stages: a fully connected layer followed by a 3D U-Net. In this two-stage structure, the reconstruction is performed only using neural networks which are trained end-to-end to learn the direct mapping between the magnitude of the unknown reflectivity image and the radar measurements. The only difference from the first reconstruction approach (Deep2S) is that we use neural network-based computation in the first stage instead of an analytical one.

This approach has been motivated by the CNN-based reconstruction approach, ReconNet,~\cite{kulkarni2016reconnet} proposed for a blockwise 2D CS recovery problem. Instead of using a single fully connected layer structure for direct inversion, we introduce a second stage consisting of the 3D U-Net. The main reason is to improve the reconstruction performance using an additional CNN structure and also decrease the training time of the whole network. The total number of parameters in this two-stage architecture is 76,013,778, which corresponds to 26 fold increase compared to Deep2S.

For training, the simulated radar measurements and the corresponding ground-truth reflectivity magnitudes are used as input and output, respectively. The two DNNs are trained end-to-end. After training, we provide our radar measurements in the test dataset to the 
trained network to directly reconstruct the reflectivity magnitude of the unknown scene. This approach has similar computational complexity as the first one. In what follows, we provide the details of the approach. 

\subsubsection{First Stage (Fully Connected Layer)}

The first stage contains a fully connected layer that takes the radar measurements as input and outputs a 3D intermediate reconstruction for the magnitude of the unknown reflectivity image. Since the measurements are complex-valued, its real and imaginary parts are concatenated and provided as input to the fully connected layer. This fully connected network aims to learn a mapping from the complex-valued measurement space to the real-valued 3D reconstruction space (where the magnitude of the reflectivity image lives). The output of the fully connected layer is of size $25 \times 25 \times 25$ in $x$, $y$, and $z$ directions. Subsequently, the resulting datacube is upsampled by two along the $z$ direction using zero-order hold and then cropped by one to achieve an output size of $25 \times 25 \times 49$. The resulting 3D intermediate reconstruction is fed into the second stage. 

\subsubsection{Second Stage (3D U-Net)}

In the second stage, a deep neural network is employed to improve this 3D intermediate reconstruction as before. For this purpose, same 3D U-Net architecture is used as given in Fig. \ref{fig:U_Net} to perform a fair comparison with the first approach. 

\section{Experimental Results}
\label{section:Numerical Results}

In this section, we demonstrate the effectiveness of the developed methods using both simulated data and experimental measurements. We first analyze the performance of the developed methods under various observation scenarios using simulated data, and compare their performance with backprojection~\cite{liu2017mimo,anadol2018uwb} and sparsity-based CS~\cite{oktem2019sparsity} reconstructions. We also provide an analysis on the effect of measurement SNR and different network architectures. 
Furthermore, we investigate the resolution achieved with our reconstruction approach at compressive MIMO imaging settings and compare this to the expected theoretical resolution for the conventional (non-compressive) settings.
Lastly, we illustrate the performance with experimental data to demonstrate applicability to real-world measurements.

\subsection{Synthetic Dataset Generation}
\label{subsection:Synthetic Scene Generation}

\begin{figure}[tbh!]
	\centering
	\includegraphics[width=1\linewidth]{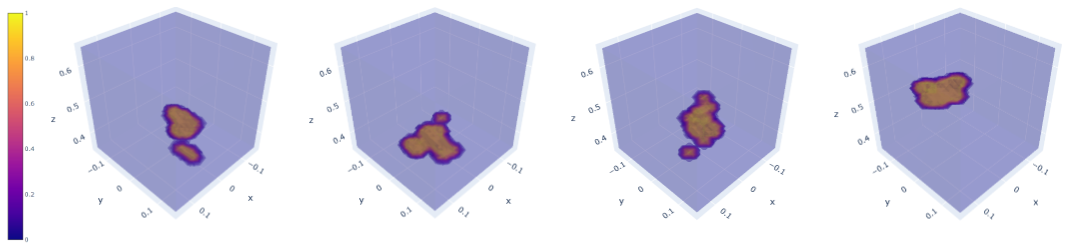}
	\caption{Samples of Synthetically Generated Dataset (The units of x, y, and z-axis are meters).}
	\label{fig:dataset}
\end{figure}

A large experimental dataset is not available in many radar imaging applications. As a result, in deep-learning based approaches, neural networks are often trained using synthetically generated datasets~\cite{alver2021plug}.
In this work, to enable generalizability to different target scenarios, 3D extended targets are randomly generated within the cube we want to infer. Fig.~\ref{fig:dataset} shows samples from our synthetically generated dataset.

\begin{figure}[tbh!]
	\centering
	\includegraphics[width=1\linewidth]{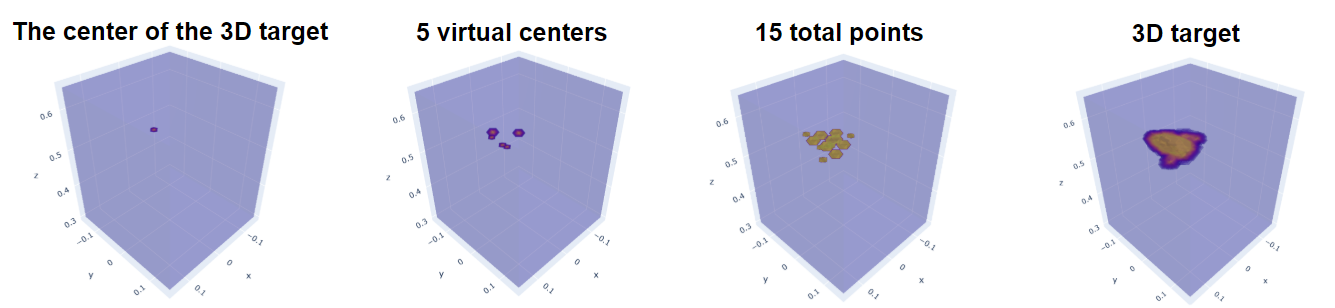}
	\caption{Synthetic Dataset Generation (unit is meters).} 
	\label{fig:synthetic_process}
\end{figure}

To randomly generate a large synthetic dataset consisting of 3D extended targets, the procedure illustrated in Fig.~\ref{fig:synthetic_process} is followed. Firstly, the center of the target is randomly chosen from a uniform distribution in the range -0.05 to 0.05~\unit{m} for the x and y-axis and, 0.41 to 0.59~\unit{m} for the z-axis. Then, around the center of the target 5 virtual centers are chosen according to a Gaussian distribution with zero mean and standard deviation of 2. Moreover, for every virtual center, 3 points are generated according to a Gaussian distribution with zero mean and standard deviation of 1.5. The variance parameter of the Gaussian distribution determines the size of the distributed target. For a large variance, generated distributed targets take up more volume within the cube compared to the selection of a smaller variance. For one synthetic scene, we totally have 15 points chosen randomly within the cube. To obtain volumetric objects, these 15 points are passed through a 3D Gaussian filter with a standard deviation of 1.3. This is then passed through the sigmoid function which performs the amplitude normalization of the generated 3D targets to a maximum value of $1$. Hence the amplitude values of the generated targets are in the range of 0 to 1.

With this procedure, we obtain different 3D extended objects that spread within the cube from a randomly chosen center. The training, validation, and test datasets contain separate 800, 100, and 100 images which were randomly generated in this way.

\subsection{Training Procedure} 
 
The implementation and training are performed with TensorFlow \cite{tensorflow2015} and Keras\cite{chollet2015keras} using Google Colaboratory which is a hosted Jupyter Notebook service that requires no setup to use and provides 12.7 GB RAM and a CPU with default Adam Optimizer. The training of the networks used in the experiments 
took about one hour for both approaches. The learning rate is chosen as $10^{-3}$, batch size equals 16, loss function is chosen as $l_2$ loss, maximum number of epochs is set to 100, and the early stopping criterion is selected as 15 consecutive epochs with no drop in validation loss. 

\subsection{Simulation Setting}

The sketch of the used simulation setting is shown in Fig. \ref{fig:RadarImagingSystem}. As a sparse MIMO array topology, we consider a Mills Cross array since it is commonly used due to its imaging performance~\cite{oktem2019sparsity,zhuge2012three}, but other proposed alternatives~\cite{zhuge2012study} could also be considered. The width of the planar array is \SI{0.3}{\meter}, which includes 12 uniformly spaced transmit antennas and 13 uniformly spaced receive antennas in a cross configuration. The target center is located approximately \SI{0.5}{\meter} away from the 2D MIMO array in its near-field. The frequency, f, is swept between 4 GHz to 16 GHz with uniform steps.
In the numerical simulations, the number of frequency steps is selected as 7, 15, and 31 respectively to investigate the performance of the developed methods at different compression levels.  

For the ideal case with non-sparse antenna array and frequency steps (which is not the case here), the expected theoretical resolution~\cite{zhuge2012three} is 2.5 cm in the cross-range dimensions, $x$ and $y$, and 1.25 cm in the down-range dimension, $z$. To reconstruct the reflectivity image in a cube of size $\SI{0.3}{\meter} \times \SI{0.3}{\meter} \times \SI{0.3}{\meter}$, we choose the voxel size as $\SI{1.25}{\cm} \times \SI{1.25}{\cm} \times \SI{0.625}{\cm}$ in $x$, $y$, and $z$ dimensions, which are equal to the half of these theoretical resolutions in each dimension. Then the reflectivity image that we want to infer contains $25 \times 25 \times 49$ voxels in $x$, $y$, and $z$ directions, respectively.

Using the synthetic dataset generation procedure described in Section \ref{subsection:Synthetic Scene Generation}, we randomly generate different 3D extended objects that spread within $25 \times 25 \times 49$ cube. By using these images and the forward model in Eqn.~\eqref{eq:ForwardModelMatrixForm}, noisy measurements are simulated with a chosen signal-to-noise ratio (SNR). Measurement SNR is defined as $10 \, \text{log}_{10}\left(\frac{\| \A \s\|^2_2}{M\cdot\sigma^2_w}\right)$ where $\sigma_w$ denotes the standard deviation of the complex-valued white Gaussian noise added to the measurements.

\subsection{Performance Analysis with Simulated Data} \label{firstAnalysis}

We now present the performance of the developed methods, namely Deep2S and DeepDI, by considering different compression levels in the observations. To compare their performance, the results are also obtained using the adjoint operation, backprojection (BP)~\cite{liu2017mimo,anadol2018uwb} and total-variation (TV) based CS reconstruction~\cite{oktem2019sparsity}. For optimum parameter selection in TV regularization, we choose the regularization parameter $\lambda$ in Eqn.~\eqref{eq:sparseRecovery} through linear search in order to maximize the average reconstruction performance in the test dataset. Based on this, $\lambda$ is set to 25 in all experiments.

We evaluate the performance of the algorithms based on the reflectivity magnitudes since in various applications its phase generally does not contain any useful information. As performance metrics, we use 3D peak-signal-to-noise ratio (PSNR) and the structural similarity index (SSIM). A higher SSIM and PSNR value indicates a better reconstruction or higher similarity between the reconstructed and the ground truth images. SSIM~\cite{wang2004image} is computed for each two-dimensional slice over the range direction and then averaged. 3D PSNR is calculated between the three-dimensional ground-truth image $s$ and the reconstructed image $\mathnormal{\hat{s}}$ using $10 \log _{10}\left(\frac{s_{max}^{2}}{MSE}\right)$ where $s_{max}$ is the maximum magnitude value of the image voxels (which is $1$ in our case) and $MSE$ is the mean square
error calculated between the magnitudes as $\sum_{x=0}^{N_x-1} \sum_{y=0}^{N_y-1} \sum_{z=0}^{N_z-1} \frac{(|s(x, y, z)|-|\hat{s}(x, y, z)|)^{2}}{N_x \cdot N_y \cdot N_z}$. For visual evaluation, we also provide 3D visualization of sample reconstructed images without applying any thresholding/truncation to the small values (unlike many earlier works).

We start our analysis for the case with 30dB measurement SNR and 15 frequency steps. Accordingly, we utilize the training dataset and the measurements simulated at this setting to train the proposed deep architectures. For the analyses at different settings, we use transfer learning to fine-tune the weights of these trained DNNs.

\begin{figure}[tbh!]
	\centering
	\includegraphics[width=0.8\linewidth]{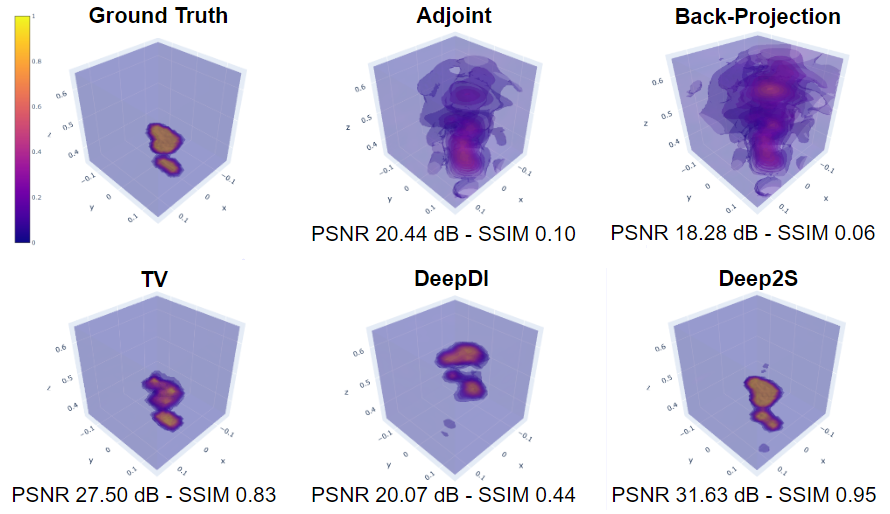}
	\caption{Reconstructions of the different algorithms for the first test image of the synthetically generated dataset at 30 dB SNR (Number of Frequency Steps: 15), (The units of x, y, and z-axis are meters).}
	\label{fig:result_different_algorithms}
\end{figure}

Figure \ref{fig:result_different_algorithms} illustrates the reconstruction performance of different methods for a sample image cube in the test dataset with 30dB measurement SNR and 15 frequency steps. As seen, both the Deep2S approach and the TV algorithm provide good
reconstruction performance while the Deep2S reconstruction is much better both visually and quantitatively. Note that Deep2S gives nearly artifact-free result, although the adjoint operation provides a poor input to the DNN. As expected, both the adjoint and back-projection yield large artifacts in the reconstruction with limited observations. The reconstruction provided by the DeepDI approach is also poor. Although DeepDI result is realistically looking, it is a false reconstruction. In fact, neither the target shape nor the target location is correct in the DeepDI result. Hence while physics-based learned direct inversion, Deep2S, provides faithful reconstructions, pure deep-learning based direct inversion, DeepDI, largely fails.

Table \ref{tab:result2_time} shows the average runtimes of different methods
for the test images. These runtimes are computed on a machine with 16 GB RAM, and an i7-7700HQ (@ 2.80GHz) CPU. From these results, TV reconstruction was computed in MATLAB, whereas the other approaches were run in Python. Note that the developed deep learning-based direct inversion methods, Deep2S and DeepDI, require less than a second to reconstruct the reflectivity field of a $25 \times 25 \times 49$ scene, while the reconstruction with 3D TV has a runtime on the order of minutes. The long runtime of iterative TV method is expected since each iteration requires multiple computations of the forward and adjoint operators, whereas Deep2S requires a single adjoint computation. (Nevertheless, note that the efficiency of the 3D TV reconstruction method (or in general, sparsity-based CS methods) also depends on the underlying optimization approach, and can slightly differ from one approach to other.) Hence the Deep2S approach not only surpasses the other approaches in terms of reconstruction performance but also is computationally more efficient except for the adjoint and backprojection which give poor performance. These results illustrate that the Deep2S approach is capable of achieving high image quality with low computational cost as desired for real-time applications.

\begin{table}[H]
\centering
\caption{Average Runtimes for 100 Test Images at 30 dB SNR (Number of Frequency Steps: 15).}
\label{tab:result2_time}
\begin{tabular}{P{0.13\linewidth}|P{0.132\linewidth}P{0.132\linewidth}P{0.132\linewidth}P{0.132\linewidth}P{0.132\linewidth}}
\toprule
 & BP &Adjoint & TV & DeepDI & Deep2S  \\
\midrule
Runtime & 0.225 s & 0.225 s & 501s & 0.223 s & 0.454 s\\
\bottomrule
\end{tabular}
\end{table}

We now investigate the average reconstruction performance of different methods under different compression levels. For this, we simulate noisy measurements at 30dB SNR for 100 test images using 7, 15, and 31 frequency steps. Hence reconstruction is performed approximately with 4\%, 8\%, and 16\% data ($M/N$) in these compressive MIMO imaging settings (which correspond to compression levels of 96\%, 92\%, and
84\%). The average reconstruction performance of different methods is given in Table \ref{tab:result1}. In all cases, the Deep2S approach significantly outperforms the other approaches in terms of PSNR and SSIM. DeepDI provides a poorer performance that is even worse than TV reconstruction. When the number of frequency steps is decreased, the reconstruction performance of all approaches starts to decrease due to the increased ill-posedness of the inverse problem. In the worst case when the number of frequency steps is 7, the Deep2S approach provides a faithful reconstruction with an average PSNR of 29.1 dB and SSIM of 0.89. As the number of frequency steps increases, the Deep2S approach passes 30 dB PSNR and 0.94 SSIM. Also, increasing the number of frequency steps from 15 to 31 does not provide a significant change in the reconstruction performance of Deep2S. Hence 15 frequency steps seem to be enough for this imaging scenario. Moreover, note that 3D PSNR appears to be a better quality metric than SSIM since it is computed for the overall 3D object while the computed average SSIM can not measure the fidelity over the range direction since it is computed for 2D slices only and then averaged. Lastly, for all compressive settings, the quality of the images (i.e. average PSNR/SSIM values) obtained with the adjoint operation is similar to (even slightly better than) the BP reconstruction. For this reason, BP is omitted in the subsequent performance comparisons.

\begin{table}[t]
	\centering
	\caption{Average PSNR and SSIM Values for Different Number of Frequency Steps at 30 dB SNR. Best results are shown in bold.}
	\label{tab:result1}
\begin{tabular}{M{0.14\linewidth}|M{0.132\linewidth}M{0.132\linewidth}M{0.132\linewidth}M{0.132\linewidth}M{0.132\linewidth}}
\toprule
Freq. steps      & BP  & Adjoint     & TV          & DeepDI      & Deep2S \\
\midrule
7            &  17.8/0.06 & 20.3/0.12 & 22.7/0.44 & 22.6/0.65 & \textbf{29.1}/\textbf{0.89} \\
15           &  20.3/0.13 & 21.8/0.19 & 25.5/0.74 & 23.2/0.61 & \textbf{30.4}/\textbf{0.94} \\
31           &  22.5/0.26 & 23/0.35   & 26.3/0.83 & 23.3/0.62 & \textbf{30.4}/\textbf{0.94} \\
\bottomrule
\end{tabular}
\end{table}

\begin{figure}[tbh!]
	\centering
	\includegraphics[width=1\linewidth]{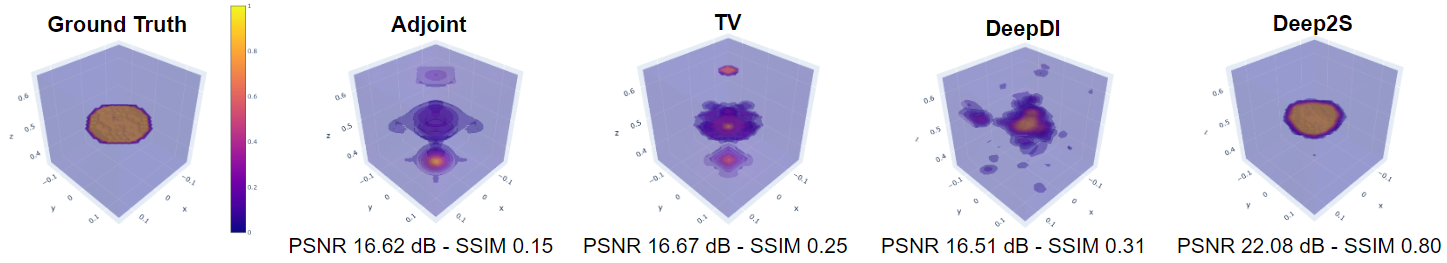}
	\caption{Reconstructions of the different algorithms for the ellipsoid test image at 30 dB SNR (15 Frequency Steps).}
	\label{fig:ellipse}
\end{figure}

The performance of learning-based direct reconstruction methods can depend on the training data since the learning mostly relies on this.
For this reason, testing the developed methods for scenes different from the training dataset is important and investigated here as illustrated in Figure~\ref{fig:ellipse}. To test the generalization ability of the methods, we introduce a 3D target image that has different characteristics than the synthetically generated training images. The 3D target image is an ellipsoid that is centered on the cube. This image takes up more volume than the training dataset images. Also, the introduced ellipsoid has a bulk large volume within the cube while the training dataset contains targets that are mostly shapeless and randomly scattered in the volume. Figure \ref{fig:ellipse} shows the results obtained with different reconstruction algorithms at 30 dB SNR and 15 frequency steps. The developed Deep2S approach remains to provide the best reconstruction performance both visually and quantitatively. The adjoint operation, DeepDI and TV reconstructions show significant volume artifacts in this case. Moreover, although not clearly seen in the figure, the TV reconstruction fails to fill the volume inside the ellipsoid. This illustrates that the developed Deep2S approach can provide robust performance for 3D targets different from the training set.

In the analyses performed so far, we have considered real-valued 3D reflectivity scenes with no random phase. This simulation setting appears to be the commonly analyzed case in the other microwave imaging works as well. Nevertheless, in addition to these analyses, we will now focus on a more realistic scenario that takes into account the complex-valued and random phase nature of scene reflectivities, which have been mostly neglected in the earlier works. Because scene reflectivities have random phase in various applications~\cite{Munson1984offset,cetin2001feature}, we add random phase to our synthetically generated dataset. Random phase is added for every voxel from a uniform distribution in the range of $[-\pi,\pi]$. From now on, we will only examine this more realistic imaging scenario that involves scene reflectivities with random phase~\cite{Munson1984offset,cetin2001feature, alver2021plug, guven2016augmented}. 
For this, noisy measurements are re-simulated with random phase added scene reflectivities using the forward model in Eqn.~\eqref{eq:ForwardModelMatrixForm}. The training of the networks is performed through transfer learning, where the weights of the previously trained networks are unfrozen and re-trained for five additional epochs with the new data obtained for this case. 

\begin{table}[b!]
	\centering
	\caption{Average PSNR and SSIM Values for Different Number of Frequency Steps and Random Phase Added Scene Reflectivities at 30 dB SNR. Best results are shown in bold.}
	\label{tab:result1phase}
\begin{tabular}{P{0.15\textwidth}|P{0.15\textwidth}P{0.15\textwidth}P{0.15\textwidth}P{0.15\textwidth}}
\toprule
        Frequency steps & Adjoint & DeepDI & Deep2S\\
\midrule
        7 & 22.6/0.26 & 22.7/0.73 & \textbf{28.4}/\textbf{0.93} \\ 
        15 & 23.7/0.49 & 23.4/0.81 & \textbf{29.2}/\textbf{0.93} \\ 
\bottomrule
\end{tabular}
\end{table}

The average reconstruction performance of different methods for 100 test images involving random phase is given in Table~\ref{tab:result1phase} for 30 dB SNR case with two different number of frequency steps. As before, in both cases, the Deep2S approach outperforms the other approaches in terms of PSNR and SSIM. Moreover, the reconstruction performance of all approaches decreases when the number of frequency steps is decreased. We observe that this table shares similar results with Table~\ref{tab:result1} which was obtained using scene reflectivities with no random phase. Hence the general performance behavior remains similar when dealing with this more realistic imaging scenario.

\begin{figure}[tbh!]
\centering
    \centering
     \begin{subfigure}[b]{0.2\textwidth}
         \centering
         \textbf{Ground Truth\vspace{7.42PT}}
         \includegraphics[width=\textwidth]{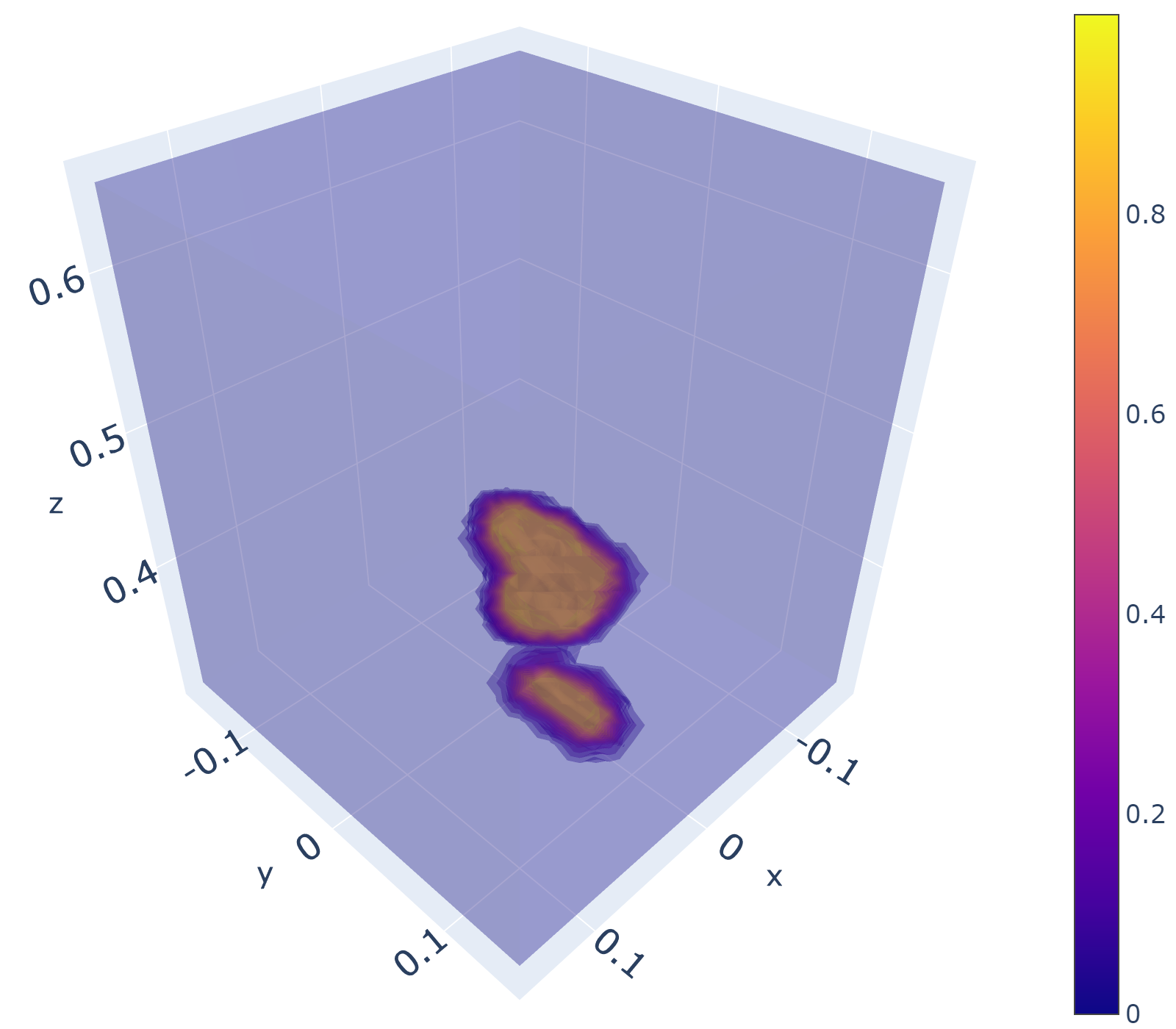}
         \vspace{2.4pt}
     \end{subfigure}
     \hfill
     \begin{subfigure}[b]{0.18\textwidth}
         \centering
         \textbf{Adjoint\vspace{5PT}}
         \includegraphics[width=\textwidth]{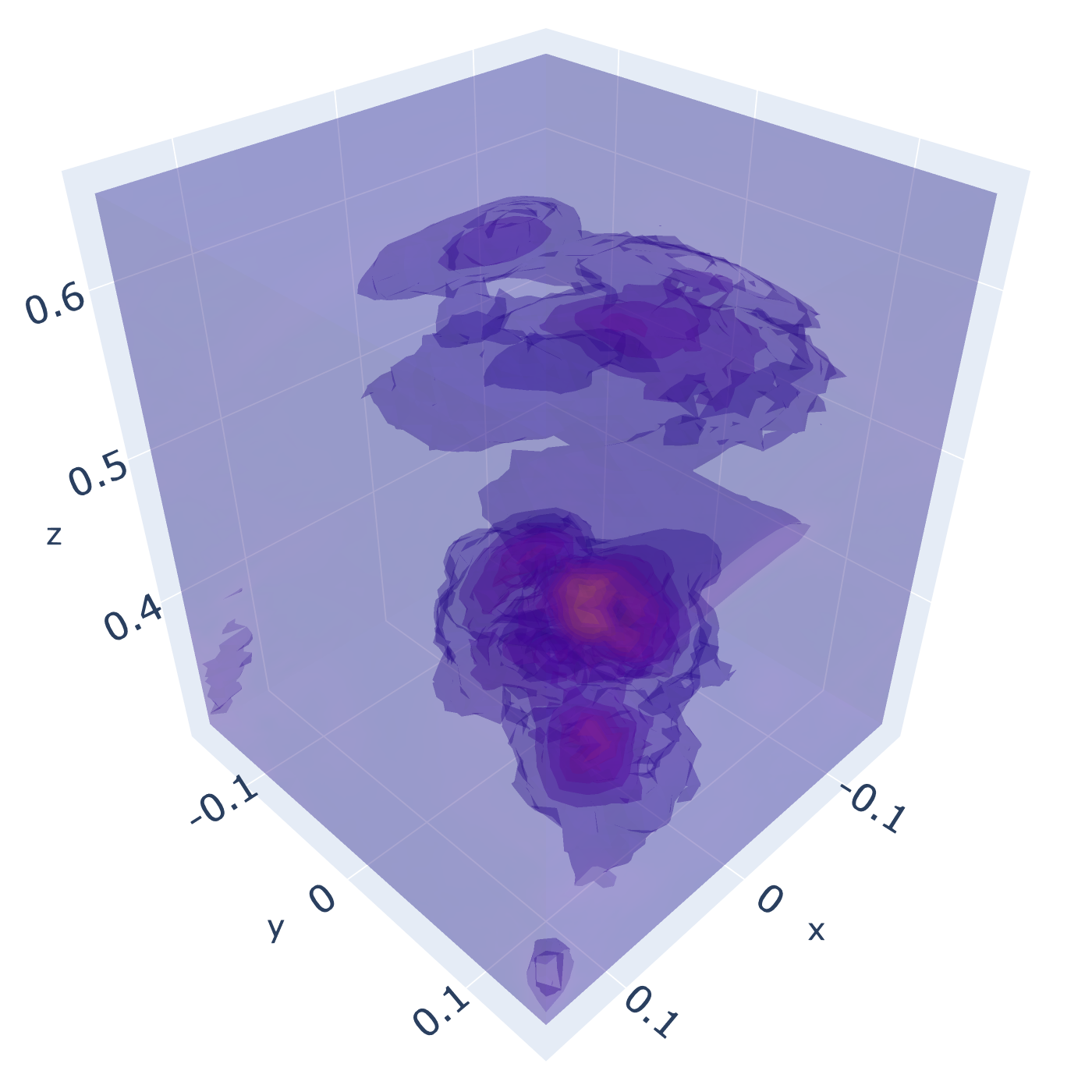}
          \scalebox{.55}{PSNR 23.48 dB - SSIM 0.32}
     \end{subfigure}
     \hfill
     \begin{subfigure}[b]{0.18\textwidth}
         \centering
         \textbf{DeepDI\vspace{5PT}}
         \includegraphics[width=\textwidth]{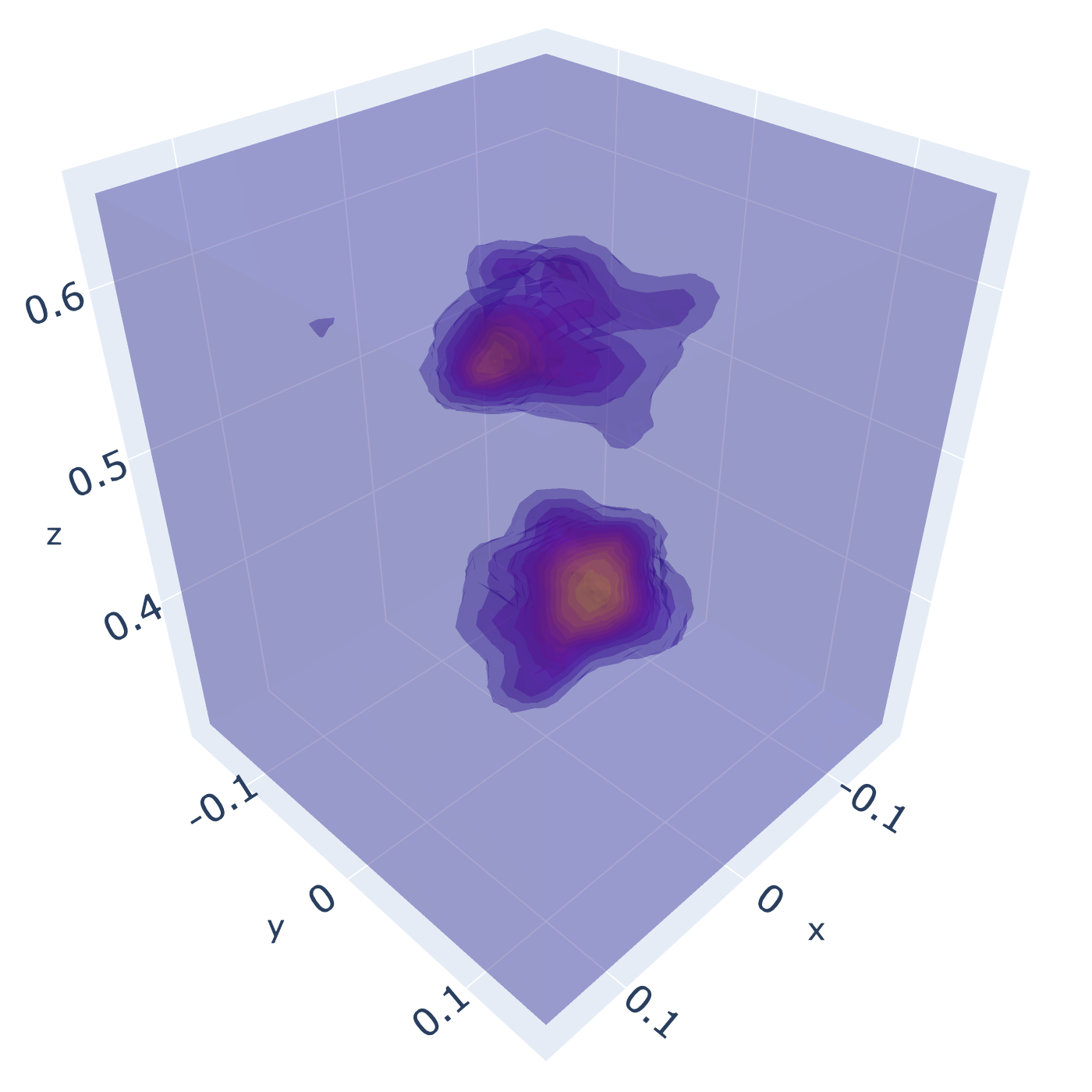}
        \scalebox{.55}{PSNR 22.58 dB - SSIM 0.66}

     \end{subfigure}
     \hfill
     \begin{subfigure}[b]{0.18\textwidth}
         \centering
         \textbf{DeepDI Expanded}
         \includegraphics[width=\textwidth]{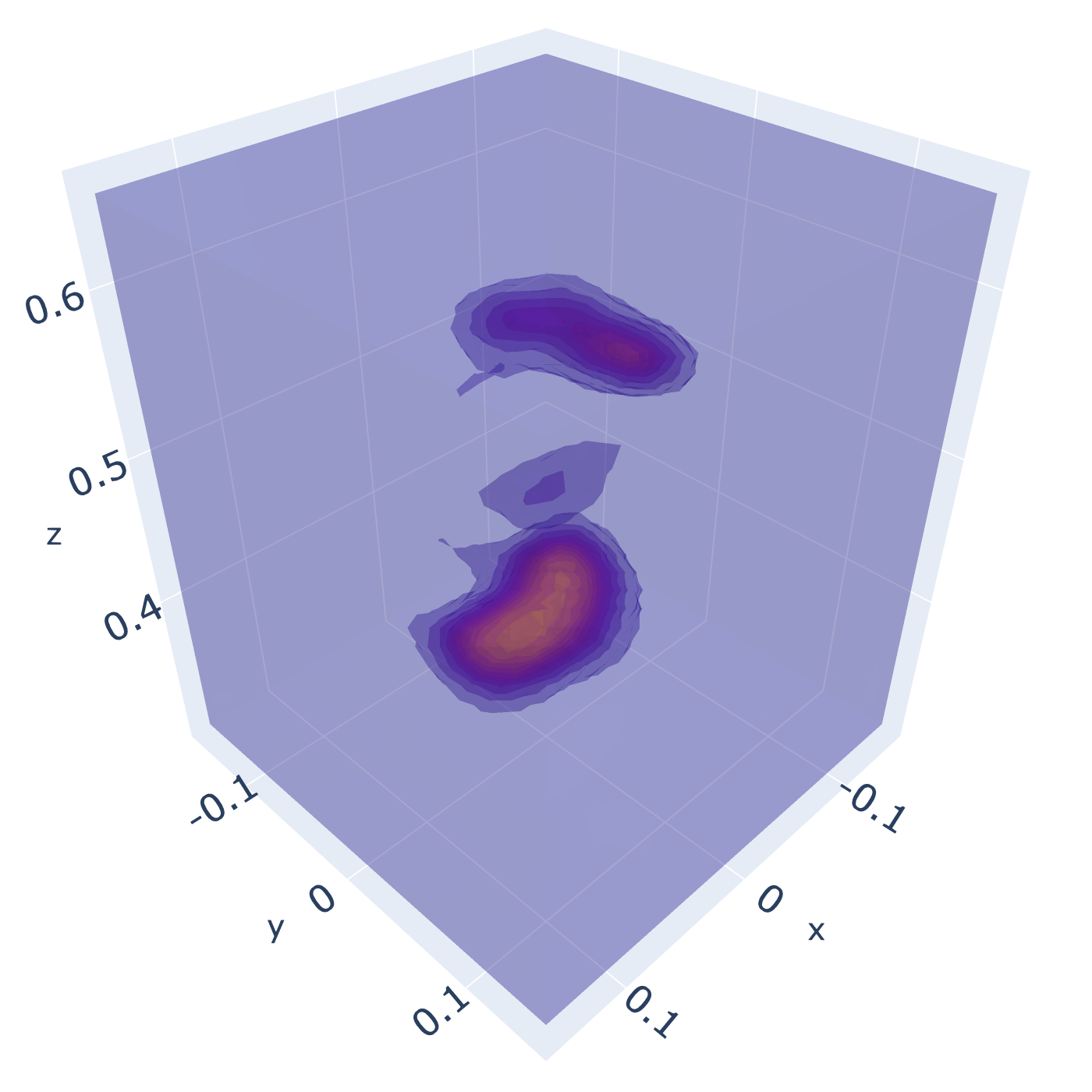}
          \scalebox{.55}{PSNR 23.15 dB - SSIM 0.66}

     \end{subfigure}
     \hfill
     \begin{subfigure}[b]{0.18\textwidth}
         \centering
         \textbf{Deep2S\vspace{5PT}}
         \includegraphics[width=\textwidth]{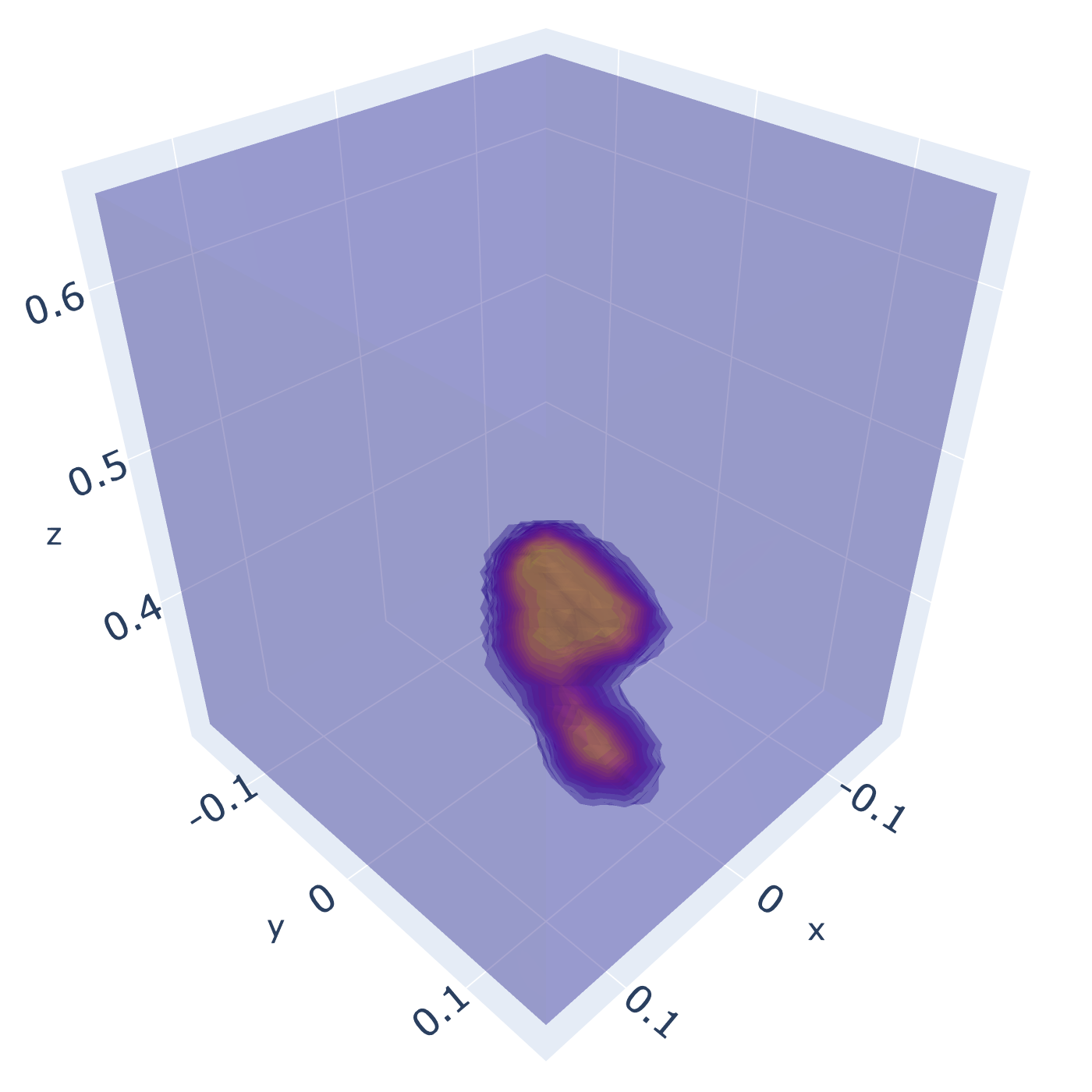}
         \scalebox{.55}{PSNR 29.50 dB - SSIM 0.94} 
     \end{subfigure}

     \begin{subfigure}[b]{0.18\textwidth}
         \centering
         \includegraphics[width=\textwidth]{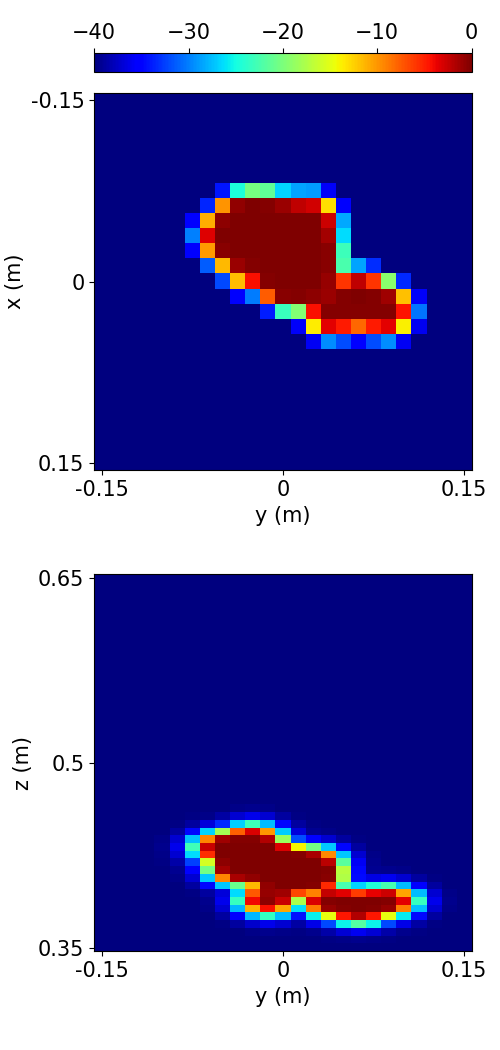}
     \end{subfigure}
     \hfill
     \begin{subfigure}[b]{0.18\textwidth}
         \centering
         \includegraphics[width=\textwidth]{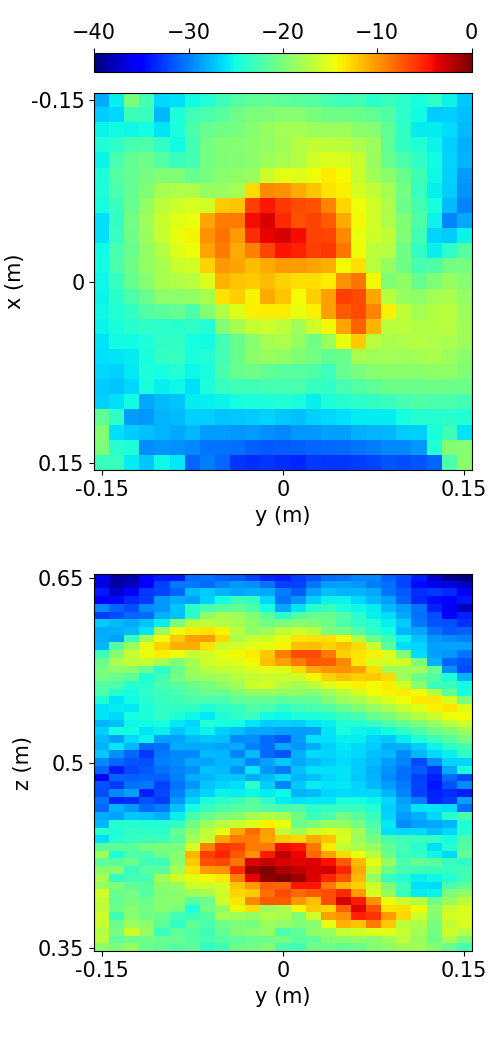}
     \end{subfigure}
     \hfill
     \begin{subfigure}[b]{0.18\textwidth}
         \centering
         \includegraphics[width=\textwidth]{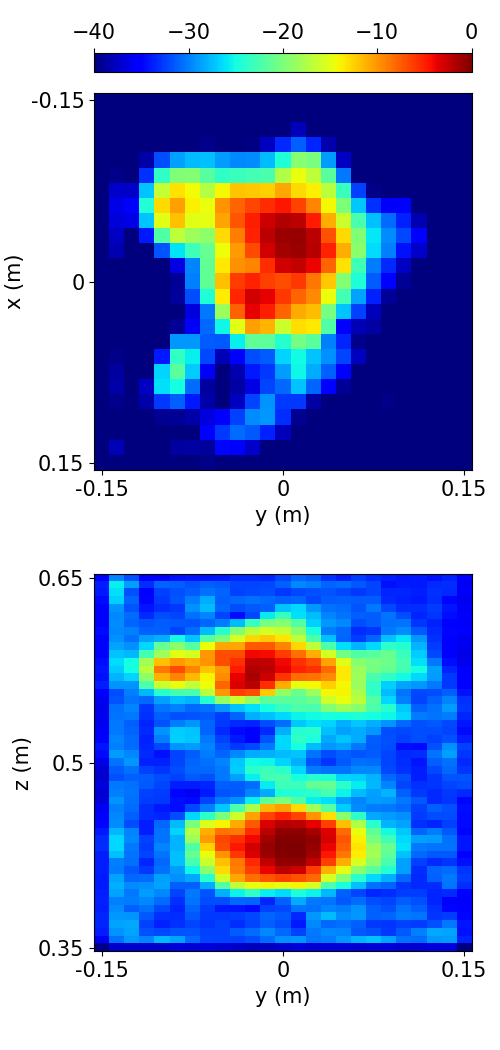}
     \end{subfigure}
     \hfill
     \begin{subfigure}[b]{0.18\textwidth}
         \centering
         \includegraphics[width=\textwidth]{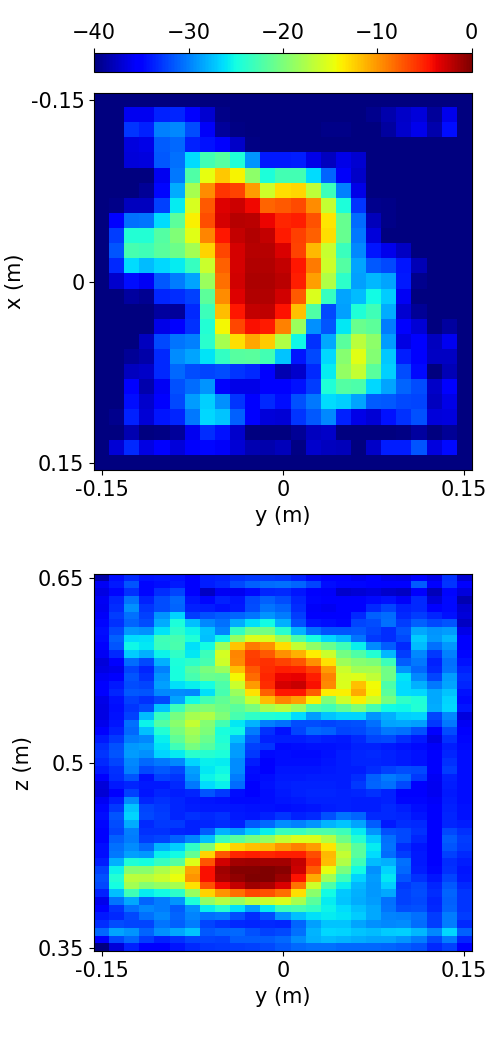}
     \end{subfigure}
     \hfill
     \begin{subfigure}[b]{0.18\textwidth}
         \centering
      \includegraphics[width=\textwidth]{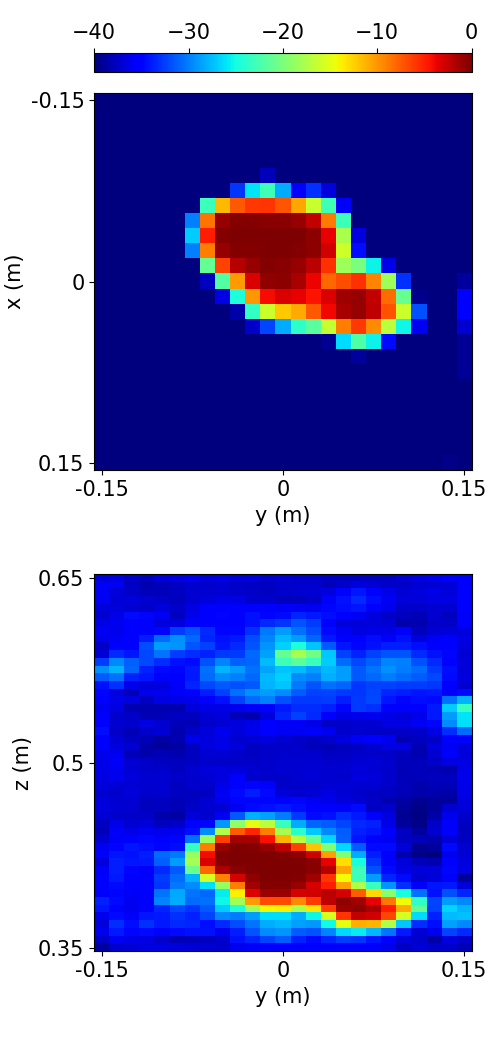}
     \end{subfigure}
 
\caption{Reconstructions of the different algorithms for the first test image of the synthetically generated dataset at 30 dB SNR after adding random phase (Number of Frequency Steps: 15), (The units of x, y, and z-axis are meters). The first row provides the 3D view of the image cube in linear scale whereas the second and third rows show front/side views in dB scale obtained by maximum projection of the image cube onto the $x-y$ and $y-z$ planes, respectively. (3D rotating views of these reconstructions can be found at \href{https://github.com/METU-SPACE-Lab/Efficient-Learned-3D-Near-Field-MIMO-Imaging}{https://github.com/METU-SPACE-Lab/Efficient-Learned-3D-Near-Field-MIMO-Imaging} as video.)}
	\label{fig:result_different_algorithms_phase}
\end{figure}

To also evaluate the performance visually, we provide sample reconstructions using 15 frequency steps (i.e. $\sim$ 8\% data) 
in Figure \ref{fig:result_different_algorithms_phase}. Note that, in the earlier analyses, DeepDI and Deep2S have been trained with the same amount of training data for fairness. However, a large volume of data is generally required for successful training of purely DNN-based approaches that aim to map the measurements directly to the image of interest by exploiting fully connected networks~\cite{kulkarni2016reconnet,zhu2018image,yao2019}. To enrich our analysis, we additionally analyze the performance of the DeepDI using an expanded training dataset. For this, the training and validation datasets are increased five-fold to include 4000 and 500 synthetically generated images, respectively. After training DeepDI with this larger dataset for the initial measurement setting with 15 frequency steps and 30 dB SNR, the average performance on the test dataset increased from $23.4$ PSNR and $0.81$ SSIM to $24.6$ PSNR and $0.84$ SSIM. Although the performance of the DeepDI is slightly improved with the expanded training data, it still does not provide reasonably good reconstructions as illustrated in Figure \ref{fig:result_different_algorithms_phase}. Both DeepDI approaches (trained with the original and expanded datasets) suffer from large errors in the reconstructions.
On the other hand, the Deep2S approach provides a faithful reconstruction although it was trained with five-fold less data compared to DeepDI Expanded. This again illustrates the significance of incorporating physics-based knowledge into the reconstruction process, whenever available.

\subsection{SNR Analysis}
\label{subsection:SNR}
We now analyze the effect of SNR on the reconstruction performance of the Deep2S method. For this, we consider the previous measurement setting with 15 frequency steps and random phase-added reflectivities. For our analysis, we vary the SNR from -10 dB to 30 dB by 10 dB steps. By using transfer learning to re-train the 3D U-Net, we obtain five different networks which are fine-tuned for different SNR values -10, 0, 10, 20, and 30 dB. 
To train our 3D U-Net architecture, the radar measurements of the generated dataset are corrupted with additive white Gaussian noise with the specified SNR values. 

\begin{figure}[tbh!]
\centering
\scalebox{.85}{\includegraphics[width=0.7\linewidth]{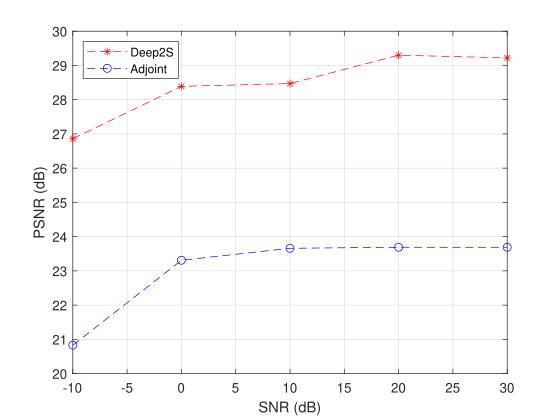}}
	\caption{Average PSNR values of adjoint operation and, Deep2S approach for 100 test images of the synthetically generated dataset versus measurement SNR.}
	\label{fig:SNR_PSNR}
\end{figure}

The average reconstruction performance of the Deep2S approach and the adjoint operation is plotted with respect to SNR in Figure \ref{fig:SNR_PSNR}. The results indicate that the average reconstruction performance of the Deep2S approach does not drop significantly until -10 dB SNR. Furthermore, even when SNR is -10 dB, the Deep2S approach provides an acceptable reconstruction performance with an average PSNR of 26.9 dB. Therefore, we can conclude that the performance of the Deep2S is mostly robust to different noise levels.

\begin{figure}[tbh!]
	\centering
	\includegraphics[width=0.9\linewidth]{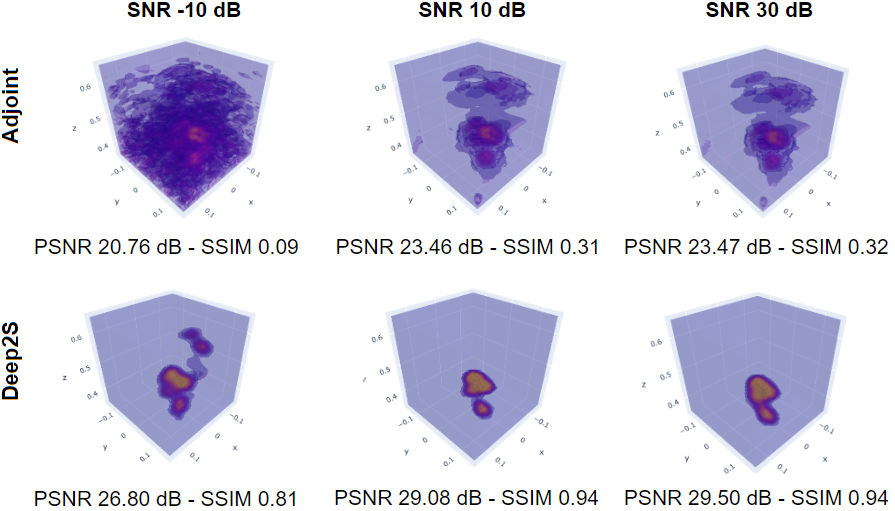}
	\caption{Reconstructions of the adjoint operation and, Deep2S approach for the first test image of the synthetically generated dataset (Number of Frequency Steps: 15).}
	\label{fig:SNR_visual}
\end{figure}

Figure \ref{fig:SNR_visual} shows sample reconstructions for the adjoint and the Deep2S approach at -10, 10, and 30 dB SNR. As seen, for all noise levels the Deep2S approach substantially reduces the artifacts on the intermediate reconstructions obtained with the adjoint operation. For example, when we inspect the adjoint and the Deep2S reconstructions at -10dB SNR, we observe that Deep2S can provide reconstructions that preserve the general shape and location of the object although the measurements are highly contaminated by noise. Moreover, for higher SNR cases with 10dB and 30dB, adjoint operation still results in artifact clusters distributed mainly at the upper voxels, which is mostly a consequence of the sparse measurements (i.e. $\sim$ 8\% data). Nevertheless, Deep2S is capable of providing high-quality reconstructions by removing most of these artifacts. 

\subsection{Performance Comparison and Improvement 
with Different Network Architectures}
\label{subsection:Different Networks}
There are various different DNN architectures developed for enhancement and denoising purpose both in the context of 2D real-valued imaging~\cite{ongie2020,jin2017deep} as well as radar imaging~\cite{cheng2020compressive,Gao2019Enhanced,jing2022enhanced,yang2020isar, mu2020deepimaging, hu2020inverse}.  
In this section, we analyze the effect of using different architectures on the performance of the developed Deep2S and DeepDI approaches.

Since Residual Net (ResNet) architectures and 2D convolutional kernels have been used in various earlier radar imaging works including 3D near-field MIMO imaging, first, we change the 3D U-Net architecture in the second stage of Deep2S with an ablated version involving 2D convolutional kernels as well as with a 3D ResNet architecture. The implementation of the 2D U-Net architecture involves replacing the 3D convolutions in Fig.~\ref{fig:U_Net} with 2D convolutions. The implementation of the ResNet architecture is based on \cite{zhang2017beyond}. Here we use a modified version of the original ResNet architecture by replacing the 2D convolutions with 3D convolutions. 
Moreover, to achieve a similar training time (complexity) with the 3D U-Net architecture, the network layer size is selected as 10. The training of both the 2D U-Net and ResNet take approximately an hour. The number of parameters in the 2D U-Net and 3D ResNet architectures are respectively 996,609 and 890,817.

\begin{table}[b]
    \centering
    \caption{Average PSNR and SSIM Values for 15 Frequency Steps at 30dB SNR for Different Network Architectures}
    \label{tab:ResNet}
    \begin{tabular}{P{3cm}P{3cm}P{3cm}}
    \toprule
         ResNet & 2D U-Net & 3D U-Net  \\ 
        \midrule
         26.1/0.66 & 24.47/0.77 & \textbf{29.2}/\textbf{0.93}\\
        \bottomrule
    \end{tabular}
\end{table}

\begin{figure}[tbh!]
	\centering
 
 \centering
     \begin{subfigure}[b]{0.2\textwidth}
         \centering
         \textbf{Ground Truth\vspace{7.42PT}}
         \includegraphics[width=\textwidth]{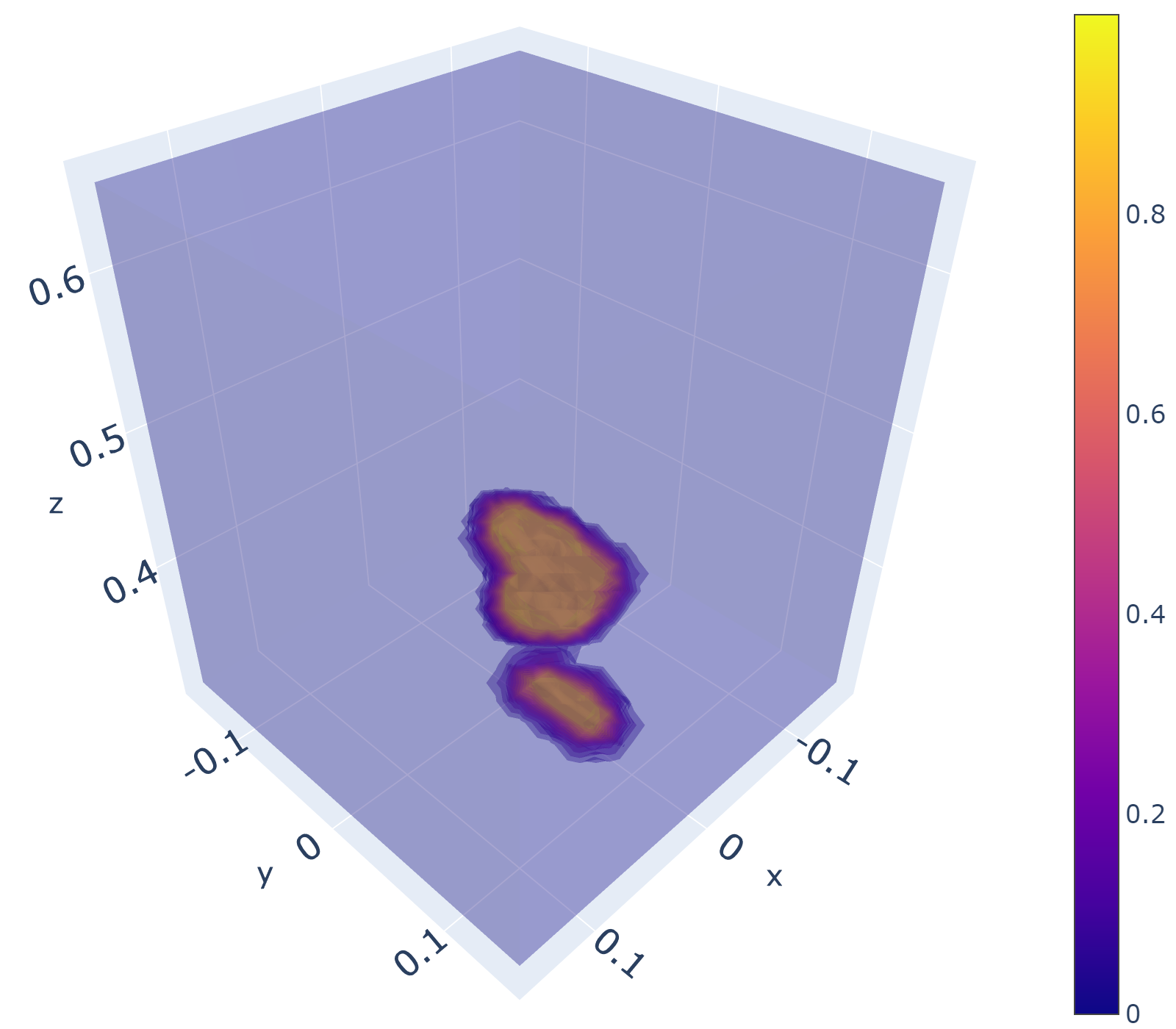}
         \vspace{2.4pt}
     \end{subfigure}
     \hfill
     \begin{subfigure}[b]{0.18\textwidth}
         \centering
         \textbf{ResNet\vspace{5PT}}
         \includegraphics[width=\textwidth]{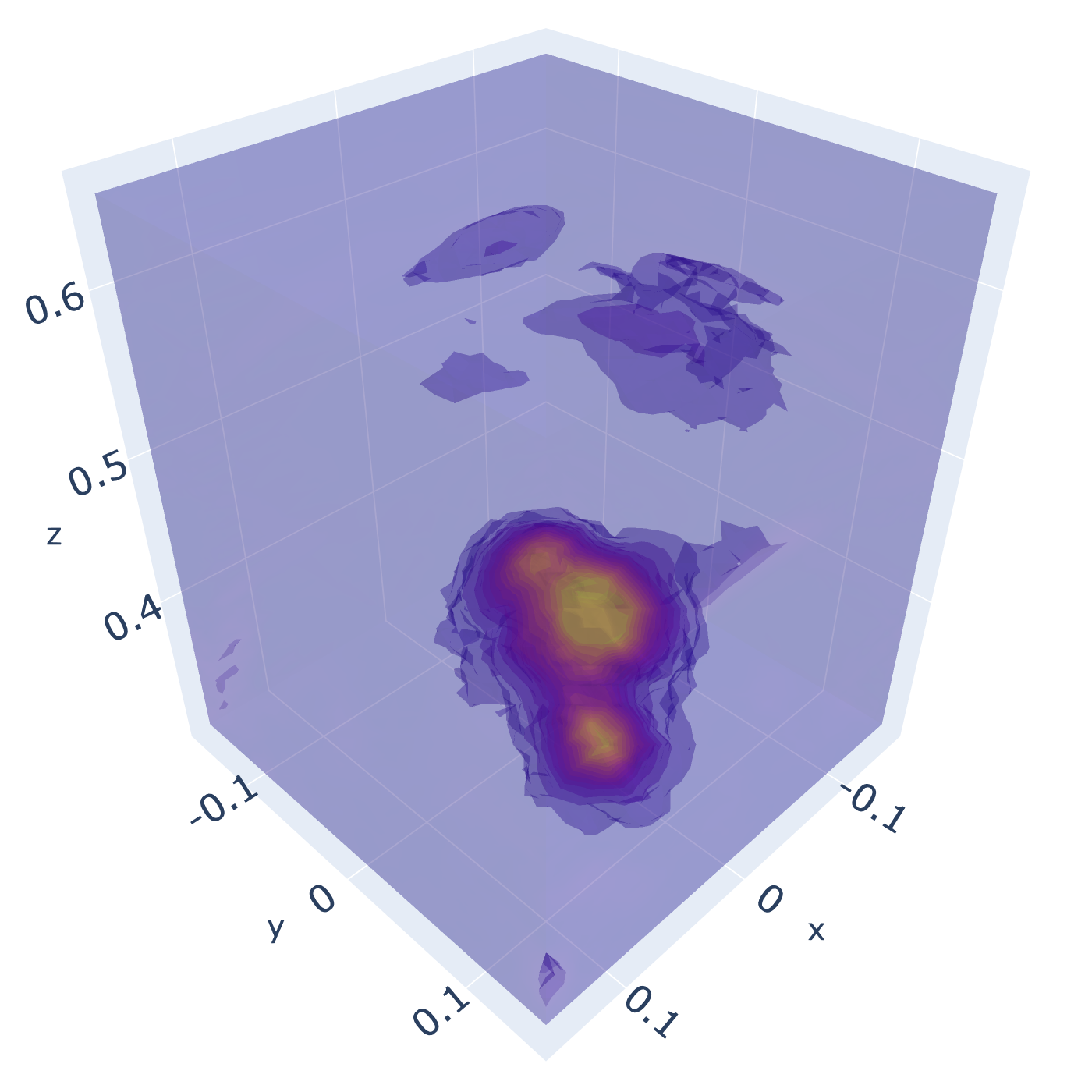}
        \scalebox{.55}{PSNR 26.64 dB - SSIM 0.55}
     \end{subfigure}
     \hfill
    \begin{subfigure}[b]{0.18\textwidth}
         \centering
         \textbf{2D U-Net\vspace{5PT}}
         \includegraphics[width=\textwidth]{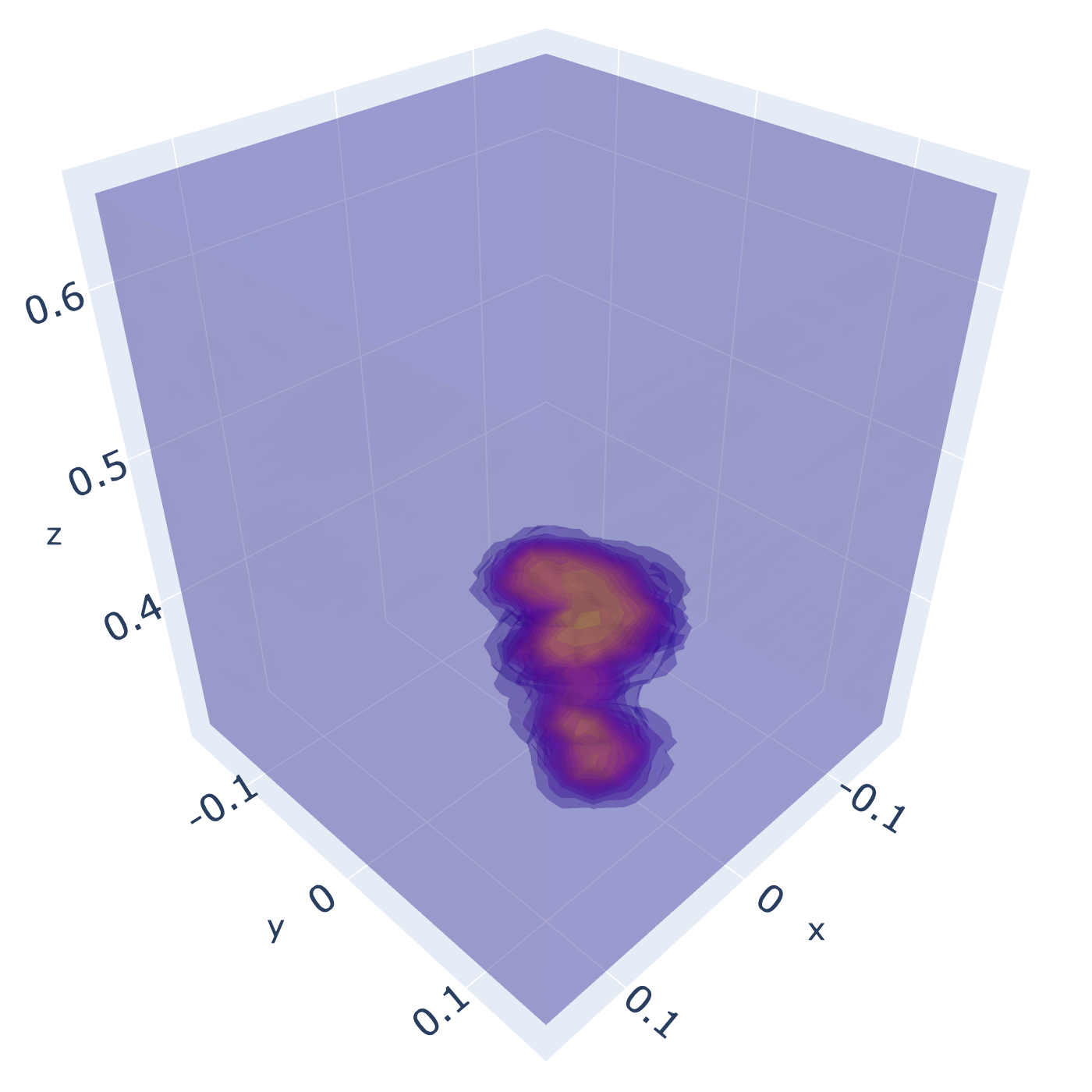}
          \scalebox{.55}{PSNR 26.76 dB - SSIM 0.78}
     \end{subfigure}
     \hfill
     \begin{subfigure}[b]{0.18\textwidth}
         \centering
         \textbf{3D U-Net\vspace{5PT}}
         \includegraphics[width=\textwidth]{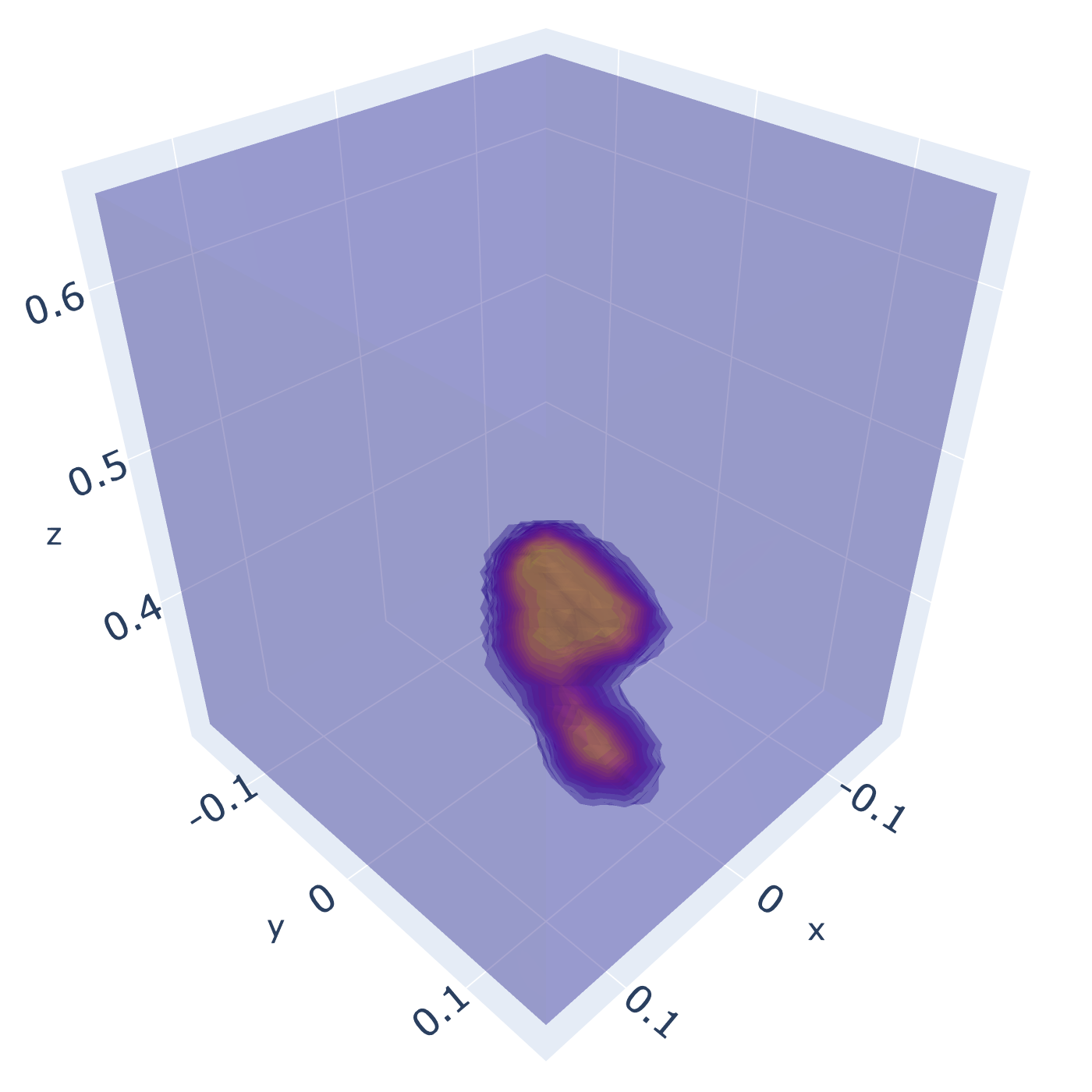}
          \scalebox{.55}{PSNR 29.50 dB - SSIM 0.94}

     \end{subfigure}

     \begin{subfigure}[b]{0.18\textwidth}
         \centering
         \includegraphics[width=\textwidth]{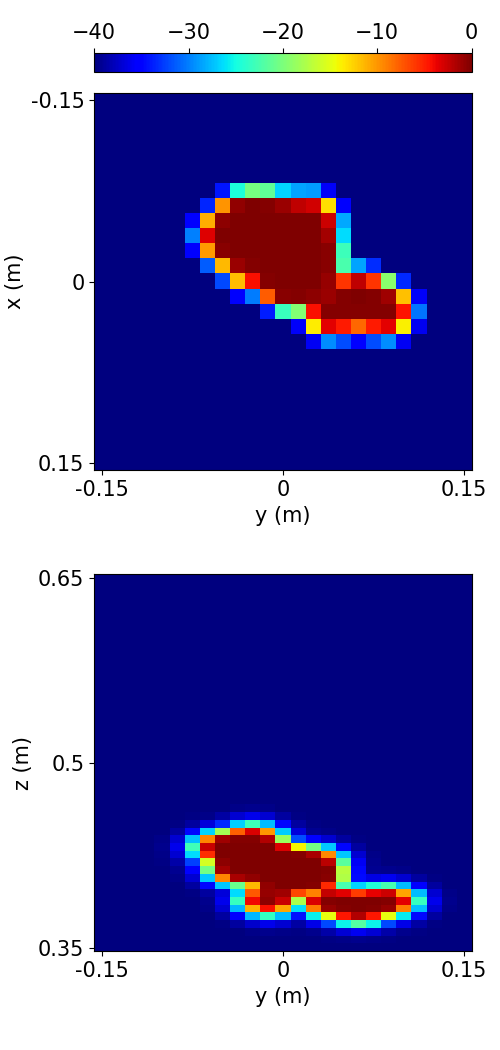}
     \end{subfigure}
     \hfill
     \begin{subfigure}[b]{0.18\textwidth}
         \centering
         \includegraphics[width=\textwidth]{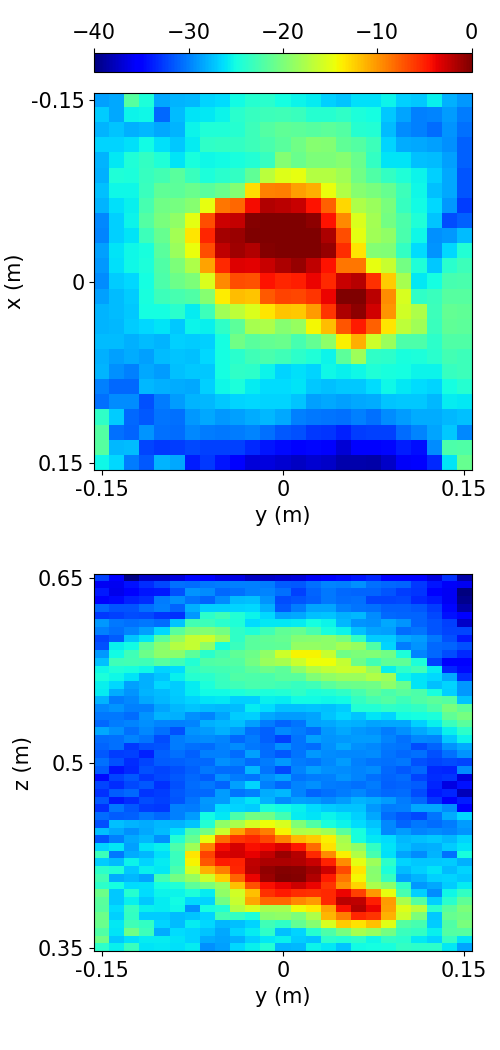}
     \end{subfigure}
     \hfill
     \begin{subfigure}[b]{0.18\textwidth}
         \centering
         \includegraphics[width=\textwidth]{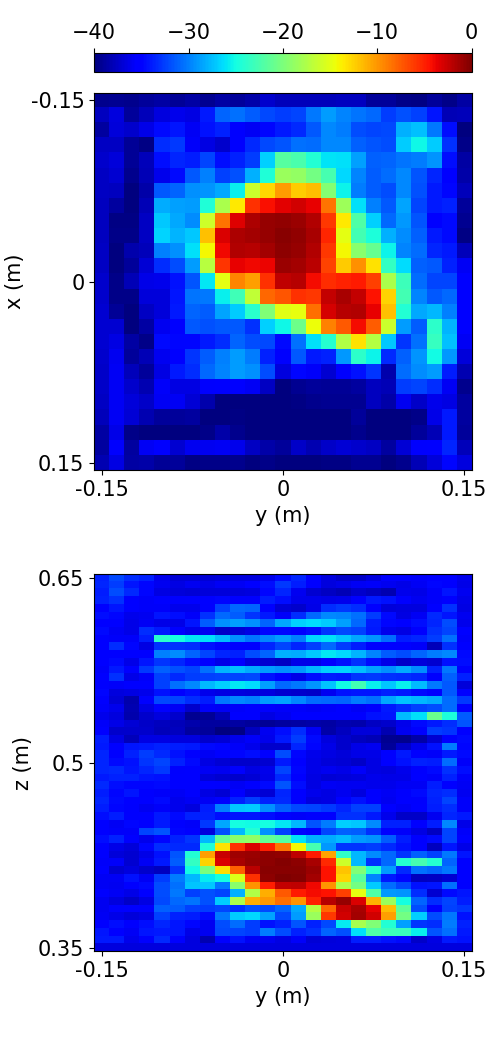}
     \end{subfigure}
     \hfill
     \begin{subfigure}[b]{0.18\textwidth}
         \centering
         \includegraphics[width=\textwidth]{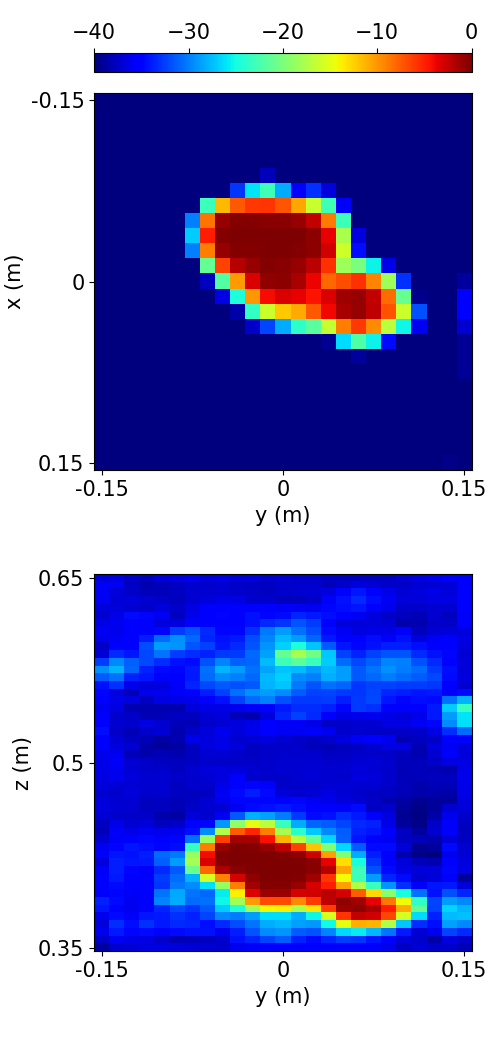}
     \end{subfigure}

	\caption{Deep2S Reconstructions using Different DNN Architectures for the first test image of the synthetically generated dataset at 30 dB SNR (Number of Frequency Steps: 15). The first row provides the 3D view of the image cube in linear scale whereas the second and third rows show front/side views in dB scale obtained by maximum projection of the image cube onto the $x-y$ and $y-z$ planes, respectively. (3D rotating views of these reconstructions can be found at \href{https://github.com/METU-SPACE-Lab/Efficient-Learned-3D-Near-Field-MIMO-Imaging}{https://github.com/METU-SPACE-Lab/Efficient-Learned-3D-Near-Field-MIMO-Imaging} as video.) }
	\label{fig:resnet}
\end{figure}

Table \ref{tab:ResNet} shows the average reconstruction performance of the developed Deep2S approach with 3D U-Net architecture in comparison to its modified versions with 3D ResNet and 2D U-Net, for the setting with 15 frequency steps and 30 dB SNR. These results demonstrate that 3D U-Net architecture that involves 3D convolutions in a multi-scale structure yields better performance than both the 3D ResNet and 2D U-Net. This illustrates the superiority of the 3D U-Net architecture for jointly exploiting range and cross-range correlations of extended targets as used in the second stage of the developed approaches. In particular, this result also validates our expectation that 3D convolution kernels are better suited for capturing 3D extended targets compared to their 2D analogs. For visual evaluation, we also provide sample reconstructions obtained using 3D ResNet and 2D U-Net architectures in Figure \ref{fig:resnet}. As seen, the ResNet and 2D U-Net yield worse reconstructions with more artifacts along both cross-range and range directions, whereas the reconstruction obtained with 3D U-Net architecture is more faithful with the ground truth. 

After analyzing the performance with different denoisers in the second stage, we now conduct another analysis by performing two modifications related to the first stage. In the first modification, we provide the complex-valued intermediate reconstructions as input to the denoiser instead of providing only their magnitudes. The goal here is to investigate whether processing the complex-valued intermediate reconstructions, as done in some of the earlier works~\cite{Gao2019Enhanced,jing2022enhanced,yang2020isar,mu2020deepimaging}, increases the imaging quality over only processing the magnitude. Due to the random phase nature of the scene reflectivities in various applications, our approaches aim to refine only the magnitude of the back-projected measurements obtained at the output of the first stage. We conjecture that the phase component of the back-projected measurements in the image domain does not constitute a useful information due to its random nature, and hence we provide magnitude-only information as input to the second stage. On the other hand, some of the earlier works focused on refining the intermediate reconstructions in their complex-valued form by either processing real-imaginary parts as two independent channels or using complex-valued CNNs (with larger number of trainable parameters).  To the best of our knowledge, there is no work that compared complex-valued processing with magnitude-only processing, and showed the superiority of one of them to the other. For this reason, we perform this comparative evaluation that has been missing in the earlier works. As the second modification, we change the adjoint operation used in the first stage of the Deep2S with a fully connected trainable layer. The goal here is to analyze whether multiplication with a learned matrix provides better performance than multiplication with the adjoint matrix ($\A^H$) to provide the intermediate reconstruction. This approach can also be considered as a modified version of our DeepDI approach to reduce its need for a substantial training data.

To process complex-valued intermediate reconstructions, we remove the magnitude operation from the first stage of Deep2S. Instead, we provide the complex-valued representation as input to the 3D U-Net after normalizing it with its magnitude. The first layer of the 3D U-Net is hence modified to accept two-channel input, which consists of the real and imaginary components of the complex-valued intermediate reconstruction. The resulting U-Net architecture contains 2,874,017 parameters (which is slightly larger than before). We train this architecture with a learning rate of $10^{-3}$ by using 800 random-phase added training scenes for the setting with 30dB measurement SNR and 15 frequency steps. This type of complex-valued refinement is often referred as RV-CNN in the earlier radar imaging works. Here we call this modified version of Deep2S as CV-Deep2S (Complex-Valued Deep2S).

As another modification of our approach, we replace the adjoint operation in the first stage with a fully connected trainable layer to perform multiplication with a learned matrix instead of adjoint matrix. This approach shares similarity with the model-based learning methods as 
the observation matrix of the imaging system is transformed into a learnable set of weights and initialized using physics-based observation model to improve training. 
Utilizing this trainable layer followed by magnitude operation as the first stage, we obtain an hybrid approach between DeepDI and Deep2S, and call it Deep2S+ since the training is performed through transfer learning from Deep2S. The first stage of Deep2S+ serves to project the complex-valued measurements to the image space by performing a linear operation as follows:
\begin{equation}
    \begin{bmatrix}
         \mathbf{v}_R \\
         \mathbf{v}_I
        \end{bmatrix} = \begin{bmatrix}
         \mathbf{P}_R & -\mathbf{P}_I\\
         \mathbf{P}_I & \mathbf{P}_R
        \end{bmatrix}
        \begin{bmatrix}
         \y_R \\
         \y_I 
        \end{bmatrix}
        \label{eqn:RealRepresetation}
\end{equation}
Here $\y_R = \Re\{\y\}, \y_I = \Im\{\y\} \in \R^M $ are real and imaginary components of the measurements, $\mathbf{P}_R,\mathbf{P}_I \in \R^{N\times M}$ represent the real and imaginary components of the applied matrix, and similarly $\mathbf{v}_R$, $\mathbf{v}_I$ are real and imaginary components of the result. This operation simply corresponds to the multiplication of the complex-valued measurements with a complex-valued matrix, and is similar to the operations used for the construction of complex-valued convolutional neural networks~\cite{Gao2019Enhanced, jing2022enhanced}. The trainable parameters of this layer are $\mathbf{P}_R$ and $\mathbf{P}_I$. The total number of parameters in Deep2S+ approach is equal to 146,198,153 (hence 2 fold increase compared to DeepDI).

Note that both DeepDI and Deep2S+ are purely DNN-based approaches. But while the first stage of DeepDI independently processes the real and imaginary components of the measurements to form an intermediate reconstruction for the image magnitude, Deep2S+ jointly processes these components (similar to Deep2S) to form a complex-valued intermediate reconstruction (whose magnitude is then input to the second stage). We train Deep2S+ with a learning rate of $10^{-3}$ by using 800 random-phase added training scenes at 30dB measurement SNR and 15 frequency steps. Here we apply transfer learning from Deep2S and initialize the 3D U-Net architecture in the second stage with the pre-trained Deep2S model of the same setting. To exploit physics-based information, we also initialize the projection layer in the first stage using $\A^H$, i.e. by setting $\mathbf{P}_R = \Re\{\A^H\}$ and $\mathbf{P}_I = \Im\{\A^H\}$. Note that this initialization provides the adjoint result as output. Therefore, at the beginning of the training of Deep2S+, we exactly have the Deep2S, and the goal is to improve both the first and second stages of the Deep2S through training (hence the name Deep2S+).

The average performance of CV-Deep2S, Deep2S and Deep2S+ are shown in Table \ref{tab:PerformanceDeep2SPlus}. As seen, the performance increase with Deep2S+ compared to DeepDI is substantial, which illustrates the significance of utilizing physics-based information to give a head-start to a fully trainable architecture. Deep2S+ also provides improvement over Deep2S with a higher PSNR and SSIM, which shows that the training improved the first and second stages of the Deep2S. On the other hand, although the average PSNR of CV-Deep2S exceeds Deep2S and Deep2S+, its average SSIM is substantially lower and even close to that of Deep2S utilizing 2D U-Net. This suggests that processing the intermediate reconstructions in complex-valued form (instead of magnitude) does not yield an improvement on the performance.

\begin{table}[b]
    \centering
    \caption{Average PSNR and SSIM Values for 15 Frequency Steps at 30dB SNR for Different Network Architectures}
    \label{tab:PerformanceDeep2SPlus}
    \begin{tabular}{P{3cm}P{3cm}P{3cm}}
    \toprule
        CV-Deep2S & Deep2S & Deep2S+ \\ 
        \midrule
       29.8/0.80 & 29.2/0.93 & 29.4/0.95 \\
        \bottomrule
    \end{tabular}
\end{table}

\begin{figure}[tbh!]

    \centering
      \begin{subfigure}[b]{0.2\textwidth}
         \centering
         \textbf{Ground Truth\vspace{7.42PT}}
         \includegraphics[width=\textwidth]{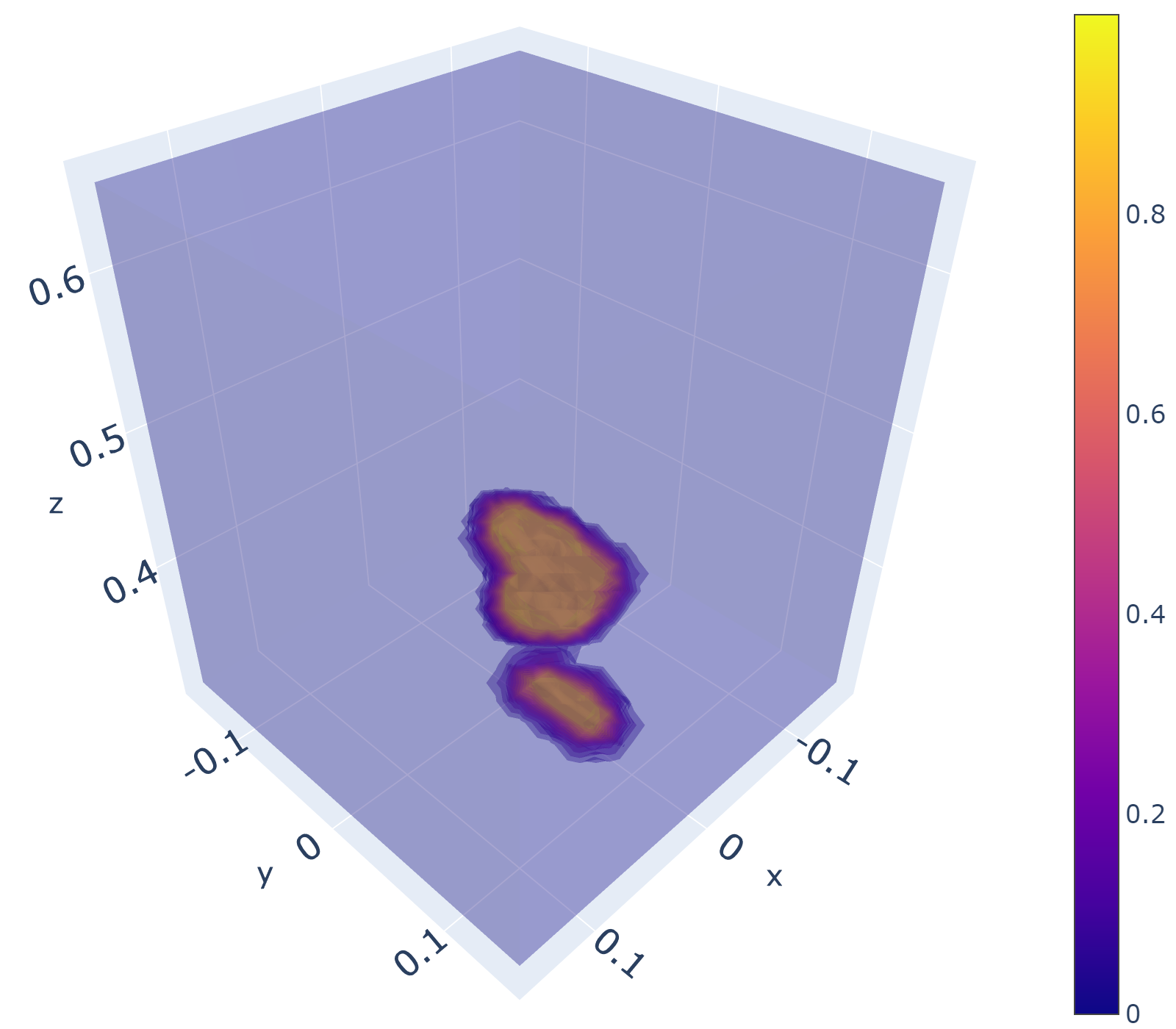}
         \vspace{2.4pt}
     \end{subfigure}     
     \hfill
    \begin{subfigure}[b]{0.18\textwidth}
         \centering
         \textbf{CV-Deep2S\vspace{5PT}}
         \includegraphics[width=\textwidth]{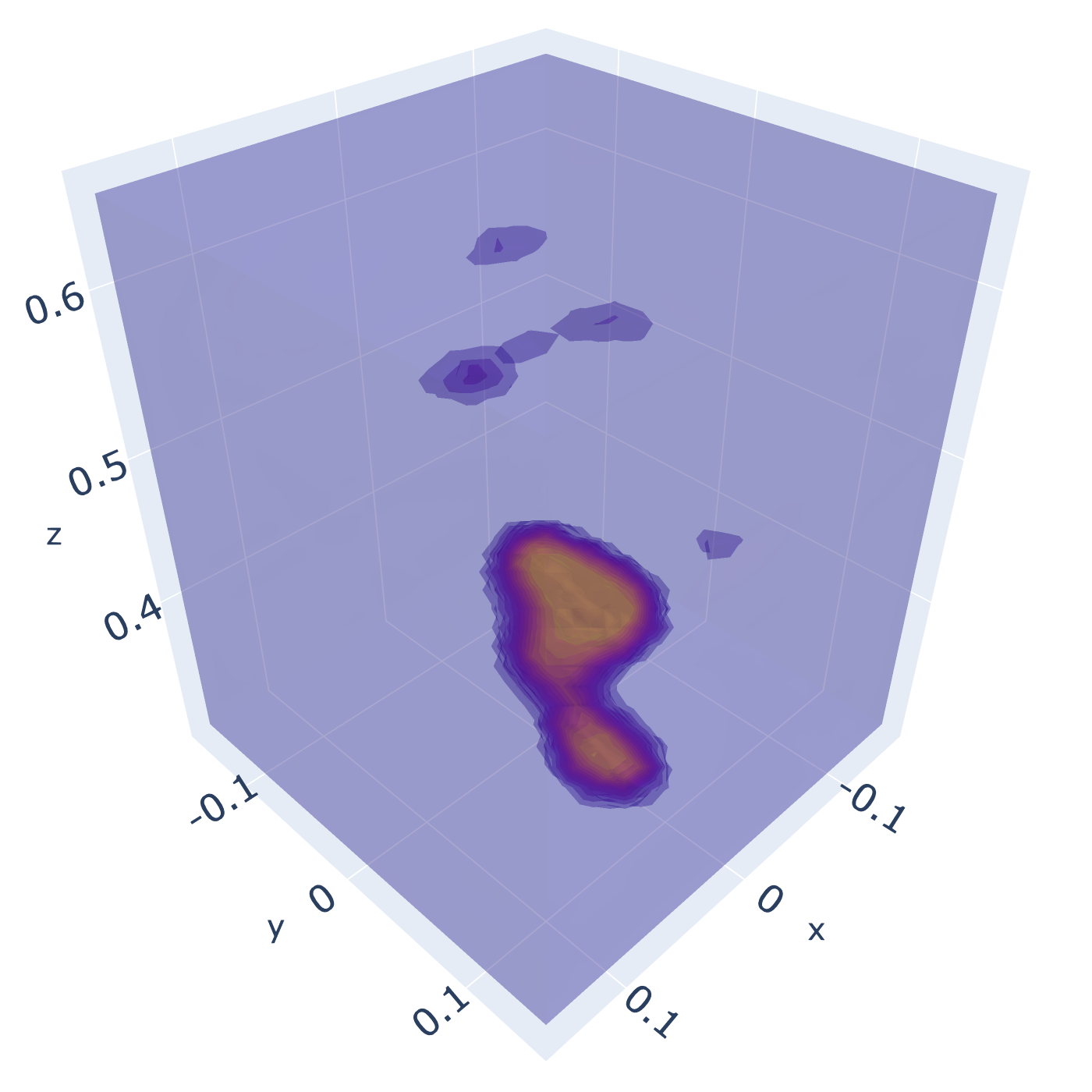}
        \scalebox{.55}{PSNR 30.90 dB - SSIM 0.79}
     \end{subfigure}
     \hfill
     \begin{subfigure}[b]{0.18\textwidth}
         \centering
         \textbf{Deep2S\vspace{5PT}}
         \includegraphics[width=\textwidth]{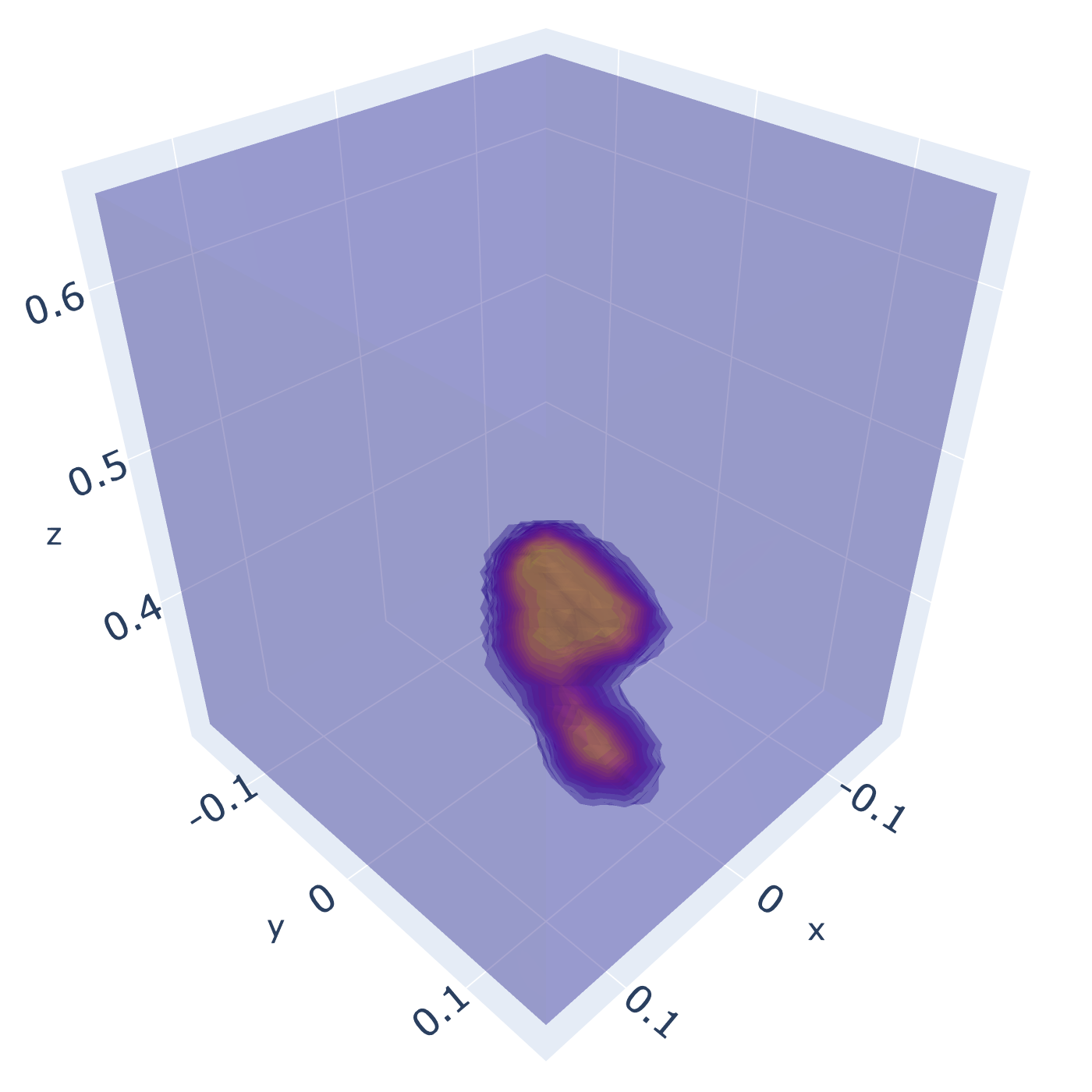}
          \scalebox{.55}{PSNR 29.50 dB - SSIM 0.94}
     \end{subfigure}
     \hfill
     \begin{subfigure}[b]{0.18\textwidth}
         \centering
         \textbf{Deep2S+\vspace{5PT}}
         \includegraphics[width=\textwidth]{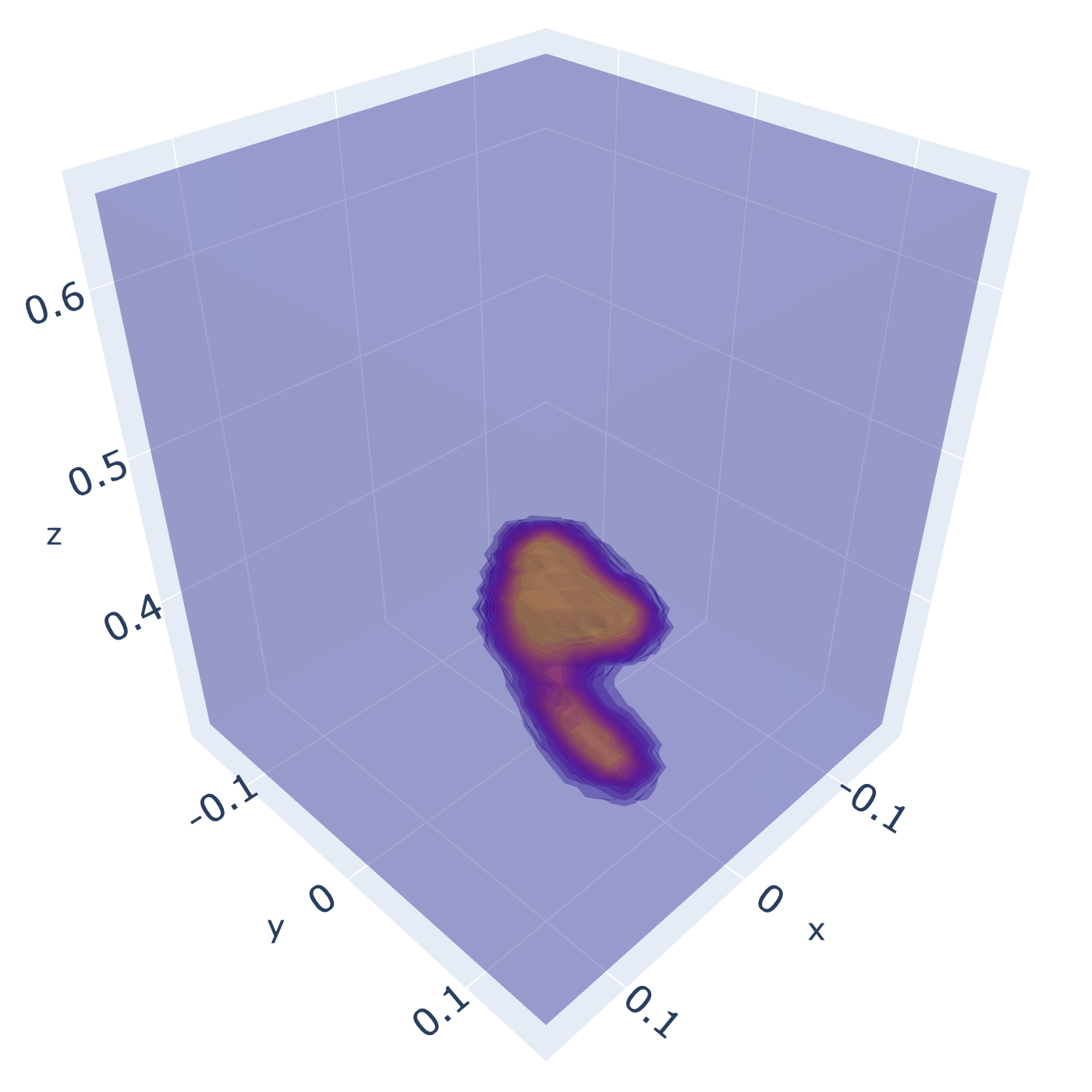}
        \scalebox{.55}{PSNR 28.88 dB - SSIM 0.94}
     \end{subfigure}

    \begin{subfigure}[b]{0.18\textwidth}
         \centering
         \includegraphics[width=\textwidth]{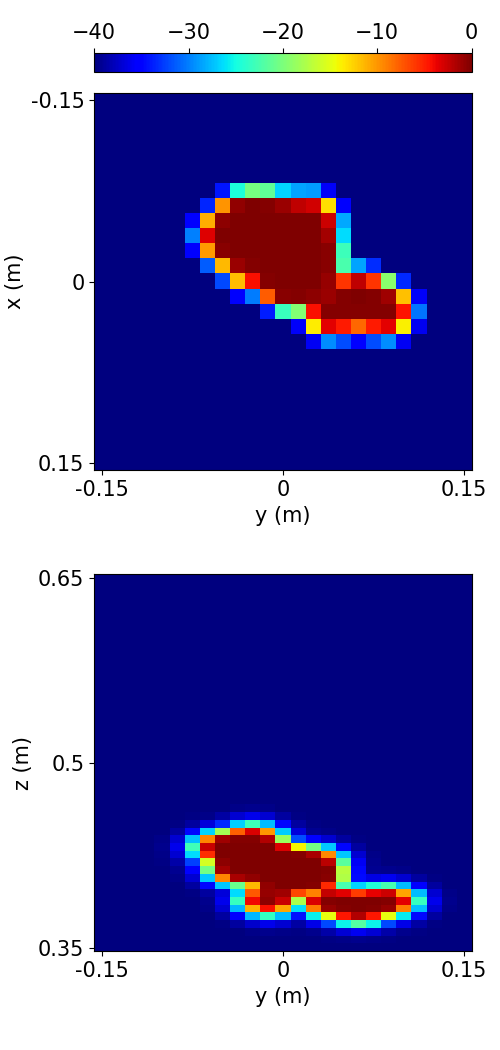}
     \end{subfigure}     
     \hfill
     \begin{subfigure}[b]{0.18\textwidth}
         \centering
         \includegraphics[width=\textwidth]{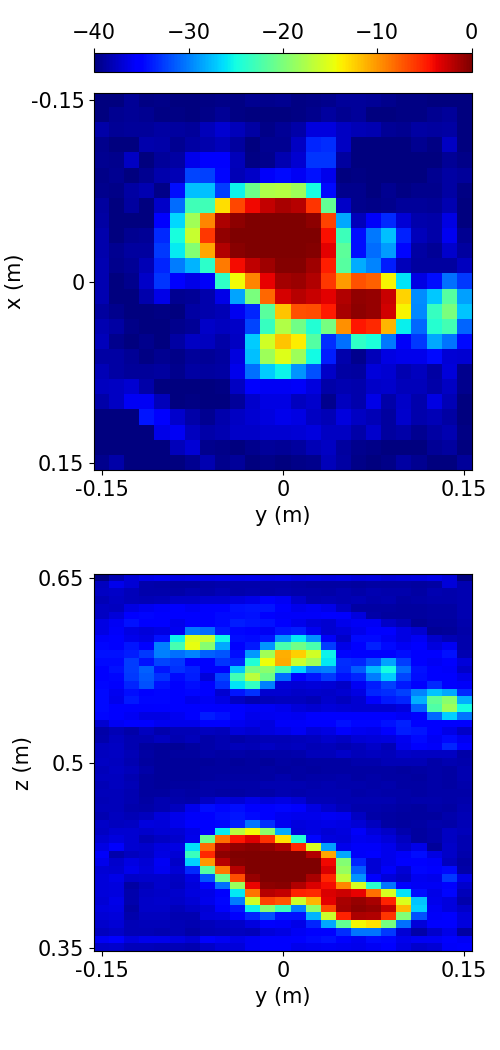}
     \end{subfigure}
    \hfill 
     \begin{subfigure}[b]{0.18\textwidth}
         \centering
         \includegraphics[width=\textwidth]{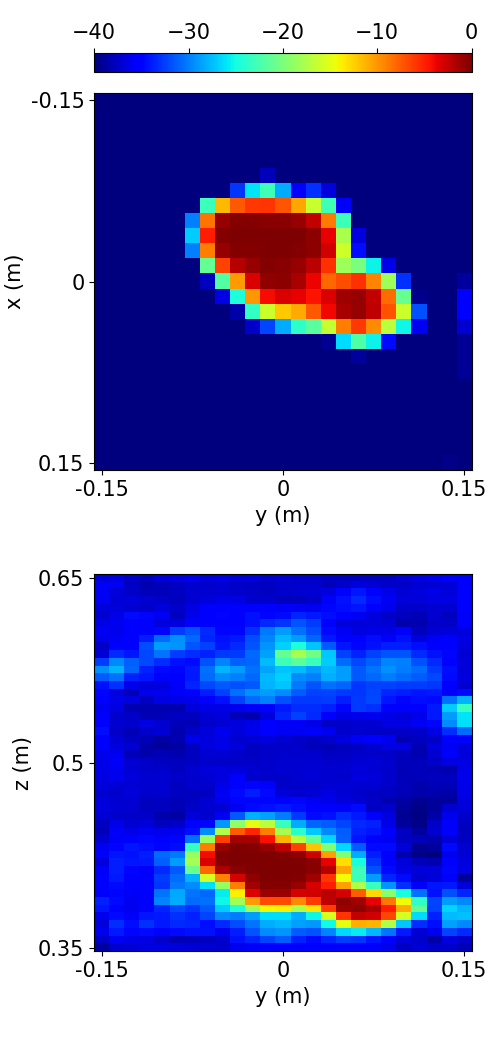}
     \end{subfigure}
     \hfill
     \begin{subfigure}[b]{0.18\textwidth}
         \centering
         \includegraphics[width=\textwidth]{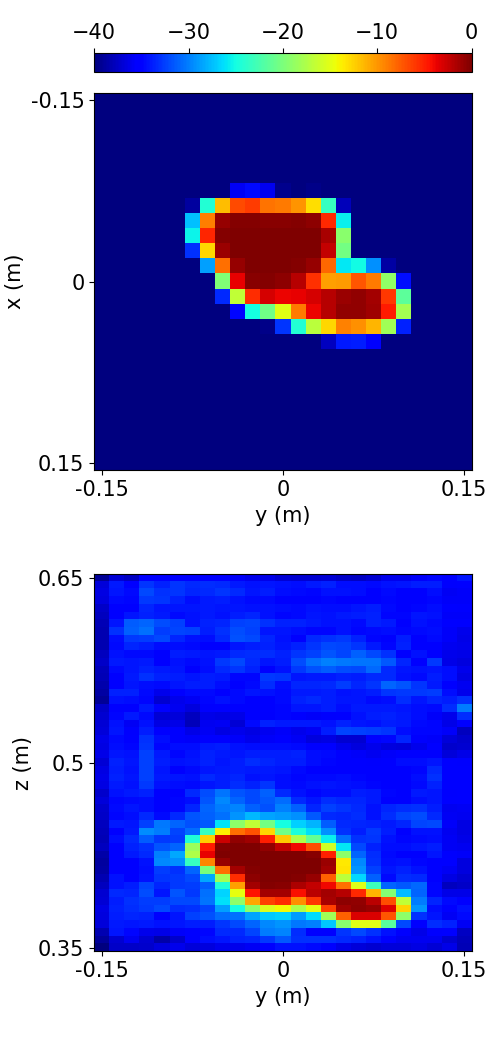}
     \end{subfigure}

    \caption{Deep2S Reconstructions of the first test image of the synthetically generated dataset at 30 dB SNR (Number of Frequency Steps: 15) using Different Architectures for the First Stage. The first row provides the 3D view of the image cube in linear scale whereas the second and third rows show front/side views in dB scale obtained by maximum projection of the image cube onto the $x-y$ and $y-z$ planes, respectively. (3D rotating views of these reconstructions can be found at \href{https://github.com/METU-SPACE-Lab/Efficient-Learned-3D-Near-Field-MIMO-Imaging}{https://github.com/METU-SPACE-Lab/Efficient-Learned-3D-Near-Field-MIMO-Imaging} as video.)}
    \label{fig:Deep2S+}
\end{figure}

To also visually evaluate the performance, CV-Deep2S, Deep2S and Deep2S+ reconstructions are illustrated in Fig.~\ref{fig:Deep2S+} for a sample image cube from the test dataset. As seen, CV-Deep2S reconstruction contains artifact clusters, especially around $z=0.57$ plane, while Deep2S and Deep2S+ reconstructions are mostly artifact-free. While the highest PSNR is achieved by CV-Deep2S for this sample image, SSIM is substantially lower compared to the other two (similar to the average performance in Table~\ref{tab:PerformanceDeep2SPlus}). Compared to Deep2S and CV-Deep2S, Deep2S+ better reduces the artifacts along the z-axis arising due to under-sampling (as seen in the maximum projection onto the $y-z$ plane), but slightly suffers from over-smoothing (which might be due to being fully adapted to the training data). These results, along with Table~\ref{tab:PerformanceDeep2SPlus}, suggest that processing the complex-valued intermediate reconstructions does not yield increased image quality as compared to magnitude-only processing. 

\subsection{Resolution Analysis}

The resolution of a near-field MIMO imaging system is characterized in a conventional observation setting for a non-sparse antenna array and non-sparse frequency sampling~\cite{zhuge2012three}. However, as well-known, the reconstruction quality and resolution degrade in compressive settings with limited data. This is the case here since the data is acquired with a sparse MIMO array at few frequency steps. Here we investigate the resolution achieved with our reconstruction approach at compressive settings and compare this to the expected theoretical resolution for the conventional (non-compressive) setting. For this, firstly, we perform conditioning-based resolution analysis (similar to \cite{oktem2021High-Resolution}) using only the observation matrix, which is independent of the reconstruction method. Secondly, we consider multiple separated point targets as sample scenes and demonstrate the resolving capability of our approach by analyzing the reconstructed images.

\begin{figure}[tbh!]
\centering
	\includegraphics[width=0.52\linewidth]{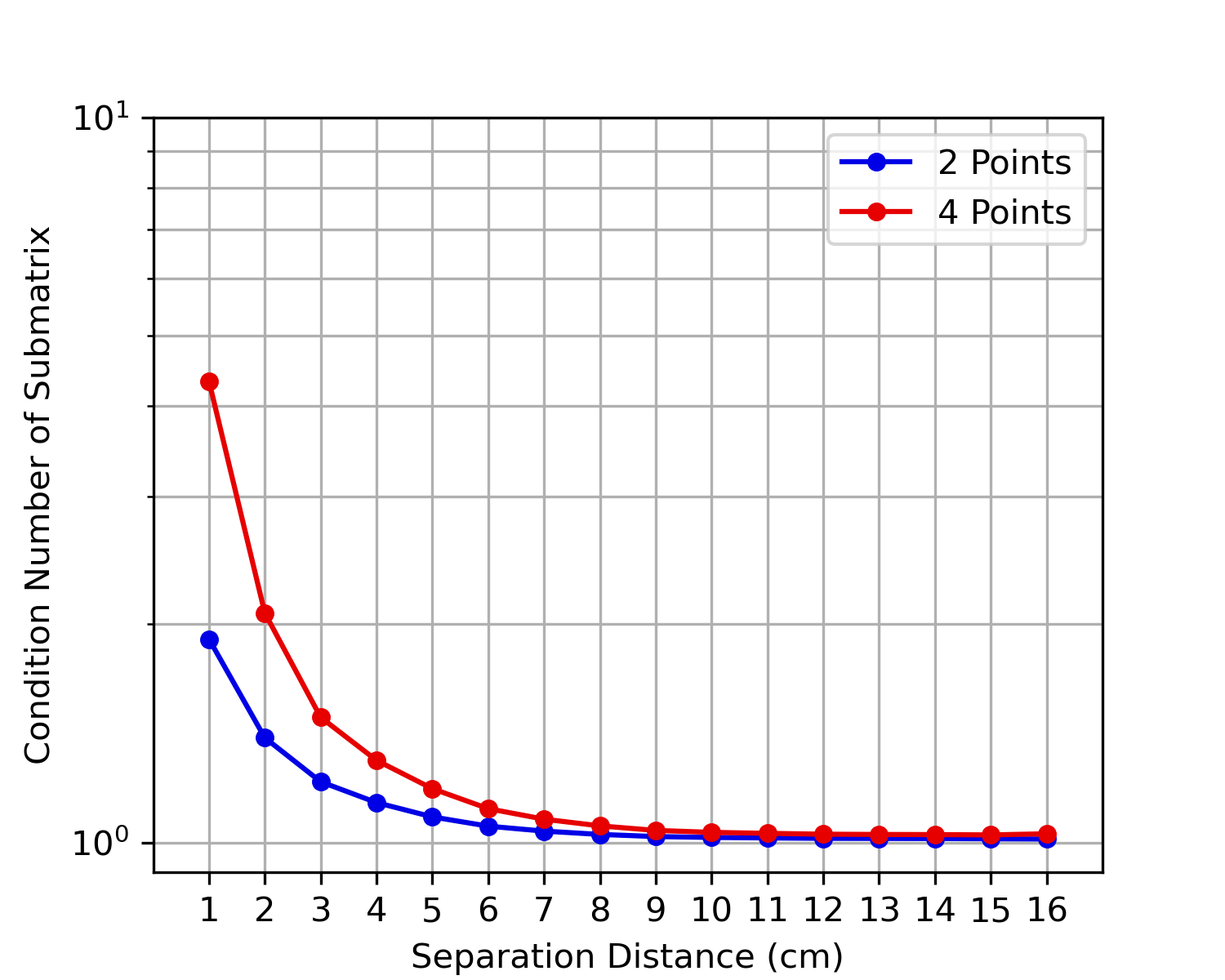}
	\caption{Conditioning of the relevant submatrices of $\A$ for
different number of point sources and separation distances.}
	\label{fig:Condition Number}
\end{figure}

In the first analysis, we consider the conditioning of the inverse problem where the goal is to estimate the values of multiple separated point targets whose locations are known when 15 frequency steps are used. Note that if this problem cannot be properly solved (owing to high condition number), the original reconstruction task of estimating both the value and location of the point targets will also not be possible to solve. Based on this fact, we investigate the condition number of the mentioned inverse problem, which is characterized by the condition number of the submatrix obtained from the observation matrix $\A$ by only keeping the columns associated with the locations of the considered point targets. For this, we consider scenes containing 2 or 4 point targets with different separation distances. The targets lie on the middle slice of the image cube along the range direction, and are placed with same distance to the center of the $x-y$ plane (as an example, see Fig.~\ref{fig:MultipleTargetsCrossRange}). The horizontal and vertical separation of the targets are changed from 1cm to 20cm with 1cm steps. Figure~\ref{fig:Condition Number} shows the condition number of the submatrices of $\A$ for these cases. As expected, conditioning gets worse as the number of point targets increases, or their separation distance decreases. More importantly, condition number appears to saturate when the separation distance is increased to 5-6 cm, which suggests that achievable resolution is close to these values. This corresponds to almost 2-fold worse resolution in our compressive setting with only $\sim$ 8\% data compared to the expected theoretical resolution of the conventional non-compressive setting (i.e. 2.5 cm).

\newcommand{\addpic}[1]{\adjustbox{valign=m}{\includegraphics[width=0.9\linewidth]{#1}}}
\begin{figure}[tbh!]
\centering
\begin{subfigure}[t]{0.45\textwidth}
\centering
\begin{tabular}{m{0.7cm}|M{0.35\linewidth}M{0.35\linewidth}}
           $\delta_{x,y}$ & {\centering Ground Truth} & {\centering Deep2S} \\
           \toprule
\rotatebox{90}{\hspace{-1cm}2.50 cm} & \addpic{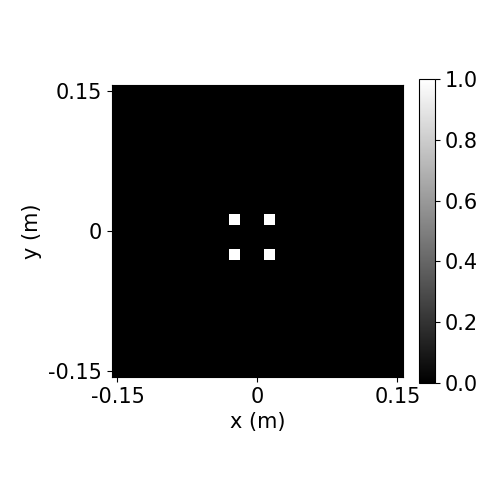} & \addpic{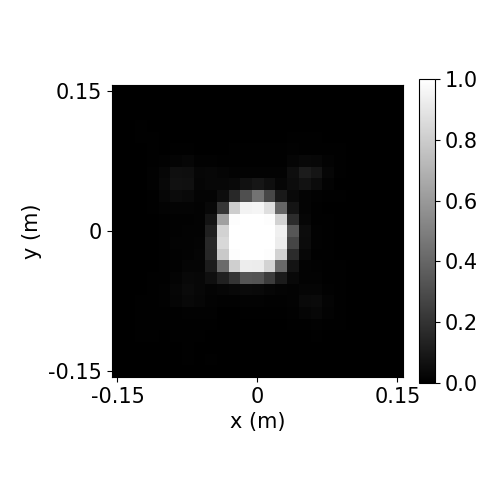}  \\
\rotatebox{90}{\hspace{-1cm}3.75 cm} & \addpic{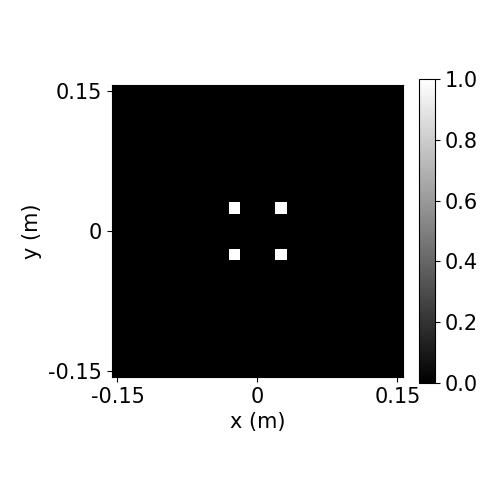} & \addpic{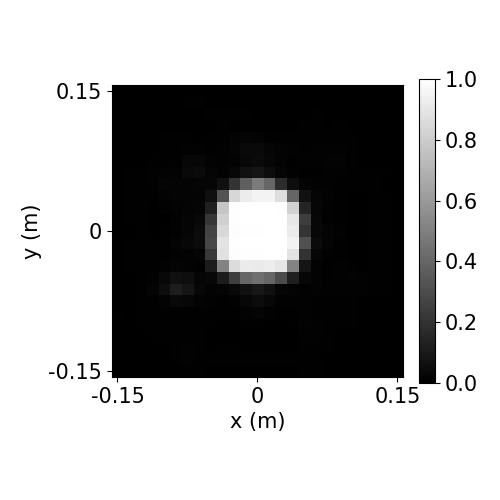}  \\
\rotatebox{90}{\hspace{-1cm}5.00 cm} & \addpic{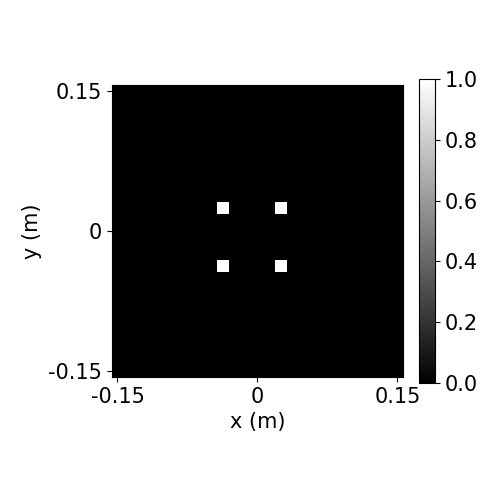} & \addpic{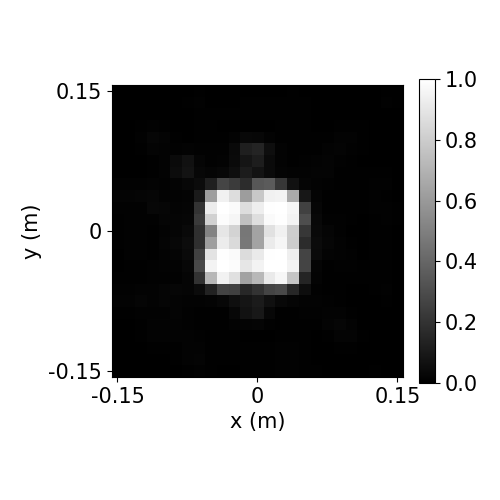}  \\
\rotatebox{90}{\hspace{-1cm}6.25 cm} & \addpic{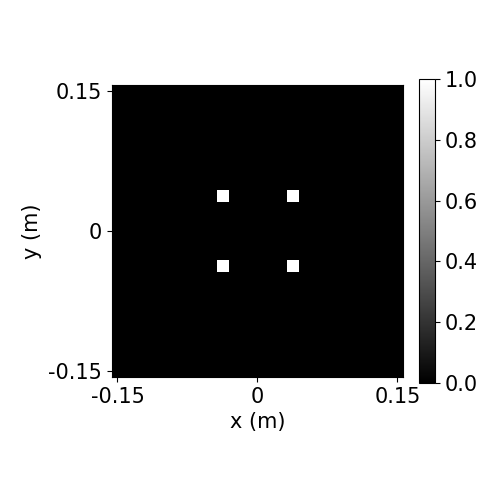} & \addpic{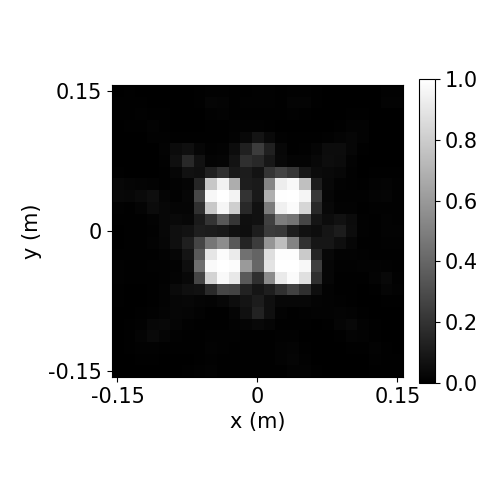}  
\end{tabular}
\caption{Cross-Range Resolution Analysis at $z=50$cm plane}
\label{fig:MultipleTargetsCrossRange}
\end{subfigure}
\begin{subfigure}[t]{0.45\textwidth}
\centering
\begin{tabular}{m{0.7cm}|M{0.35\linewidth}M{0.35\linewidth}}
           $\delta_{z}$ & {\centering Ground Truth} & {\centering Deep2S} \\
           \toprule
\rotatebox{90}{\hspace{-1cm}1.250 cm} & \addpic{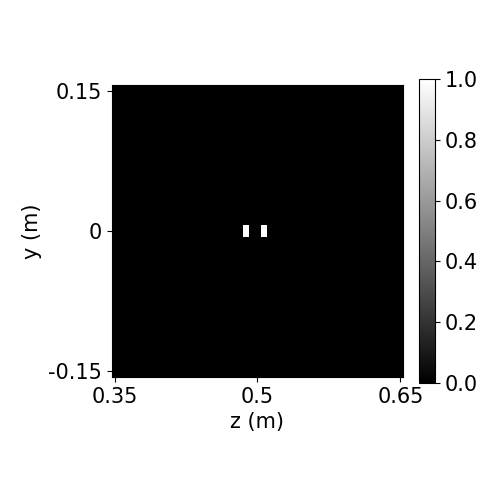} & \addpic{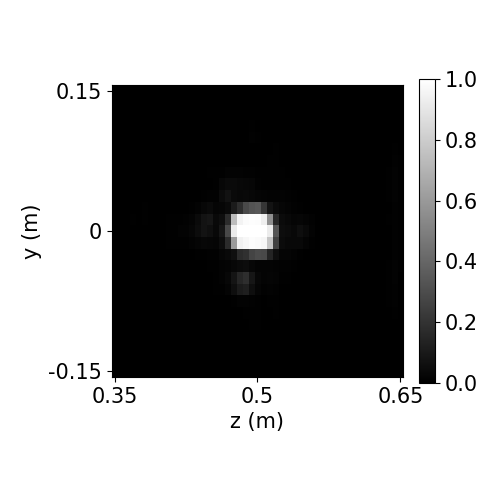}  \\
\rotatebox{90}{\hspace{-1cm}1.875 cm} & \addpic{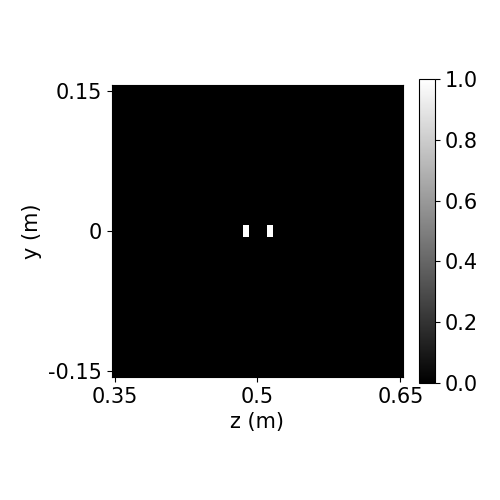} & \addpic{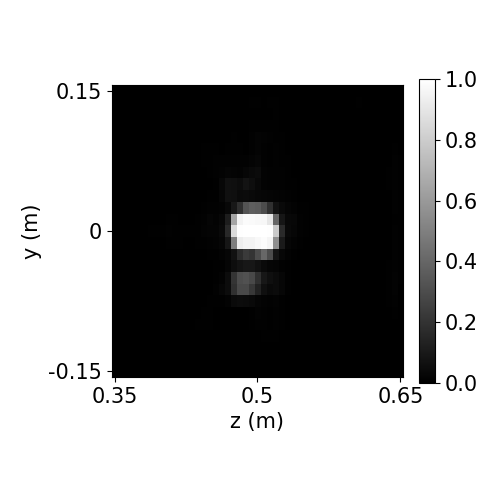}  \\
\rotatebox{90}{\hspace{-1cm}2.500 cm} & \addpic{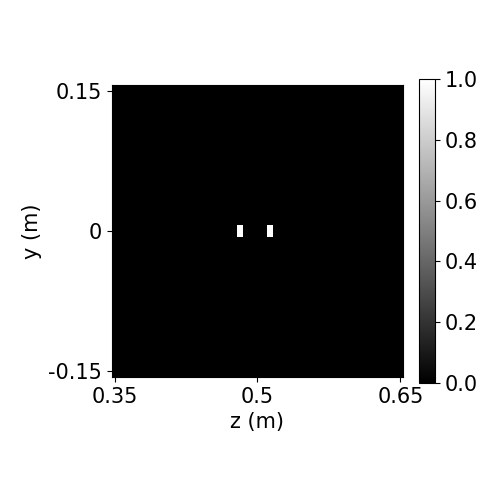} & \addpic{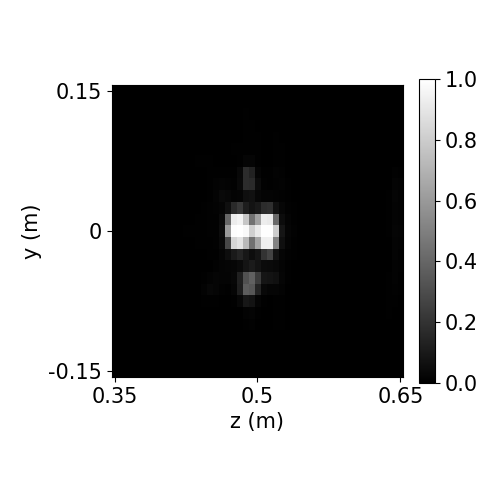}  \\
\rotatebox{90}{\hspace{-1cm}3.125 cm} & \addpic{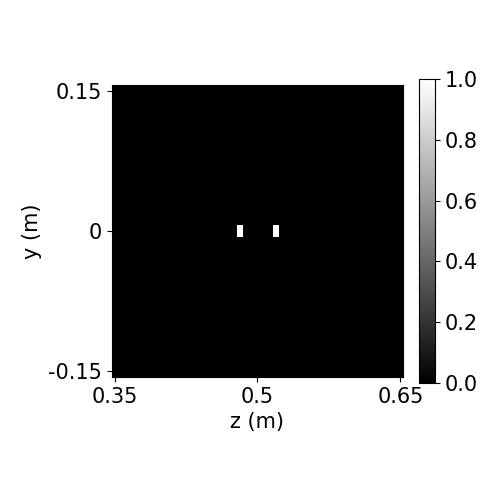} & \addpic{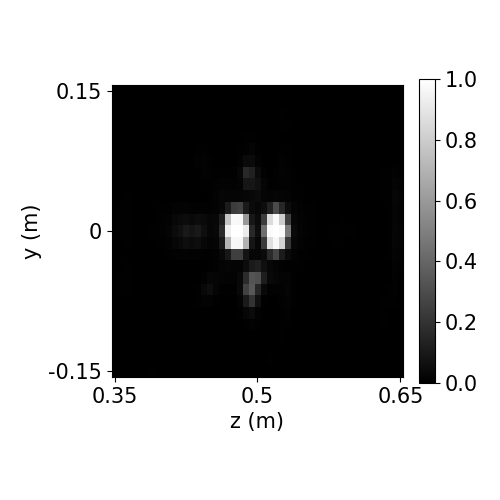}  
\end{tabular}
\caption{Range Resolution Analysis at $x,y=0$cm line.}
\label{fig:MultipleTargetsRange}\end{subfigure}
\caption{Demonstration of resolution using point targets 
at SNR $=$ 30 dB (Number of Frequency Steps: 15). Separation distance along $x$/$y$ directions ($\delta_{x,y}$) and along $z$ direction ($\delta_{z}$) are indicated on the left for each row. Front/side views are shown, which are obtained by maximum projection of the image cube onto the x-y and y-z planes, respectively.}
\label{fig:MultipleTargets}
\end{figure}

In the second analysis, we investigate the resolving capability of our approach by analyzing the reconstructed images for scenes containing multiple point targets. For this, we place 4 point targets onto the middle $z=50$cm plane, and 2 point targets along the z axis, with different separation distances as shown in Fig.~\ref{fig:MultipleTargets}.  The reconstructions obtained for the observation setting with 30 dB SNR and 15 frequency steps are also shown in the same figure together with the maximum value projected 2D plots of these scenes. As seen, our approach manages to resolve targets separated by a distance of nearly 5cm along the cross range directions, and 2.5cm along the range direction. Hence, similar to the conditioning-based analysis, there is almost 2-fold worse resolution in this compressive setting with only $\sim$ 8\% data compared to the expected theoretical resolution of the conventional non-compressive and noise-free setting. However, note that these results are obtained purely based on unrealistic scenes containing point targets, which do not exist in our training dataset that only contains extended 3D targets to mimic the real-world objects. We feel the results of this analysis may change without much change in the achievable resolution for real-world objects by enriching the training dataset with scenes containing point targets.

\subsection{Performance Analysis with Experimental Data}
\begin{figure}[tbh!]
    \centering
\hfill
    \begin{subfigure}[]{0.49\textwidth}
    \centering
    \vspace{1pt}
    \includegraphics[height=5cm]{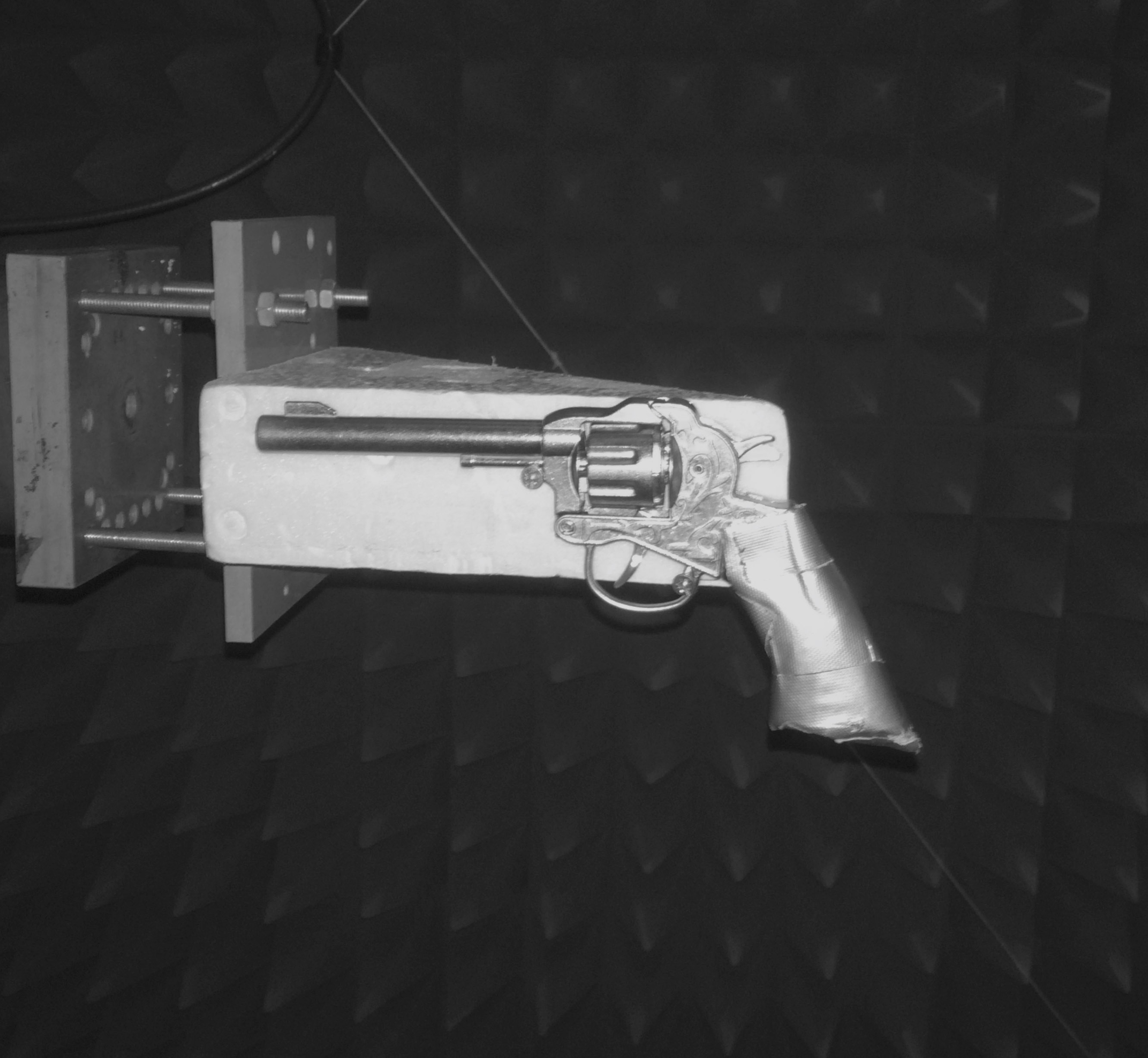}    
    \label{fig:Revolver}
    \caption{Toy Revolver \cite{Wang2020Short-Range}}
    \end{subfigure}
\hfill
\begin{subfigure}[]{0.49\textwidth}
    \includegraphics[height=5cm]{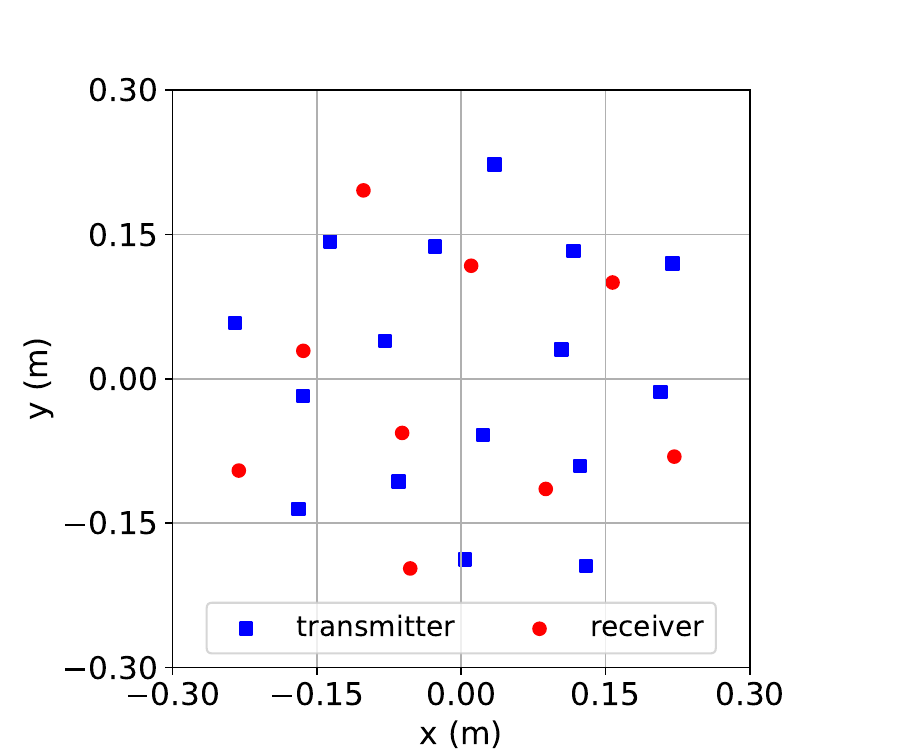}
    \caption{Spiral MIMO Array.}
    \end{subfigure}
\hfill

    \caption{Photograph of the toy revolver \cite{Wang2020Short-Range} and the spiral MIMO array.}
        \label{fig:YarovoyArray}

\end{figure}

After our comprehensive analysis using simulated data, we now demonstrate the performance of the Deep2S approach with experimental measurements that were available online~\cite{Wang2020EMData,Wang2020Short-Range}. These experimental measurements were acquired from a scene containing a toy revolver which was placed approximately 50 cm away from a sparse MIMO array~\cite{Wang2020Short-Range}. The photograph of the scene that contains the toy revolver is given in Fig.~\ref{fig:YarovoyArray} together with the used MIMO array configuration. We aim to infer the reflectivity distribution within an image cube of size 0.3m $\times$ 0.3m $\times$ 0.3m that contains the revolver. Hence the relative position of the antenna plane and the scene are the same as the previous simulated setting. On the other hand, the MIMO array used for these experimental measurements is different from the Mills Cross array used for the simulated data. In this case, the MIMO array consists of 16 transmit and 9 receive Vivaldi antennas which are distributed in a spiral configuration on the 2D antenna plane as shown in Fig.~\ref{fig:YarovoyArray}~\cite{Wang2020Short-Range}. 

For this experimental setting, we choose the voxel size same as the previous simulated setting (i.e. 1.25cm$\times$1.25cm$\times$0.625cm). Then the reflectivity image that we want to infer consists of $25\times25\times49$ voxels in x, y, and z directions, respectively. For measurements, 15 uniform frequency steps between 4 GHz and 16 GHz are used as before.  As a result, the goal is to reconstruct the unknown reflectivity image with $\sim 7\%$ data ($M/N$) in this compressive MIMO imaging case.

We obtain reconstructions with the backprojection method, adjoint operation, Deep2S, Deep2S+ and  CV-Deep2S approaches. We train CV-Deep2S by utilizing the 800 random phase added training images and simulating measurements with the forward model for the used spiral array at 30dB SNR. We also re-train the Deep2S and Deep2S+ models for this measurement setting using transfer learning from the models trained for the cross-array setting. To additionally investigate the performance of the Deep2S with a model mismatch, we use the trained Deep2S model for the cross-array setting without re-training for the spiral array case and call this Deep2S*.

\begin{figure}[tbh!]
    \centering

    \hfill
    \begin{subfigure}[b]{0.14\textwidth}
         \centering
         \vspace{-35pt}
         \textbf{\footnotesize Back-Projection}
         \includegraphics[width=1.2\textwidth]{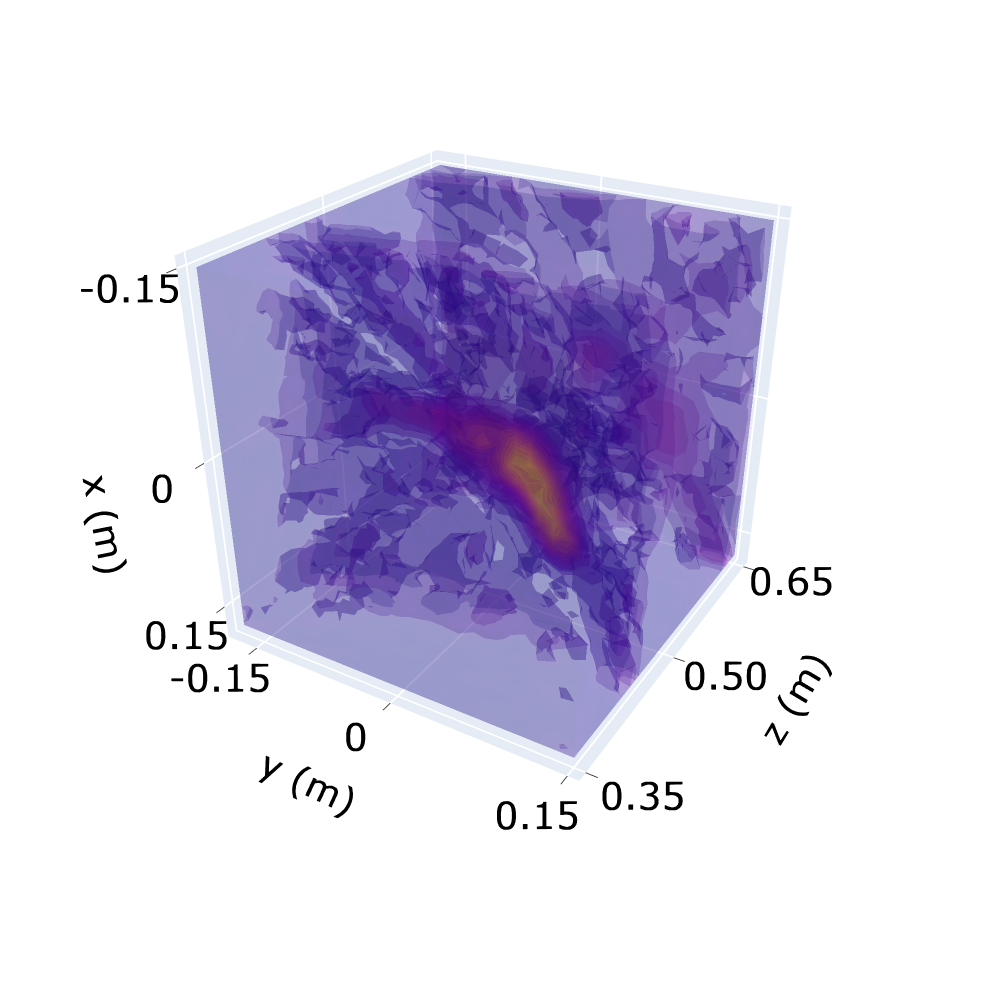}
         \vspace{2.4pt}
     \end{subfigure}     
     \hfill
     \begin{subfigure}[b]{0.14\textwidth}
         \centering
        \vspace{-35pt}
         \textbf{\footnotesize Adjoint}
         \includegraphics[width=1.2\textwidth]{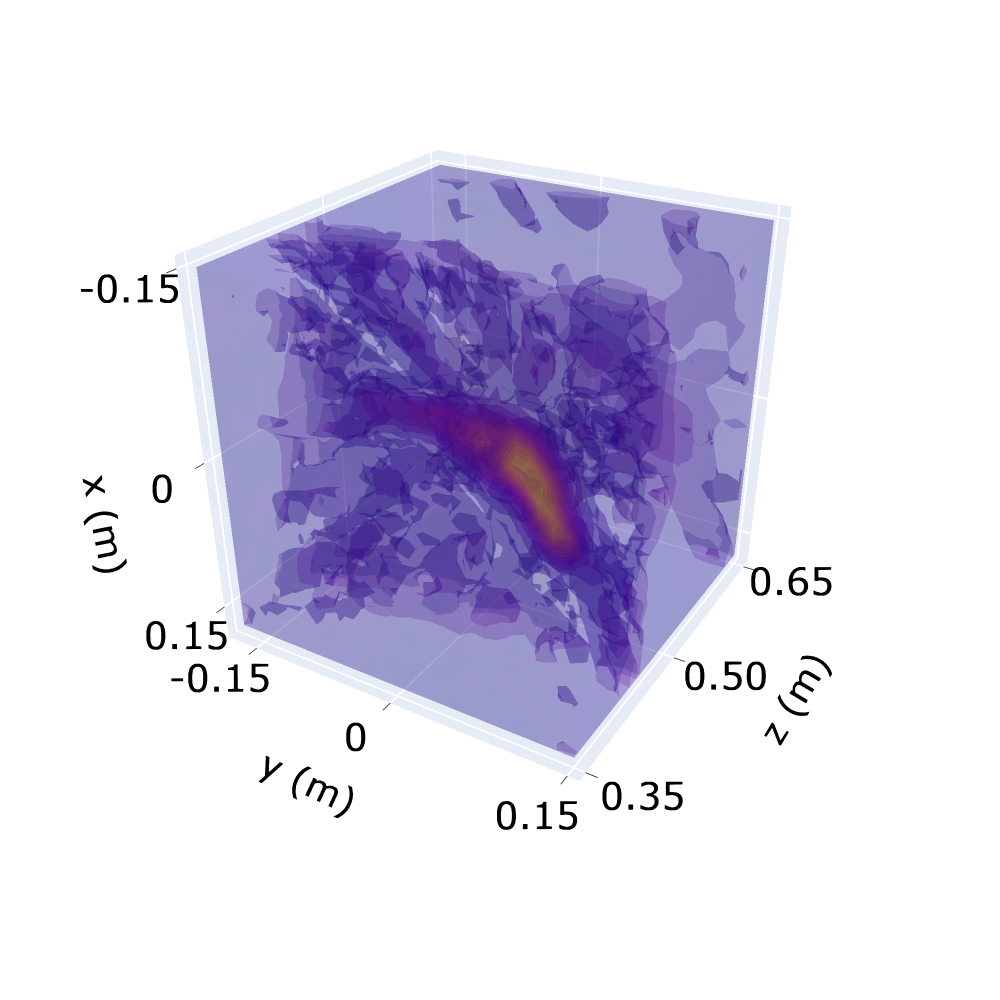}
         \vspace{2.4pt}
     \end{subfigure}     
     \hfill
    \begin{subfigure}[b]{0.14\textwidth}
         \centering
         \textbf{\footnotesize CV-Deep2S}
         \includegraphics[width=1.2\textwidth]{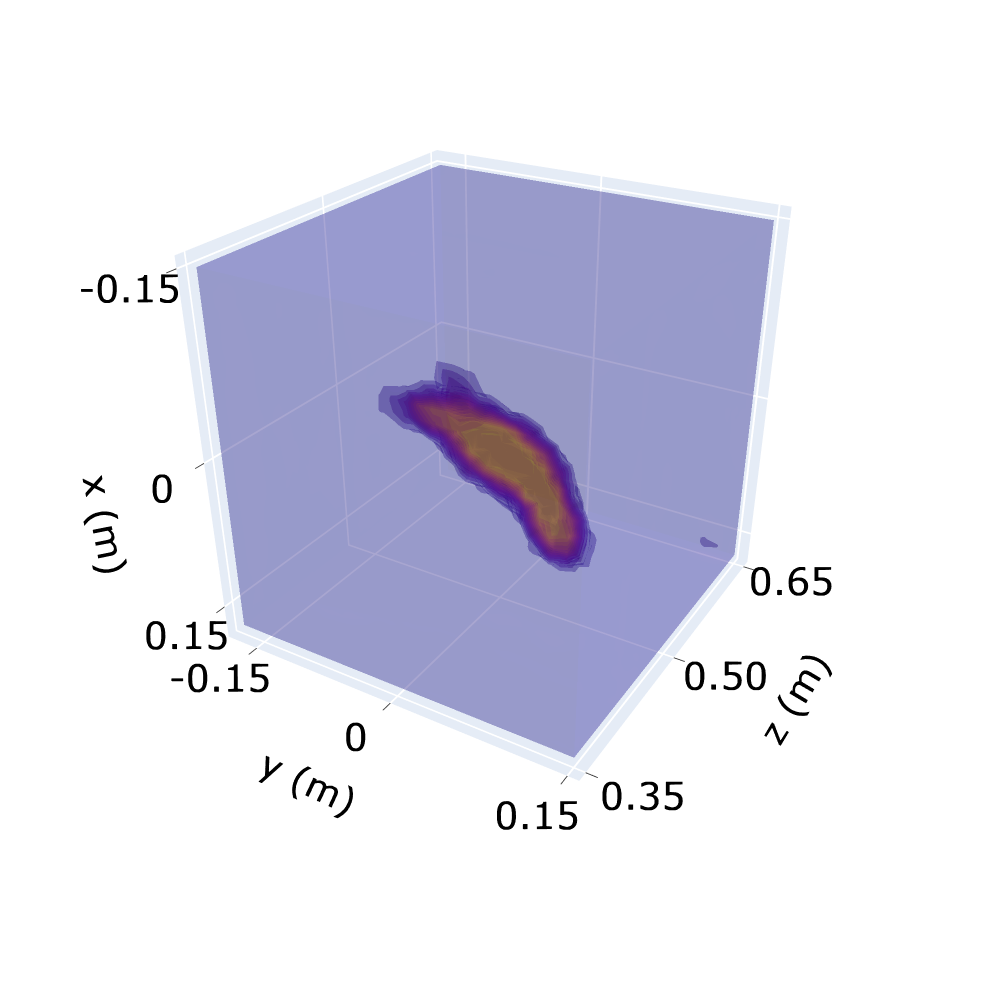}
         \vspace{2.4pt}
     \end{subfigure}     
     \hfill
     \begin{subfigure}[b]{0.14\textwidth}
         \centering
         \textbf{\footnotesize Deep2S*}
         \includegraphics[width=1.2\textwidth]{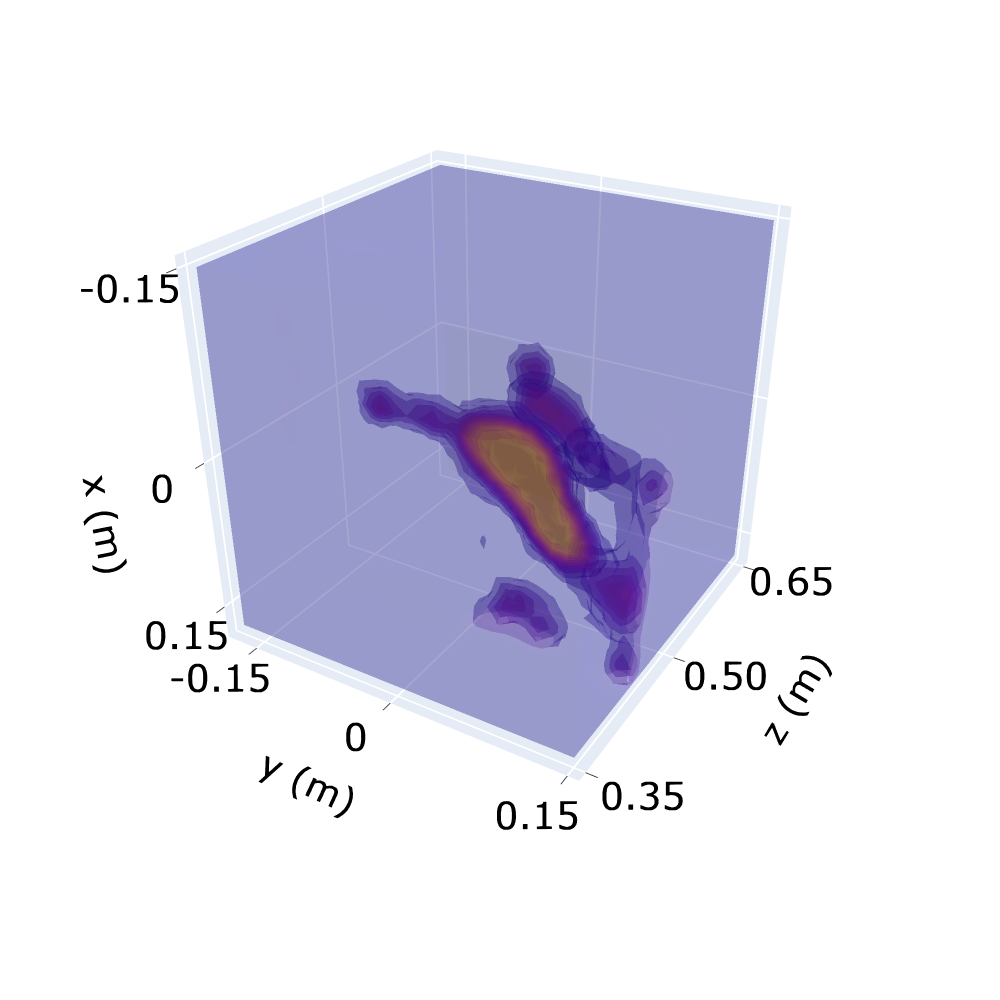}
         \vspace{2.4pt}
     \end{subfigure}     
     \hfill
     \begin{subfigure}[b]{0.14\textwidth}
         \centering
         \textbf{\footnotesize Deep2S}
         \includegraphics[width=1.2\textwidth]{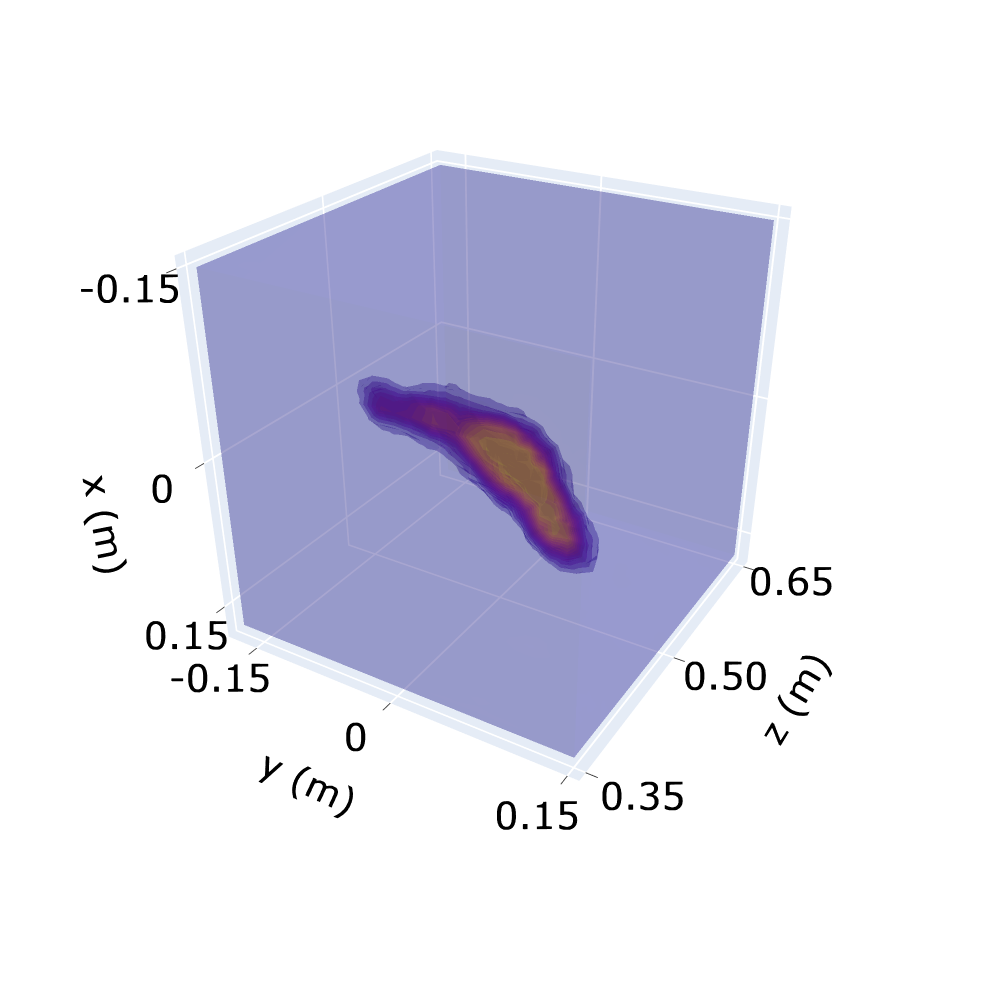}
         \vspace{2.4pt}
     \end{subfigure}  
    \hfill
     \begin{subfigure}[b]{0.14\textwidth}
         \centering
         \textbf{\footnotesize Deep2S+}
         \includegraphics[width=1.2\textwidth]{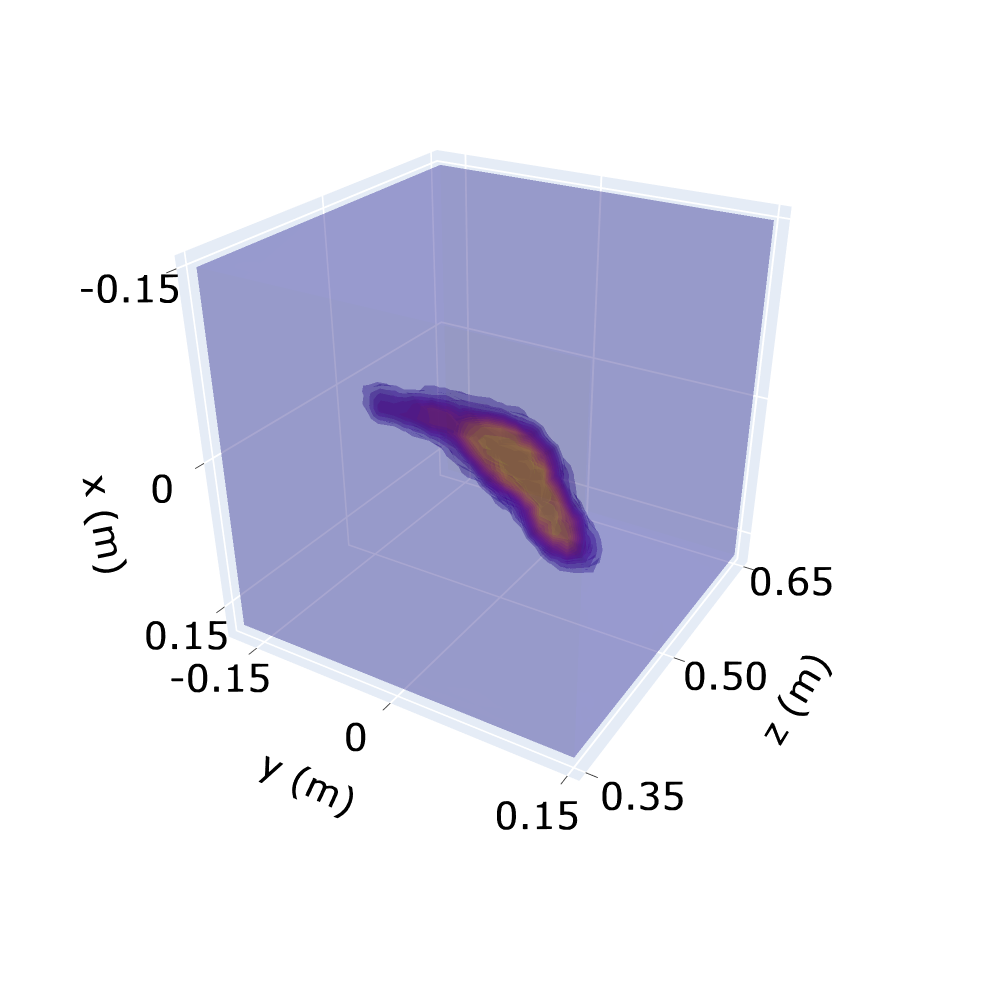}
         \vspace{2.4pt}
     \end{subfigure}  
     \hfill 
     
    \hfill
    \begin{subfigure}[b]{0.14\textwidth}
         \centering
         \vspace{-25pt}
         \includegraphics[width=1\textwidth]{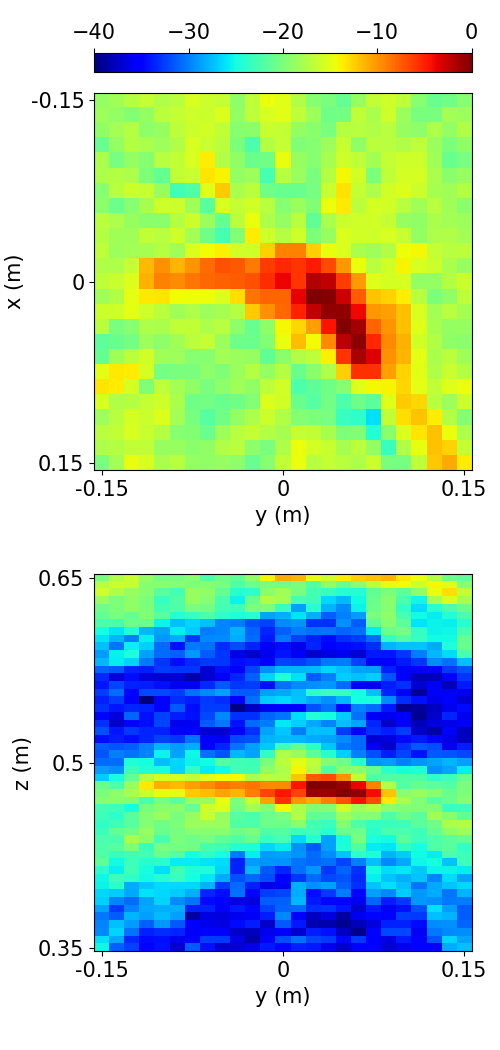}
         \vspace{2.4pt}
     \end{subfigure}     
     \hfill
     \begin{subfigure}[b]{0.14\textwidth}
         \centering
         \vspace{-25pt}
         \includegraphics[width=\textwidth]{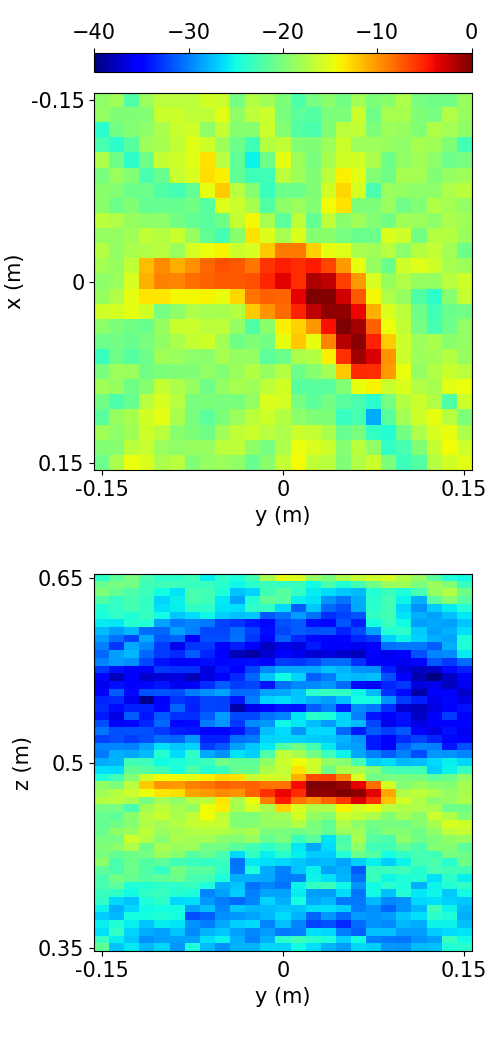}
         \vspace{2.4pt}
     \end{subfigure}     
     \hfill
     \begin{subfigure}[b]{0.14\textwidth}
         \centering
         \vspace{-25pt}
         \includegraphics[width=\textwidth]{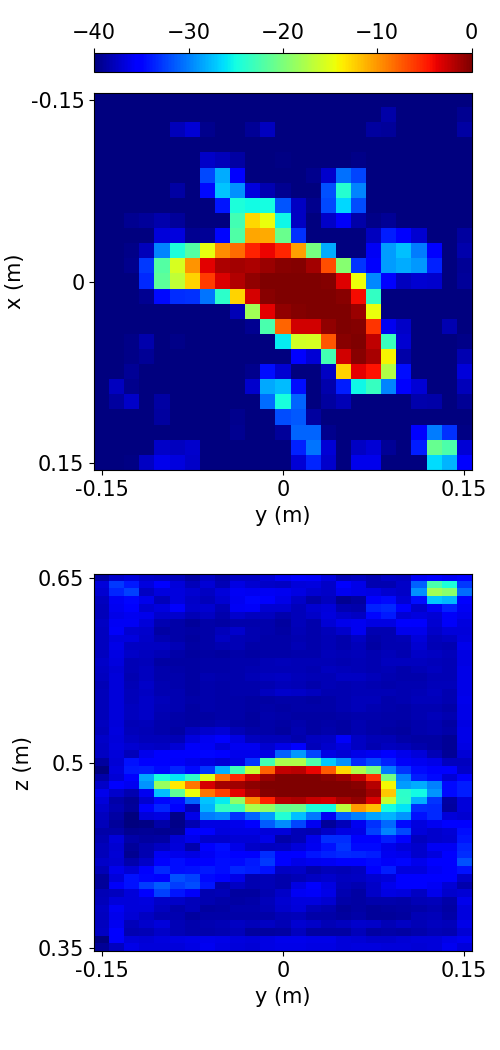}
         \vspace{2.4pt}
     \end{subfigure}     
     \hfill
     \begin{subfigure}[b]{0.14\textwidth}
         \centering
         \vspace{-25pt}
         \includegraphics[width=\textwidth]{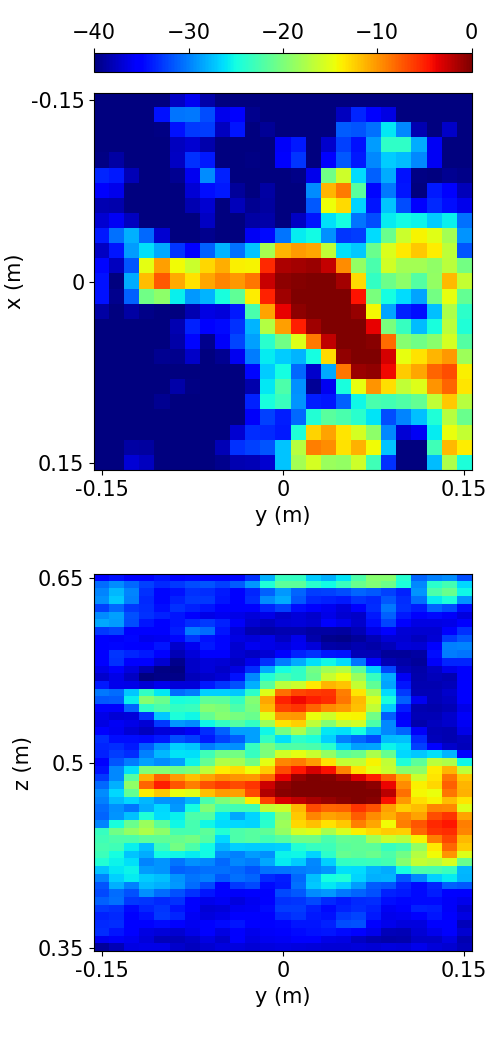}
         \vspace{2.4pt}
     \end{subfigure}
         \hfill
     \begin{subfigure}[b]{0.14\textwidth}
         \centering
         \vspace{-25pt}
         \includegraphics[width=\textwidth]{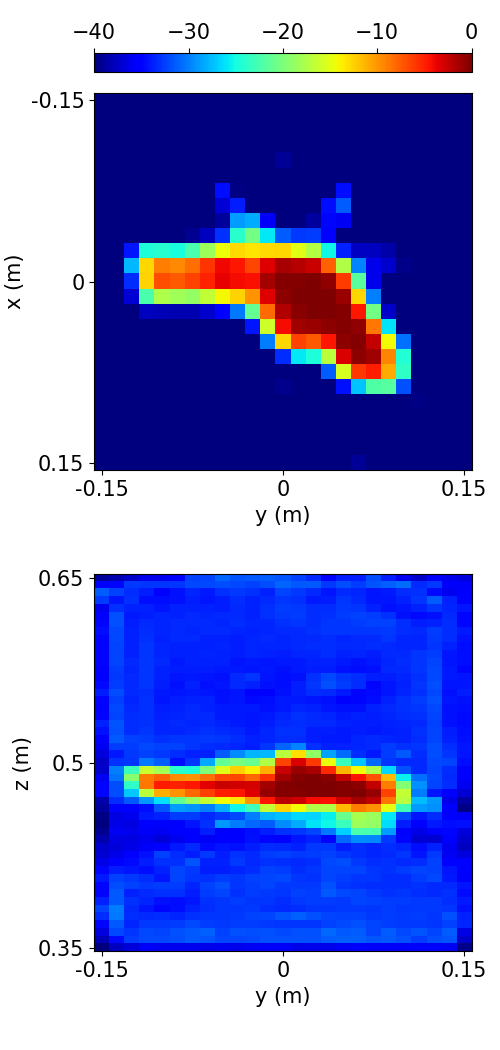}
         \vspace{2.4pt}
     \end{subfigure}
    \hfill
     \begin{subfigure}[b]{0.14\textwidth}
         \centering
         \vspace{-25pt}
         \includegraphics[width=\textwidth]{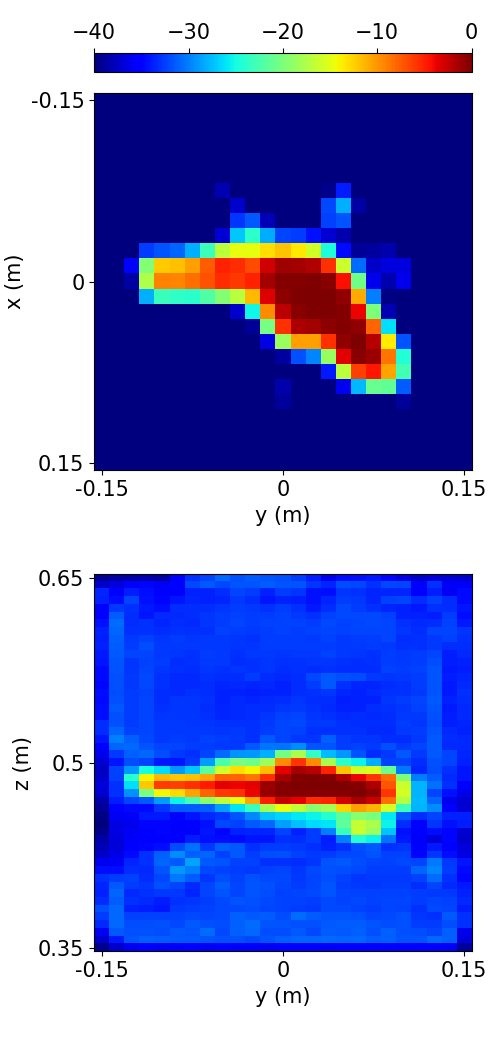}
         \vspace{2.4pt}
     \hfill

     \end{subfigure}
     
\caption{Reconstructed reflectivity magnitudes using 15 frequency steps. Deep2S* shows the reconstruction with model mismatch, i.e. the Deep2S model that utilizes the 3D U-Net trained on the Mill's Cross array. For all reconstructions, the first row provides the 3D view of the image cube in linear scale whereas the second and third rows show front/side views in dB scale obtained by maximum projection of the image cube onto the $x-y$ and $y-z$ planes, respectively. (3D rotating views of these reconstructions can be found at \href{https://github.com/METU-SPACE-Lab/Efficient-Learned-3D-Near-Field-MIMO-Imaging}{https://github.com/METU-SPACE-Lab/Efficient-Learned-3D-Near-Field-MIMO-Imaging} as video.)}
\label{fig:YarovoyResults}
\end{figure}

The experimental results are shown in Fig.~\ref{fig:YarovoyResults}. As seen, the reconstructions of backprojection and adjoint contain widespread volume artifacts. The artifacts distributed over the cross-range dimensions are mainly due to under-sampling with the sparse MIMO array, which is related to spatial aliasing~\cite{Wang2020Short-Range}. More importantly, both backprojection and adjoint results contain significant artifacts along the range direction (especially around the $z=0.65$m and $z=0.55$m planes), which is mainly related to sparsely sampled 4-16 GHz frequency band. Compared to the backprojection result, these artifacts are slightly less in the adjoint result, which is similar with our observation for the simulated setting. On the other hand, all of the deep learning based approaches substantially reduce these widespread artifacts present in the adjoint reconstruction.

In particular, with Deep2S, the reconstructed revolver is clearly separated from the scene background as desired. It provides a clean and sharp image with a very high dynamic range and contrast, and hence yields the highest imaging quality among the tested methods. Compared to the adjoint operation and Deep2S* involving model mismatch, both the side-lobe artifacts and range artifacts are substantially reduced through re-training for the spiral array setting with simulated data. This illustrates that Deep2S has some flexibility to be used in experimental settings that are different from the trained one; but when the amount of artifacts in the adjoint result significantly changes compared to the trained setting, it benefits from re-training for the new observation setting, as expected. 
Moreover, compared to CV-Deep2S, Deep2S appears to better preserve the shape of the revolver and also yields less artifacts. Hence similar to the simulated case, we again observe with experimental data that directly processing the complex-valued intermediate reconstructions does not provide improvement over only processing the magnitudes, and even seems to reduce the generalizability of the approach. Lastly, although Deep2S+ reconstruction is nearly identical to that of Deep2S, similar to the simulated setting, we also provide its reconstruction for completeness.

\section{Conclusions} 
\label{section:Conclusions}
In this paper, we have developed three novel deep learning-based reconstruction approaches for 3D near-field MIMO radar imaging. The main motivation was to reconstruct the 3D reflectivity magnitude of the scene with high image quality and low computational cost so that it can be used in real-time imaging applications. For this reason, we focused on learned direct reconstruction techniques due to their feedforward (non-iterative) nature. We structured the developed methods to consist of two stages. The first stage maps complex-valued measurements to the 3D image domain. Resulting intermediate reconstruction is then refined in the second stage by a DNN to obtain the final reflectivity magnitude of the scene. Here we process only the magnitudes of the intermediate reconstructions due to the random phase nature of the reflectivities in various applications. It must also be noted that in real practical scenarios we are interested in imaging 3D extended targets which generally have correlations along both range and cross-range directions. Unlike the 2D architectures in the literature, the DNNs used in the second stage of the developed approaches utilize 3D convolutional layers to jointly exploit correlations along range and cross-range directions, which is particularly useful for reconstructing extended 3D targets. Our approaches also use less amount of data for network training compared to the other approaches thanks to the exploited U-Net architecture involving multi-resolution processing.

From the developed approaches, the Deep2S method exploits the physics-based knowledge by utilizing the adjoint of the forward model in its first stage. The adjoint operation has the benefit of fast computation due to its non-iterative nature. For comparison, the DeepDI approach is also developed which replaces the physics-based first stage with a fully connected neural network. In this two-stage structure, the reconstruction is
performed directly from the radar measurements using only neural networks, and the observation
model is not used. Moreover, to investigate the effect of physics-based initialization on the training of a purely learning-based method, we also developed another approach, Deep2S+, that replaces the adjoint operation in the first stage with a fully connected layer to perform multiplication with a trainable matrix which is learned through transfer learning from the adjoint matrix. All adjustable parameters of the developed approaches are learned end-to-end, which avoids the difficulties of parameter tuning that exist in sparsity-driven CS recovery methods. 

We evaluate the performance of the developed methods comparatively with the commonly used approaches using both simulated and experimental data. As expected, pure DNN-based DeepDI approach was outperformed by the physics-aware Deep2S and Deep2S+ methods. This result is in accordance with the existing literature which states the significance of incorporating physics-based knowledge into DNN-based reconstruction whenever possible; otherwise, a substantial volume of data is required for training purely DNN-based approaches. Our comparative results also demonstrate that processing the intermediate reconstructions in complex-valued form (instead of magnitude) does not yield an improvement on the performance. It has been observed that both Deep2S and Deep2S+ methods are capable of providing high-quality reconstructions even with highly under-sampled data while enabling fast reconstruction, which show promise for real-time compressive imaging applications. 

In order to train the proposed deep learning-based methods, we generated a large synthetic dataset consisting of 3D extended targets. Naturally, a real-world dataset would be better for training. However, a large dataset is not available in many radar imaging applications including near-field MIMO radar imaging. Our approach successfully generates different 3D distributed objects that spread within the reconstruction cube from a randomly chosen center. We also add random phase to our synthetically generated dataset to take into account the complex-valued and random phase
nature of scene reflectivities, which have been mostly neglected in the earlier works. Nevertheless the performance of the developed methods on experimental data might be further 
improved by extending this synthetic dataset generation procedure to obtain more realistic and task-oriented training data. 

Lastly we note that although the developed Deep2S method is quite fast with a runtime on the order of milliseconds on a standard computer, the adjoint computation in its first stage can be further accelerated using Fourier-based methods~\cite{miran2021sparse,marks2017fourier}. Moreover, exploring the performance of the developed methods with other types of 3D network architectures (such as those mimicking transforms) may improve the reconstruction quality and is a topic for future study. 
Additionally, although plug-and-play and unrolling-based deep-learning reconstruction methods are generally difficult to use for real-time computational imaging requiring large-scale 3D reconstruction, it would still be interesting to study these type of methods for near-field MIMO radar imaging since they can offer higher image quality at the cost of longer reconstruction time. Developing such methods for 2D/3D complex-valued radar imaging problems is subject to various difficulties depending on the specific radar imaging problem at hand~\cite{alver2021plug}. For near-field MIMO radar imaging, these difficulties include handling randomness in the phase of the reconstructed images, large-scaleness of the 3D reconstruction problem, substantial requirements for memory usage, computational complexity as well as training time due to computationally intensive computations for the forward and adjoint operators at every iteration. But if developed, comparing their performance with the proposed deep learning-based direct reconstruction methods in this paper could reveal the trade-off between the imaging quality and the reconstruction time, and the applicability of such methods to real-time applications.

\section{Acknowledgments}
This work was supported by the Scientific and Technological Research Council of Turkey (TUBITAK) under Grant 120E505 (1001 Research Program).

\end{document}